\documentclass[]{article}

\usepackage{graphicx}
\usepackage{float}
\usepackage{url}
\usepackage{authblk}

\title{\textbf{Expected Sensitivity to Invisible Higgs Boson Decays at the ILC with the SiD Detector}}
\author[1,2]{Chris Potter}
\author[3]{Amanda Steinhebel}
\author[1,2]{Jim Brau}
\author[4]{Austin Pryor}
\author[4]{Andy White}
\affil[1]{Department of Physics, University of Oregon}
\affil[2]{Institute for Fundamental Science, University of Oregon}
\affil[3]{NASA Goddard Space Flight Center}
\affil[4]{Department of Physics, University of Texas at Arlington}

\date{\today}

\begin{document}

\maketitle

\begin{abstract}
In the Standard Model (SM) of particle physics, the branching ratio for Higgs boson decays to a final state which is invisible to collider detectors, $H \rightarrow ZZ^{\star} \rightarrow \nu \bar{\nu} \nu \bar{\nu}$, is order 0.10\%. In theories beyond the SM (BSM), this branching ratio can be enhanced by decays to undiscovered particles like dark matter (DM). At the Large Hadron Collider (LHC), the current best upper limit on the branching ratio of invisible Higgs boson decays is 11\% at 95\% confidence level. We investigate the expected sensitivity to invisible Higgs decays with the Silicon Detector (SiD) at the International Linear Collider (ILC). We conclude that at $\sqrt{s}=250$~GeV with 900 fb$^{-1}$ integrated luminosity each for $e_{L}^-e_{R}^+$ and $e_{R}^-e_{L}^+$ at nominal beam polarization fractions, the expected upper limit is 0.16\% at 95\% confidence level. 
\end{abstract}

\section{Introduction}

\subsection{Invisible Higgs at the ILC}

The discovery of the Higgs boson in 2021 at the Large Hadron Collider (LHC) \cite{Aad:2012tfa,Chatrchyan:2012ufa} provides a new window into particle physics. The International Linear Collider (ILC) \cite{Behnke:2013xla,Baer:2013cma,Phinney:2007gp,Behnke:2013lya} is an $e^+e^-$ collider proposed by the international community to exploit that new window with precision measurements of the Higgs boson properties. Current community planning efforts in the field of particle physics include the ILC as a viable and potentially richly rewarding next step for the field \cite{fujii2021ilc,Adachi:2022zmw}.

In the Standard Model (SM) the properties of this particle are predicted with high precision. Any measured deviation from these properties suggests new physics beyond the SM (BSM). Collider \emph{invisible} decay is decay to particles which do not interact with the detector material and are either stable, or unstable but decay outside the effective sensitive volume of the detector. Thus some examples of invisible Higgs boson decays are Higgs to dark matter (DM) particles, Higgs to unknown longlived particles (LLP), and Higgs to neutrinos. In the SM the branching ratio of $H \rightarrow ZZ^{\star} \rightarrow \nu \bar{\nu} \nu \bar{\nu}$ is approximately 0.10\%. Thus the SM invisible Higgs decay is out of range for the LHC, which expects even at the high luminosity LHC (HL-LHC) to reach an upper limit of 2.5\% at 95\% confidence level \cite{Cepeda:2650162}. The current limits from the LHC are 11\% (18\%) at 95\% confidence level from ATLAS (CMS) \cite{ATLAS-CONF-2020-052,CMS:2022qva}. The ILC may improve on the HL-LHC expected limit by an order of magnitude or more \cite{kato2020probing}.

In the Higgstrahlung process $e^+e^- \rightarrow ZH$ with invisible Higgs boson decay, several channels are defined by the $Z$ decay. In the hadron channel $Z \rightarrow q\bar{q}$, accounting for 70\% of signal events, producing missing energy and jets with high particle multiplicity after hadronization of the quarks. In the electron channel $Z \rightarrow e^+ e^-$, 3.4\% of signal events, the signature is missing energy and an $e^+e^-$ pair reconstructing to the $Z$ mass. In the muon channel $Z \rightarrow \mu^+ \mu^-$, also 3.4\% of signal events, the signature is missing energy and a $\mu^+\mu^-$ pair reconstructing to the $Z$ mass. The tau channel $Z \rightarrow \tau^+ \tau^-$ accounts for 3.4\% of signal events but is not considered here. The neutrino channels $Z \rightarrow \nu_{e} \bar{\nu}_{e},\nu_{\mu} \bar{\nu}_{\mu},\nu_{\tau} \bar{\nu}_{\tau},$ account for 20\% of signal events and are also not considered here. Below, the \emph{lepton} channel refers only to the electron and muon channels.

\subsection{ILC Beams and SiD Detector}

\begin{table}[t]
\begin{center}
\begin{tabular}{|c|c|c|} \hline
Process  & $\sigma_{LR}$ [pb]  & $\sigma_{RL}$ [pb]\\ \hline \hline
$e^+ e^- \rightarrow WW$ & 37.5 & 2.58\\
$e^+ e^- \rightarrow e^{\pm} \nu W^{\mp}$ & 10.2  & 1.09  \\
 $e^+ e^- \rightarrow e^+ e^- Z$ & 3.17  & 2.00  \\
 $e^+ e^- \rightarrow ZZ$ & 1.80  & 0.827 \\
$e^+ e^- \rightarrow \nu \bar{\nu} Z$ & 0.220  & 0.013  \\ \hline
 $e^+ e^- \rightarrow ZH$ & 0.313  & 0.211 \\ \hline
\end{tabular}
\caption{Cross sections for signal and background processes at $\sqrt{s}=250$~GeV. Electron beams are 80\% polarized and positron beams are 30\% polarized. The effect of initial state radiation (ISR) is included but beamstrahlung is not. Obtained with Whizard 2.6.4 \cite{Kilian:2007gr}.}
\label{tab:ilcxsec}
\end{center}
\end{table}

The ILC design is detailed comprehensively in the ILC Technical Design Report (TDR) Volume 3 \cite{Phinney:2007gp}. Electron and positron beams are accelerated to high energy in linacs made up of superconducting RF cavities. The nominal center-of-mass energy in the TDR is $\sqrt{s}=500$~GeV, but this reverts to $\sqrt{s}=250$~GeV in the ILC Machine Staging Report \cite{Evans:2017rvt} with possible upgrade to $\sqrt{s}=500$~GeV by extension of the linacs. The maximum cross section for the Higgstrahlung process $e^+ e^- \rightarrow ZH$ occurs near $\sqrt{s}=250$~GeV. See Table \ref{tab:ilcxsec} for the cross sections of processes relevant to this study.

One important design feature of the ILC is the ability to produce polarized electron and positron beams. The composition of beams with fraction $P_{e^-}$ electron polarization and $P_{e^+}$ positron polarization is $\frac{1}{4}(1\mp P_{e^-})(1 \pm P_{e^+})$ for opposite polarization cases $e_{L}^{-}e_{R}^{+}$ and $e_{R}^{-}e_{L}^{+}$, and $\frac{1}{4}(1\pm P_{e^-})(1 \pm P_{e^+})$ for same polarization cases $e_{R}^{-}e_{R}^{+}$ and $e_{L}^{-}e_{L}^{+}$. The nominal assumption for polarization fraction in the TDR is $P_{e^-}=$80\% polarized electrons and $P_{e^+}=$30\% polarized positrons, though it is hoped that the positron polarization fraction can be made higher.

The ILC instantaneous luminosity at $\sqrt{s}=250$~GeV will depend critically on the beam parameters, but $L=1.8 \times 10^{34}$~cm$^{-2}$s$^{-1}$ is expected. Beamstrahlung, the radiation of photons from electrons or positrons in one colliding bunch due to the field produced by the oncoming colliding bunch, also depends critically on the beam parameters. The luminosity sharing between polarization cases in the staging report is assumed to split equally between $e_{L}^{-}e_{R}^{+}$ and $e_{R}^{-}e_{L}^{+}$ for an integrated luminosity of 900~fb$^{-1}$ each, with 100~fb$^{-1}$ each for $e_{R}^{-}e_{R}^{+}$ and $e_{L}^{-}e_{L}^{+}$. The total integrated luminosity in this scenario, which we assume for this study, is then $\int dt \mathcal{L}=$2ab$^{-1}$.

The Silicon Detector (SiD), one of two detectors proposed for the ILC, is described in detail in the ILC TDR Volume 4 \cite{Behnke:2013lya}. The other detector is the International Large Detector (ILD), also documented in \cite{Behnke:2013lya}. The SiD barrel comprises a five-layer Silicon pixel Vertex Detector, a five-layer Silicon strip Tracker, a dodecahedral electromagnetic calorimeter (ECal) with Lead absorber layers and 20 (20) sensitive thin (thick) layers of sensitive Tungsten, a hadronic calorimeter (HCal) with 11 Steel absorber layers and scintillator sensitive layers, a 5T solenoid, and a dodecahedral muon detector of Iron layers alternating with scintillator sensitive layers. The barrel radii for these subdetectors are $r_{in}=$1.4, 21.7,126.5,141.7,259.1, and 340.2 cm. The barrel is capped by endcaps with similar subdetector technology.

SiD was designed to take advantage of the particle flow technique for particle identification. Tracks reconstructed in the Vertex Detector and Tracker are extrapolated through the magnetic field produced by the solenoid to the ECal and the HCal and associated to nearby calorimeter clusters. Those tracks associated to an ECal cluster are assumed to be electrons while those associated to an HCal cluster are assumed to be charged hadrons, and those matching to hits in the muon detector are assumed to be muons. Clusters in the ECal unassociated to a track are assumed to be photons while clusters in the HCal unassociated to a track are assumed to be neutral hadrons.

\subsection{Signal and Background Simulation}

The event generation, full simulation of the SiD detector, and object reconstruction for the simulated data samples in this study are documented in \cite{Potter:2021dkr}, but are briefly summarized below. See Appendix A for a complete list of generator samples and brief descriptions of how they have been used in this study.

The signal samples are generated with Whizard 2.6.4 with Pythia6 for hadronization and decay. The Higgstrahlung process $e^+e^- \rightarrow ZH$ with fully inclusive $Z$ decays and SM invisible Higgs decay $H \rightarrow ZZ^{\star} \rightarrow \nu \bar{\nu} \nu \bar{\nu}$ is specified. The beams are polarized according the nominal fractions for the ILC, and initial state radiation is turned on. Beamstrahlung is not.

Background samples were generated with Whizard 1.4 during the Detailed Baseline Design (DBD) exercise, which was incorporated into and described in the TDR. These represent a full set of SM backgrounds with pure polarized beams, which can be mixed to reproduce any required polarization fractions. In the \texttt{all\_SM\_background} samples produced by SiD, they were mixed weighted by cross section and the required polarization fractions for the nominal ILC design.

\begin{table}[t]
\begin{center}
\begin{tabular}{|c|c|c|c|} \hline
Process & Intermediate States & $N_{ev}/\sigma_{LR}$ [fb$^{-1}$] & $N_{ev}/\sigma_{RL}$ [fb$^{-1}$] \\ \hline \hline

$e^+ e^- \rightarrow e^+ e^- \nu \bar{\nu}$ & $eeZ,\nu\nu Z,e\nu W, ZZ, WW$ & 1000 & 1000 \\
$e^+ e^- \rightarrow \mu^+ \mu^- \nu \bar{\nu}$ & $\nu\nu Z, ZZ, WW$ & 1000 & 1000  \\
$e^+ e^- \rightarrow q \bar{q} \nu \bar{\nu}$ & $\nu \nu Z, ZZ$  & 1000 & 1000 \\
$e^+ e^- \rightarrow q \bar{q}^{\prime} e \bar{\nu}$ & $e\nu W, WW$ & 100 & 800\\
$e^+ e^- \rightarrow q \bar{q}^{\prime} \mu/\tau \bar{\nu}$ & $WW$ & 100 & 800 \\ 
$e \gamma \rightarrow e q \bar{q}, \nu q \bar{q}^{\prime}$ & $eZ,\nu W$  & 1800 & 2000 \\ \hline
$e^+ e^- \rightarrow f\bar{f} \nu \bar{\nu} \nu \bar{\nu}$ & ZH  & 10000 & 10000 \\ \hline
\end{tabular}
\caption{Signal and background samples generated at $\sqrt{s}=250$~GeV for this study and their equivalent integrated luminosities. All samples are normalized to 900 fb$^{-1}$ for analysis.  See the text for the generator level requirements. A sum over lepton neutrino flavors is implied if allowed.}
\label{tab:mcsamples}
\end{center}
\end{table}

For the final sensitivity evaluation, however, dedicated background samples were newly generated with Whizard 2.6.4 with the same conditions as with signal: beam polarization and ISR but no beamstrahlung. Broadly, the backgrounds have final states $e^+e^- \nu \bar{\nu}$ (electron channel), $\mu^+ \mu^- \nu \bar{\nu}$ (muon channel), and $q\bar{q} \nu \bar{\nu}, q\bar{q}^{\prime} \ell \nu, eq\bar{q}, \nu q\bar{q}^{\prime}$ (hadron channel).  Requirements are imposed on the $Z$ and $H$ candidate masses and, for the three-fermion processes, the $p_T$ of the $Z$ candidate:
\begin{itemize}

\item candidate $Z$ mass: $60 \leq m_{f\bar{f}} \leq 120$~GeV (4f $e^+e^- \rightarrow e^+ e^-\nu\bar{\nu}, \mu^+ \mu^- \nu \bar{\nu}, q\bar{q} \nu \bar{\nu}$)

\item candidate $H$ mass: $90 \leq m_{\nu \bar{\nu}} \leq 170$~GeV (4f $e^+e^- \rightarrow e^+ e^-\nu\bar{\nu}, \mu^+ \mu^- \nu \bar{\nu}, q\bar{q} \nu \bar{\nu}$)

\item candidate $Z$ $p_T$: $20 \leq p_{T}^{q\bar{q}} \leq 60$~GeV (3f $e^+ e^- \rightarrow eq\bar{q},\nu q\bar{q}^{\prime}$)

\end{itemize}

\noindent  For the $e^+ e^- \nu \bar{\nu}$ samples the $m_{f\bar{f}}$ requirement is tightened by 10~GeV, and for the three-fermion processes it is tightened by 15~GeV. For the $e^+ e^- \nu \bar{\nu}$ samples the $m_{\nu \bar{\nu}}$ requirement is tightened by 10~GeV. See Table \ref{tab:mcsamples} for a summary of these samples.

Signal and background samples were then fully simulated in ILCSoft v02-00-02 using the compact SiD description \texttt{SiD\_o2\_v03.xml} and reconstructed with Marlin and PandoraPFA for particle flow.

\section{Cut-Based Analysis}

\begin{figure}[t]
\begin{center}
\includegraphics[width=6.in]{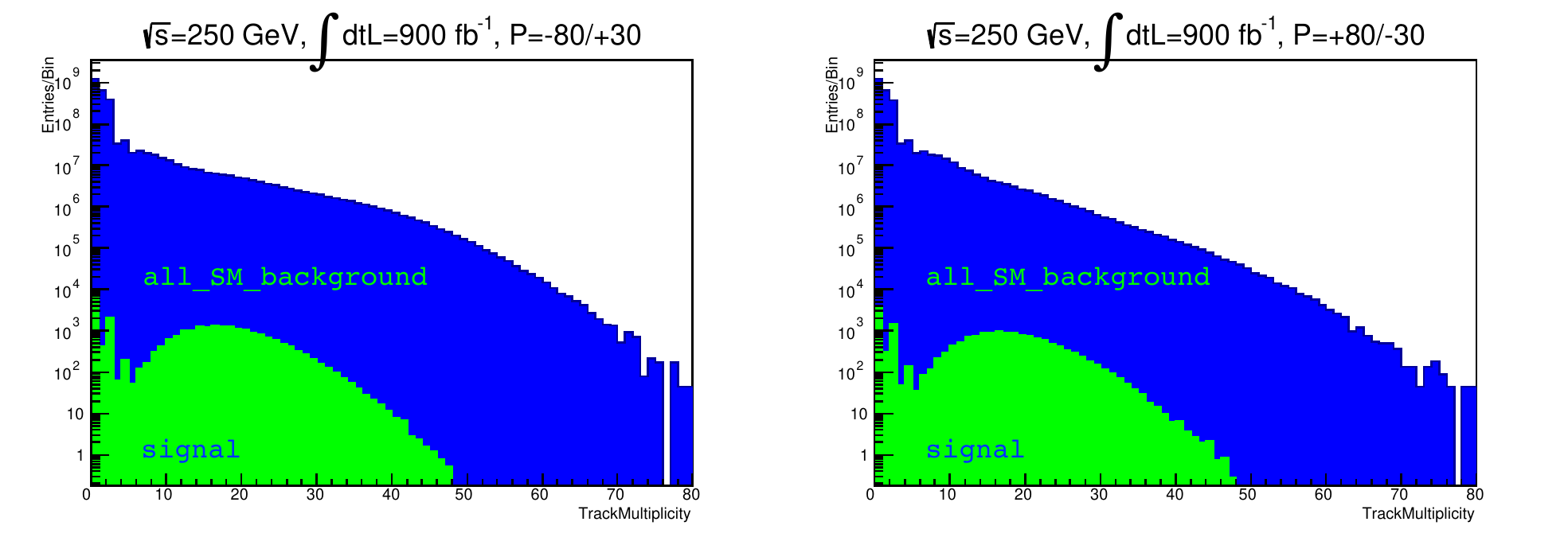}
\end{center}
\caption{Track multiplicity before any selection requirements are imposed. Signal selection requires $N_{trk}=2$ ($6 \leq N_{trk} \leq 32$) for the lepton (hadron) channels. Background \texttt{all\_SM\_background} (blue) is stacked on top of signal (green). The signal branching ratio is assumed to be 10\%. The integrated luminosity is 900 fb$^{-1}$ for each polarization case.}
\label{fig:trkmult}
\end{figure}

\subsection{Lepton Channel}

For the electron and muon channels the signal signature is similar and therefore the signal selection is similar. An $e^+e^-$ or $\mu^+ \mu^-$ pair is selected and required to be consistent with the signal $Z$ decay opening angle, momentum and mass. For both cases the momentum is measured from the tracking rather than the calorimetry or muon detector. The mass in recoil from the candidate $Z \rightarrow \ell^+ \ell^-$,

\begin{equation}
m_{rec}^2 =  s-2\sqrt{s}E_{\ell^+ \ell^-}-m_{Z}^2 \\
\end{equation}

\noindent must be consistent with the Higgs boson mass. The lepton channel signal selection is as follows:

\begin{itemize}

\item Exactly two reconstructed tracks and exactly two PFO leptons $\ell=e,\mu$. ($N_{trk}=N_{\ell}=2$)

\item Lepton pair signal consistency: 

\begin{itemize}

\item same flavor, opposite sign leptons  ($N_{e}=2$ or $N_{\mu}=2$ and  $q^{1}_{trk}+q^{2}_{trk}=0$)

\item separation consistent with production from signal $Z$ decay. ($-0.9 \leq \cos \theta_{\ell^+ \ell^-} \leq -0.2$)

\end{itemize}

\item Transverse momentum of $Z \rightarrow \ell^+ \ell^-$ candidate consistent with signal. ($20 \leq p_T^{vis} \leq 70$~GeV)

\item Mass of $Z \rightarrow \ell^+ \ell^-$ candidate consistent with $Z$ mass. ($75 \leq m_{vis} \leq 105$~GeV)

\item Recoil mass consistent with Higgs boson mass. ($110 \leq m_{rec} \leq 150$~GeV)

\end{itemize}

\begin{figure}[t]
\begin{center}
\includegraphics[width=6.in]{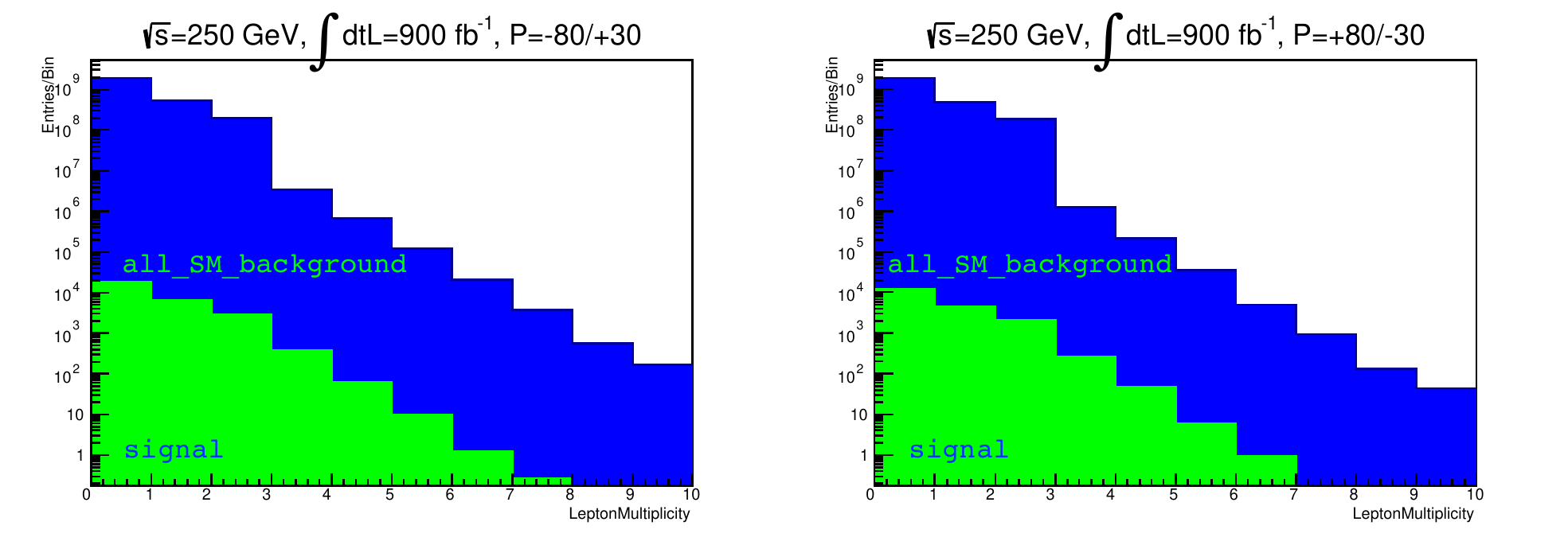}
\end{center}
\caption{Lepton  multiplicity $N_{\ell}=N_{e}+N_{\mu}$ before any selection requirements are imposed. Signal selection requires $N_{\ell}=2$ ($N_{\ell}=0$) for the lepton (hadron) channels. Background \texttt{all\_SM\_background} (blue) is stacked on top of signal (green). The signal branching ratio is assumed to be 10\%. The integrated luminosity is 900 fb$^{-1}$ for each polarization case.}
\label{fig:lepmult}
\end{figure}

\begin{figure}[p]
\begin{center}
\framebox{\textbf{Electron Channel}}

\includegraphics[width=6.in]{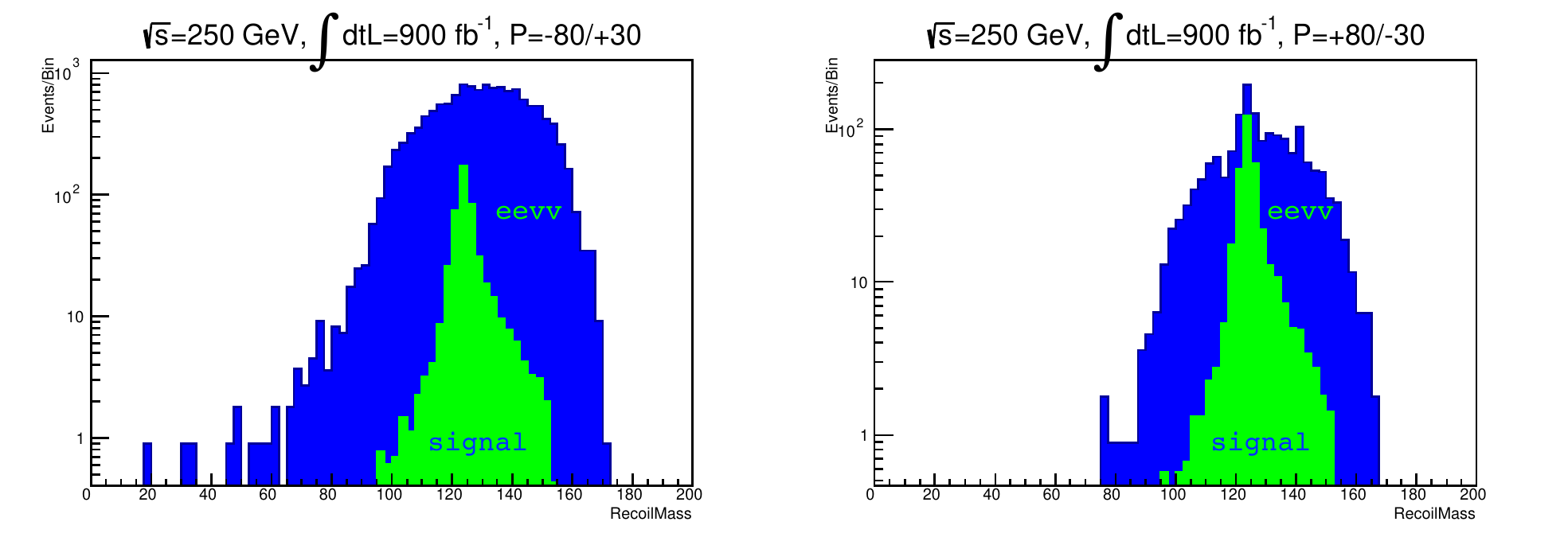}

\framebox{\textbf{Muon Channel}}

\includegraphics[width=6.in]{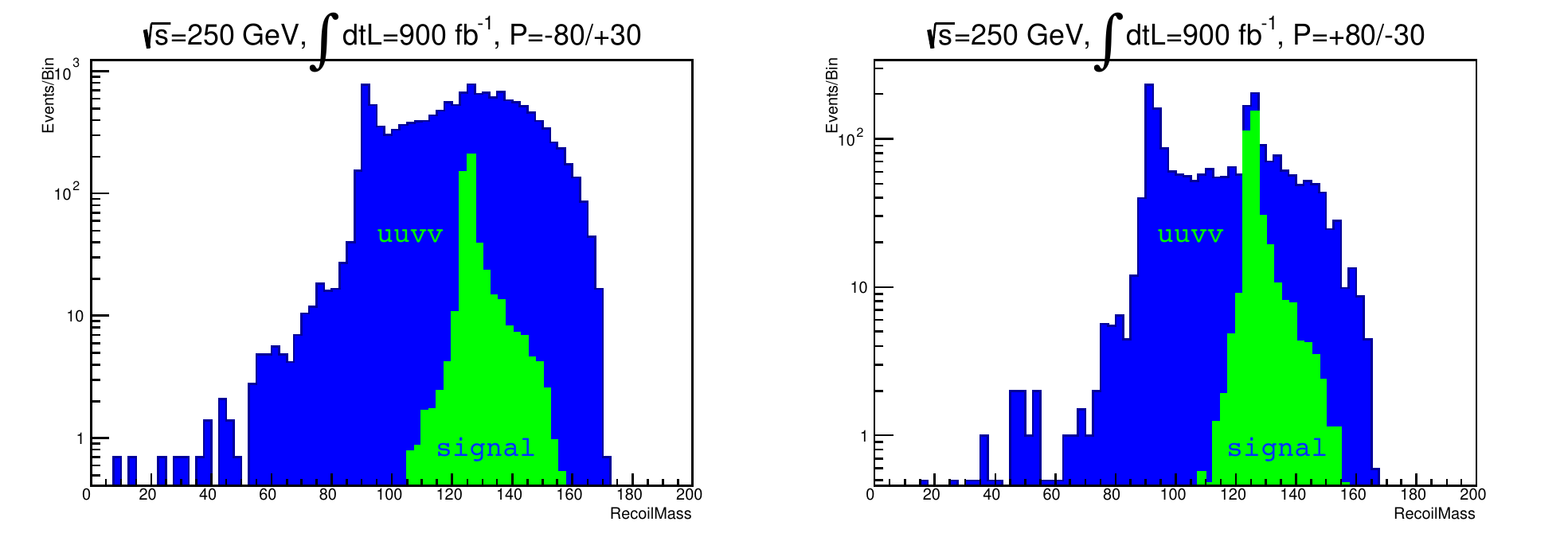}

\framebox{\textbf{Hadron Channel}}

\includegraphics[width=6.in]{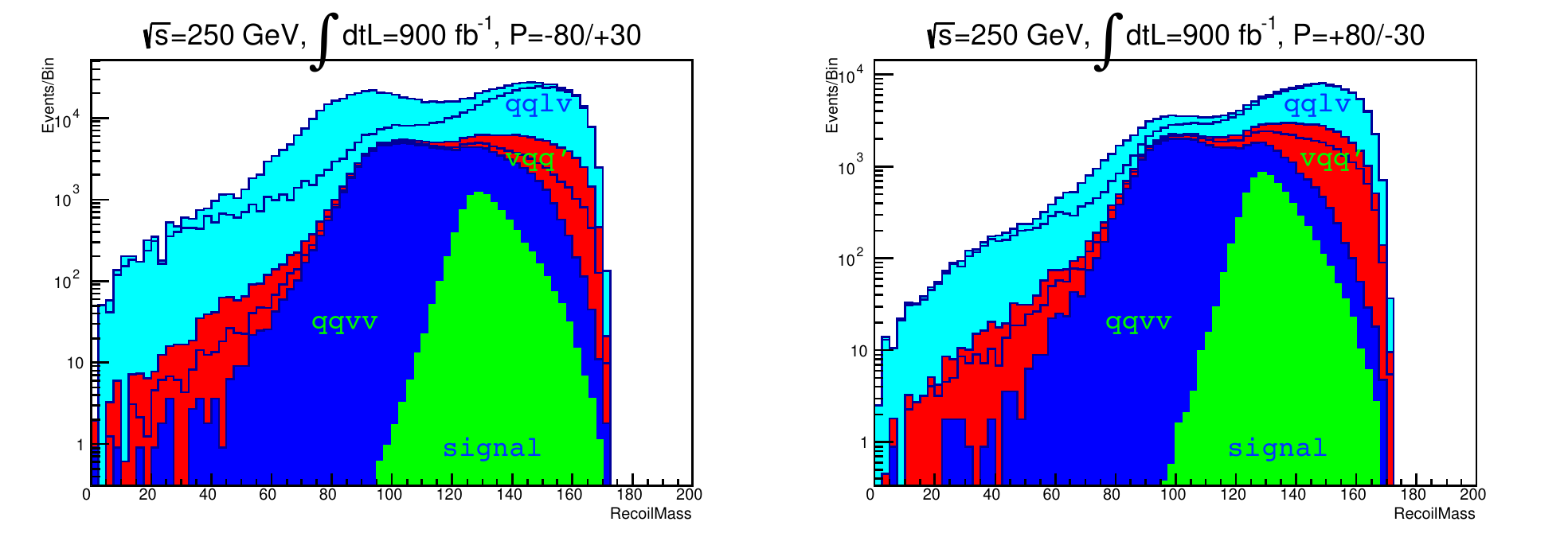}
\end{center}
\caption{Recoil mass $m_{rec}$ after selection up to the recoil mass selection for the electron, muon and hadron  channels. Backgrounds $ee\nu \nu, \mu \mu \nu \nu, qq \nu \nu$ (blue) are stacked on top of signal (green). Additional backgrounds in the hadron channel $\nu q q^{\prime}$ (red) and $qql\nu$ (cyan) are also stacked. The signal branching ratio is assumed to be 10\%. The integrated luminosity is 900 fb$^{-1}$ for each polarization case.}
\label{fig:recoil}
\end{figure}

See Figure \ref{fig:trkmult} for the track multiplicity in signal and background prior to signal selection requirements. See Figure \ref{fig:lepmult} for the lepton multiplicity in signal and background prior to signal selection requirements.  See Figures \ref{fig:recoil} and \ref{fig:masses} for the $m_{rec}$ and $m_{vis}$ distributions after full selection. See Table \ref{tab:yields} for signal and background yields, together with signal significance, after each lepton channel requirement above is imposed.

\begin{table}[p]
\begin{center}

\framebox{\textbf{Electron Channel}}
\vspace{0.1in}

\begin{tabular}{|c|c|c|c|c|c|c|} \hline
 & \multicolumn{3}{|c|}{80\% $e_{L}^{-}e_{R}^{+}$ 30\%} & \multicolumn{3}{|c|}{80\% $e_{R}^{-}e_{L}^{+}$ 30\%} \\ \hline
Requirement & Signal & Background& $\frac{S}{\sqrt{S+B}}$  & Signal & Background & $\frac{S}{\sqrt{S+B}}$ \\ \hline
All Events & $2.79\times 10^{4}$ & $2.6\times 10^{9}$ & $0.547$ & $1.89\times 10^{4}$ & $2.54\times 10^{9}$ & $0.375$ \\
Track/Lepton Mult. & $614$ & $7.63\times 10^{7}$ & $0.0702$ & $435$ & $7.64\times 10^{7}$ & $0.0498$ \\
Elec. Pair Consistency & $549$ & $3.6\times 10^{7}$ & $0.0915$ & $391$ & $3.61\times 10^{7}$ & $0.0650$ \\
$20 \leq p_{T}^{vis} \leq 70$ & $512$ & $1.56\times 10^{5}$ & $1.29$ & $362$ & $2.43\times 10^{4}$ & $2.30$ \\
$75 \leq m_{vis} \leq 105$~GeV  & $482$ & $3.19\times 10^{4}$ & $2.68$ & $344$ & $5.13\times 10^{3}$ & $4.65$ \\
$110 \leq m_{rec} \leq 150$~GeV & $471$ & $2.01\times 10^{4}$ & $3.29$ & $337$ & $2.16\times 10^{3}$ & $6.75$ \\ 
BDT$>0$ (Loose) & $380 \pm 6$ & $3690 \pm 410$ & $5.96$ & $273 \pm 5$ & $585 \pm 162$ & $9.32$ \\ \hline \hline
BDT$>0$ (Loose) & $380 \pm 6$ & $1980 \pm 44$ & $7.82$ & $273 \pm 5$ & $220 \pm 15$ & $12.3$ \\
BDT$>X_{opt}$ (Tight) & $261$ & $647$ & $8.65$ & $261$ & $181$ & $12.4$ \\ \hline 
\end{tabular}

\vspace{0.1in}

\framebox{\textbf{Muon Channel}}
\vspace{0.1in}

\begin{tabular}{|c|c|c|c|c|c|c|} \hline
 & \multicolumn{3}{|c|}{80\% $e_{L}^{-}e_{R}^{+}$ 30\%} & \multicolumn{3}{|c|}{80\% $e_{R}^{-}e_{L}^{+}$ 30\%} \\ \hline
Requirement & Signal & Background& $\frac{S}{\sqrt{S+B}}$  & Signal & Background & $\frac{S}{\sqrt{S+B}}$ \\ \hline
All Events & $2.79\times 10^{4}$ & $2.60\times 10^{9}$ & $0.547$ & $1.89\times 10^{4}$ & $2.54\times 10^{9}$ & $0.375$ \\
Track/Lepton Mult. & $637$ & $1.01\times 10^{8}$ & $0.0634$ & $460$ & $1.01\times 10^{8}$ & $0.0457$ \\
Muon Pair Consistency & $569$ & $4.36\times 10^{7}$ & $0.0861$ & $417$ & $4.48\times 10^{7}$ & $0.0623$ \\
$20 \leq p_{T}^{vis} \leq 70$ & $530$ & $6.17\times 10^{4}$ & $2.12$ & $392$ & $1.73\times 10^{4}$ & $2.94$ \\
$75 \leq m_{vis} \leq 105$~GeV & $507$ & $1.82\times 10^{4}$ & $3.71$ & $377$ & $3.64\times 10^{3}$ & $5.95$ \\
$110 \leq m_{rec} \leq 150$~GeV & $500$ & $1.06\times 10^{4}$ & $4.75$ & $372$ & $1.76\times 10^{3}$ & $8.07$ \\ 
BDT$>0$ (Loose) & $453 \pm 7$ & $1980 \pm 298$ & $9.18$ & $335 \pm 6$ & $450 \pm 142$ & $12.0$ \\ \hline \hline
BDT$>0$ (Loose) & $453 \pm 7$ & $3090 \pm 56$ & $7.61$ & $335 \pm 6$ & $480 \pm 22$ & $11.7$ \\ 
BDT$>X_{opt}$ (Tight) & $301$ & $538$ & $10.4$ & $250$ & $132$ & $12.8$ \\ \hline
\end{tabular}

\vspace{0.1in}

\framebox{\textbf{Hadron Channel}}
\vspace{0.1in}

\begin{tabular}{|c|c|c|c|c|c|c|} \hline
 & \multicolumn{3}{|c|}{80\% $e_{L}^{-} e_{R}^{+}$ 30\%} & \multicolumn{3}{|c|}{80\% $e_{R}^{-}e_{L}^{+}$ 30\%} \\ \hline
Requirement & Signal & Background& $\frac{S}{\sqrt{S+B}}$  & Signal & Background & $\frac{S}{\sqrt{S+B}}$ \\ \hline
All Events & $2.79\times 10^{4}$ & $2.6\times 10^{9}$ & $0.548$ & $1.89\times 10^{4}$ & $2.53\times 10^{9}$ & $0.376$ \\
Lepton Veto & $1.82\times 10^{4}$ & $1.88\times 10^{9}$ & $0.419$ & $1.2\times 10^{4}$ & $1.86\times 10^{9}$ & $0.278$ \\
Track/PFO Multiplicity & $1.06\times 10^{4}$ & $1.27\times 10^{8}$ & $0.944$ & $7.55\times 10^{3}$ & $1.06\times 10^{8}$ & $0.735$ \\
$20 \leq p_{T}^{vis} \leq 60$~GeV & $9.72\times 10^{3}$ & $4.06\times 10^{6}$ & $4.82$ & $6.9\times 10^{3}$ & $6.33\times 10^{5}$ & $8.62$ \\
$75 \leq m_{vis} \leq 105$~GeV & $9.1\times 10^{3}$ & $1.35\times 10^{6}$ & $7.82$ & $6.45\times 10^{3}$ & $2.43\times 10^{5}$ & $12.9$ \\
$N_{jet}=2$ & $9.1\times 10^{3}$ & $1.35\times 10^{6}$ & $7.82$ & $6.45\times 10^{3}$ & $2.43\times 10^{5}$ & $12.9$ \\
$-0.9 \leq \cos \theta_{jj} \leq -0.2$ & $8.48\times 10^{3}$ & $6.66\times 10^{5}$ & $10.3$ & $6\times 10^{3}$ & $1.39\times 10^{5}$ & $15.8$ \\
$110 \leq m_{rec} \leq 140$~GeV & $8.16\times 10^{3}$ & $2.68\times 10^{5}$ & $15.5$ & $5.77\times 10^{3}$ & $6.3\times 10^{4}$ & $22$ \\
BDT$>0$ (Loose) & $6630 \pm 23$ & $51500 \pm 1522$ & $27.5$ & $4640 \pm 19$ & $10400 \pm 684$ & $37.8$ \\ \hline \hline
BDT$>0$ (Loose) & $6630 \pm 23$ & $58600 \pm 242$ & $26.0$ & $4640 \pm 19$ & $15600 \pm 125$ & $32.6$ \\
BDT$>X_{opt}$ (Tight) & $4620$ & $21100$ & $28.8$ & $3760$ & $8490$ & $34.0$ \\ \hline
\end{tabular}

\caption{Signal yields $S$, background yields $B$, and significance $S/\sqrt{S+B}$ in the electron, muon, and hadron channel selections. The assumed signal branching ratio is 10\%. The integrated luminosity is 900 fb$^{-1}$ for each polarization case. In the final two rows of each table $B$ is estimated from the dedicated samples described in Table \ref{tab:mcsamples}. In all other rows $B$ is estimated from the \texttt{all\_SM\_background} sample. Statistical uncertainties are suppressed except for the loose BDT selection yields.}
\label{tab:yields}
\end{center}
\end{table}

\subsection{Hadron Channel}

For the hadron channel the signal signature is missing energy and two hadronic jets from quark pair hadronization. The jets are found with the Durham algorithm as implemented in the LCD Physics Tools\footnote{\url{ftp://ftp.slac.stanford.edu/groups/lcd/Physics_tools/}}. The event is forced to two jets by varying the $y_{cut}$ jetfinding parameter, which is an effective threshold for jet separation. The jet pair momentum is required to be consistent with the signal $Z$ decay opening angle, momentum and mass. The momentum of jet constituents is measured from the tracking for charged particles and the calorimetry for neutral particles. Then the mass in recoil from the candidate $Z \rightarrow jj$,

\begin{equation}
m_{rec}^2 =  s-2\sqrt{s}E_{jj}-m_{Z}^2 \\
\end{equation}

\noindent must be consistent with the Higgs boson mass. The hadron channel signal selection is as follows:

\begin{itemize}

\item Lepton $\ell=e,\mu$ veto. ($N_{\ell}=0$)

\item Track and PFO multiplicity consistent with signal. ($6 \leq N_{trk} \leq 32$ and $12 \leq N_{pfo} \leq 70$)

\item Transverse momentum of $Z \rightarrow q \bar{q}$ candidate consistent with signal. ($20 \leq p_T^{vis} \leq 70$~GeV)

\item Mass of $Z \rightarrow q \bar{q}$ candidate consistent with $Z$ mass. ($75 \leq m_{vis} \leq 105$~GeV)

\item Jet multiplicity: successful force to two jets by varying $y_{cut}$. ($N_{jet}=2$)

\item Jet pair consistency. Jet separation consistent with signal $Z$ decay. ($-0.9 \leq \cos \theta_{jj} \leq -0.2$)

\item Recoil mass consistent with Higgs boson mass. ($110 \leq m_{rec} \leq 150$~GeV)

\end{itemize}

\begin{figure}[p]
\begin{center}
\framebox{\textbf{Electron Channel}}

\includegraphics[width=6.in]{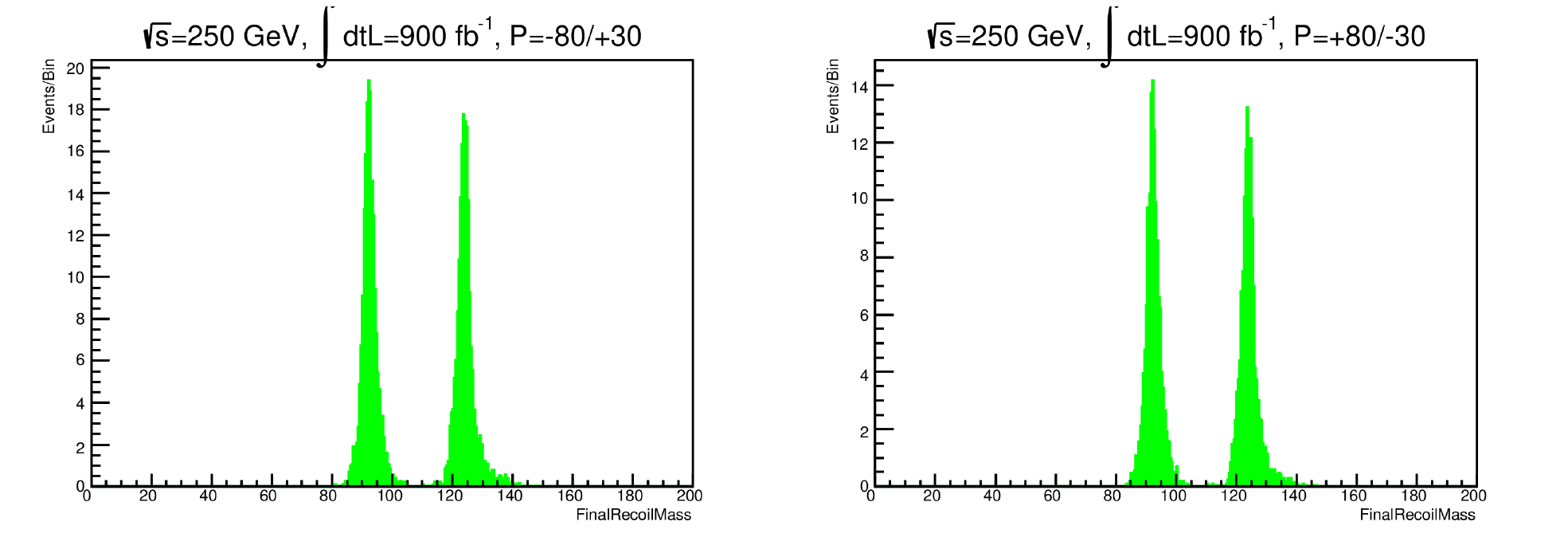}

\framebox{\textbf{Muon Channel}}

\includegraphics[width=6.in]{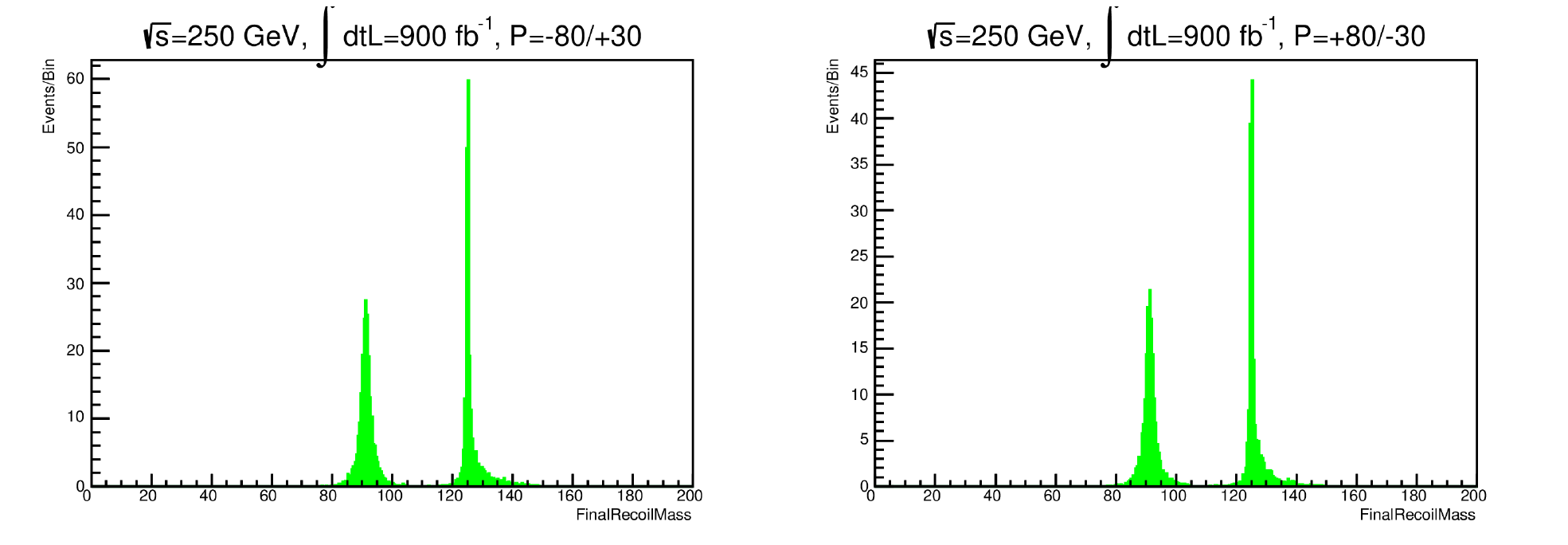}

\framebox{\textbf{Hadron Channel}}

\includegraphics[width=6.in]{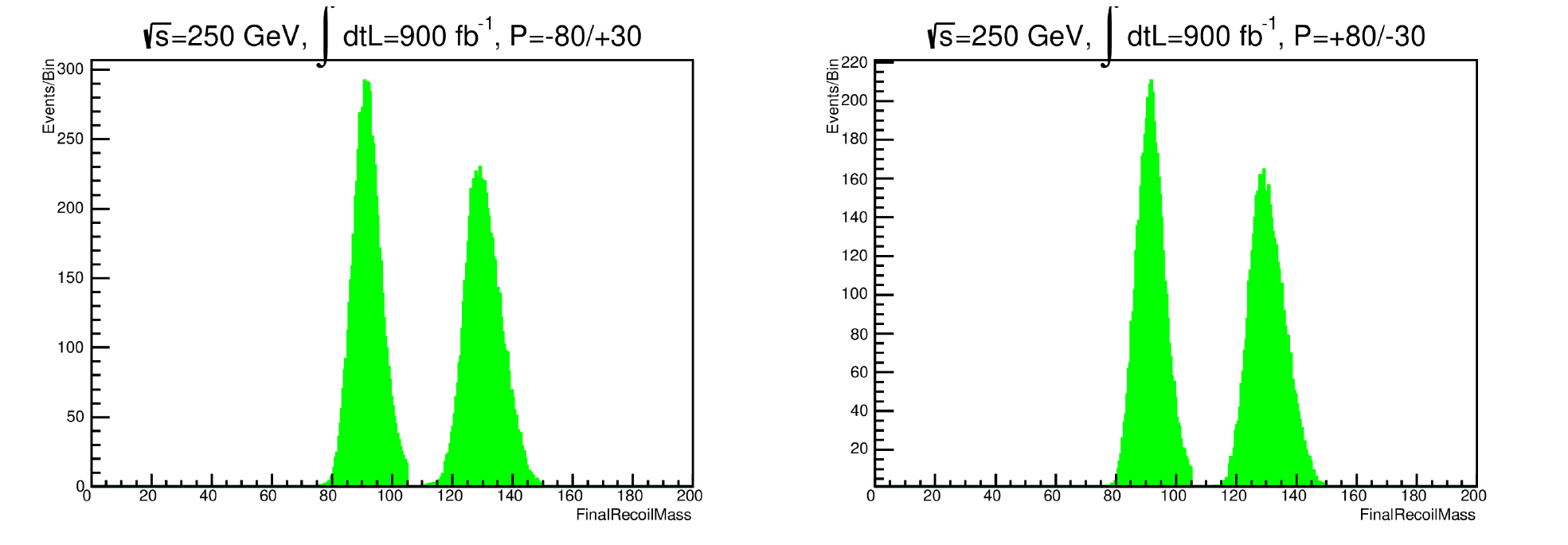}
\end{center}
\caption{Candidate $Z$ mass $m_{vis}$ and recoil mass $m_{rec}$ in the signal samples after full selection up to and including the loose BDT requirement for the electron, muon and hadron  channels.  The signal branching ratio is assumed to be 10\%. The integrated luminosity is 900 fb$^{-1}$ for each polarization case.}
\label{fig:masses}
\end{figure}

See Figure \ref{fig:trkmult} for the track multiplicity in signal and background prior to signal selection requirements. See Figure \ref{fig:lepmult} for the lepton multiplicity in signal and background prior to signal selection requirements. See Figures \ref{fig:recoil} and \ref{fig:masses} for the $m_{rec}$ and $m_{vis}$ distributions after full selection. See Table \ref{tab:yields} for signal and background yields, together with signal significance, after each hadron channel requirement above is imposed.

\subsection{Background Processes}

See Table \ref{tab:bg} for the background composition in the \texttt{all\_SM\_background} sample after full signal selection up to and including the loose BDT requirement. Two-fermion ($e^+ e^- \rightarrow f \bar{f}$) constitute a background at the 1\% level in electron, muon and hadron channels. Three-fermion processes ($e\gamma \rightarrow eZ,\nu W$), initiated by processes with ISR in the initial state, constitute a substantial background for the hadron channel but not the electron or muon channel. Four-fermion backgrounds constitute the dominant background for all three channels considered here.

\begin{table}[t]
\begin{center}
\begin{tabular}{|c|c|c|c|c|c|c|} \hline
Process & \multicolumn{2}{|c|}{Electron Channel} & \multicolumn{2}{|c|}{Muon Chanel} & \multicolumn{2}{|c|}{Hadron Chanel} \\ 
 & $e^{-}_{L}e^{+}_{R}$  & $e^{-}_{R}e^{+}_{L}$ & $e^{-}_{L}e^{+}_{R}$ & $e^{-}_{R}e^{+}_{L}$ & $e^{-}_{L}e^{+}_{R}$ & $e^{-}_{R}e^{+}_{L}$ \\ \hline \hline
4f $e^- e^+ \rightarrow WW$ & 36\% & 11\% & 61\% & 20\% & 27\% & 3\% \\
4f $e^- e^+ \rightarrow e^{\pm} \nu W^{\mp}$ & 23\% & 23\%  & 0\% & 0\% & 2\% & 1\% \\
4f  $e^- e^+ \rightarrow e^+ e^- Z$ & 13\% & 23\% & 0\% & 0\% & 0\% & 0\%\\
4f $e^- e^+ \rightarrow ZZ$ & 6\% & 16\% & 9\% & 30\% & 12\% & 33\% \\
4f $e^- e^+ \rightarrow \nu \bar{\nu} Z$ & 22\% & 27\% & 27\% & 50\% & 21\% & 18\% \\ \hline 
3f $e \gamma \rightarrow eZ, \nu W$ & 0\% & 0\% & 0\% & 0\% & 36\% & 39\% \\
2f $e^- e^+ \rightarrow f\bar{f}$ & 1\% & 0\% & 2\% & 0\% & 2\% & 1\% \\ \hline
\end{tabular}
\caption{Background composition after full electron, muon and hadron channel selections, up to and including the loose BDT requirement, determined by the \texttt{all\_SM\_background} samples.}
\label{tab:bg}
\end{center}
\end{table}

For the electron and muon channels, the dominant background is $e^+e^- \rightarrow WW$, with both $W \rightarrow \ell \nu_{\ell}$.  Subdominant backgrounds are $e^+e^- \rightarrow ZZ$ with one $Z \rightarrow \nu \bar{\nu}$ and the other $Z \rightarrow \ell^+ \ell^-$, and $e^+ e^- \rightarrow Z \nu \bar{\nu}$ with $Z \rightarrow \ell^+ \ell^-$. For the electron channel, substantial backgrounds are $e^+e^- \rightarrow e^+e^- Z$ with invisible $Z \rightarrow \nu \bar{\nu}$ and $e^+ e^- \rightarrow W e \nu$ with leptonic  $W \rightarrow e \nu$.  These $WW$- and $WZ$-fusion processes, with only one boson on-shell, are not open to the muon channel and therefore partially explain why sensitivity in the electron channel is lower than in the muon channel.  

For the hadron channel, the backgrounds are democratic, shared almost equally between three-fermion $e\gamma \rightarrow eZ,\nu W$ with hadronic $Z \rightarrow q \bar{q}$ and $W \rightarrow q \bar{q}^{\prime}$ decays, $e^+e^- \rightarrow WW$ with one leptonic $W \rightarrow \ell \nu$ and one hadronic $W \rightarrow q \bar{q}^{\prime}$ decay, $e^+e^- \rightarrow ZZ$ with one invisible $Z \rightarrow \nu \bar{\nu}$ decay and one hadronic $Z \rightarrow q \bar{q}$ decay, and $e^+e^- \rightarrow Z \nu \bar{\nu}$ with hadronic $Z \rightarrow q \bar{q}$ decay. In the backgrounds with one electron or muon, the lepton is misidentified in the reconstruction or lost outside of the sensitive detector volume. 

In both the lepton and hadron channels, the order of magnitude difference between polarization cases for the process $e^+e^- \rightarrow WW$ accounts for the stronger sensitivity in the $e_{R}^-e_{L}^+$ case.

Finally, Higgstrahlung itself $e^+ e^- \rightarrow ZH$ with invisible $Z \rightarrow \nu \bar{\nu}$ decay and hadronic Higgs boson decays also presents a minor background in the hadron channel but is negligible for the lepton channel. In this case the reconstructed mass of the Higgs boson is low enough to mimic the $Z$ boson. The recoil mass is correspondingly high enough to mimic the Higgs boson. These backgrounds are omitted here because dedicated analyses for each Higgs decay channel are expected to identify and reject them from this search.

\section{Multivariate Analysis}

In order to further improve signal sensitivity, a multivariate technique is employed to exploit differences in correlations between event parameters in signal and background events. A boosted decision tree (BDT) is used with supervised training on separate signal and background samples of events which have survived all of the cut-based requirements. The inputs to the BDT, which feature a single output, are described below. 

For both the lepton and hadron channels, separate BDTs are trained for each main background against signal. For the hadron channel, each polarization case of a given process is considered a distinct background, so there are ten BDTs for four-fermion backgrounds plus two BDTs for three-fermion backgrounds, for a total of twelve background BDTs. For the lepton channel the polarization cases are combined and considered a single background. Therefore for the lepton channel there are five BDTs for four-fermion backgrounds and none for three-fermion backgrounds.

The background BDT outputs are then used as inputs to a new BDT (combined BDT of BDTs) with a single output which is trained on signal and background composed of all major backgrounds weighted by cross section. The structure and training of the individual background BDTs as well as the BDT of BDTs are identical and are described below.

\subsection{Lepton Channel}

The lepton channels have very clean signatures with a lepton pair and nothing else in the event. Moreover the kinematics of the $Z$ candidate have a distinct signature in signal events so the kinematic parameters from the cut-based selection are included as inputs. 

Because the muon backgrounds are distinct kinematically from the electron backgrounds, an additional input flags events as either the electron channel or the muon channel, thus allowing a different optimization for each. Finally, while the cut-based selection vetos extra tracks, it allows extra neutrals like bremstrahlung photons. Therefore the PFO multiplicity is also included as an input. The input parameters are as follows for the lepton channel:

\begin{itemize}

\item Parameters from the cut-based analysis: lepton pair separation $\cos \theta_{\ell^+ \ell^-}$, $Z$ candidate transverse momentum $p_{T}^{vis}$, $Z$ candidate mass $m_{vis}$, recoil mass $m_{rec}$.

\item Electron multiplicity $N_{e}$, either $N_{e}=0$ or $N_{e}=2$. This parameter flags either the muon channel $N_{e}=0$ or the electron channel $N_{e}=2$.

\item PFO multiplicity $N_{pfo}$. This parameter flags events where residual neutral energy deposits suggest the event may not be signal-like.

\end{itemize}

After BDT training (see below), the improvement in signal sensitivity is loosely (tightly) optimized by requiring the BDT output to be larger than 0 ($X_{opt}$). See Figure \ref{fig:bdt} for the BDT output distributions in the lepton channel after all cut-based requirements are imposed. See the final rows in Table \ref{tab:yields} for the impact on signal and background yields and sensitivities in the lepton channels.

\subsection{Hadron Channel}

\begin{table}[t]
\begin{center}
\begin{tabular}{|c|c|c|c|c|c|c|} \hline
Variable & $eZ,\nu W$  & $q\bar{q}$ & $e\nu W$ & $WW$ & $eeZ,\nu\bar{\nu}Z$ & $ZZ$ \\ \hline \hline
$m_{rec}$ & 0.29/0.17 & 0.29/0.38 & 0.17/0.17 & 0.19/0.13 & 0.16/0.23 & 0.23/0.23 \\
$m_{vis}$ & 0.60/0.55 & 0.45/0.45 & 0.16/ 0.16 & 0.16/0.13 & 0.04/0.09 & 0.07/0.08 \\
$p_{T}^{vis}$ & 0.43/0.05 & 0.78/0.81 & 0.13/0.13 & 0.14/0.05 & 0.08/0.09 & 0.10/0.09 \\
$p_{z}^{vis}$ & 0.64/0.09 & 0.52/0.56 & 0.12/0.12 & 0.20/0.04 & 0.08/0.18 & 0.16/0.17 \\
$\cos \theta_{jj}$ & 0.11/0.08 & 0.29/0.54 & 0.08/0.08 & 0.08/0.10 & 0.07/0.10 & 0.09/0.09 \\ \hline
$N_{trk}$ & 0.03/0.01 & 0.19/0.21 & 0.02/0.02 & 0.01/0.02 & 0.00/0.00 & 0.00/0.00 \\
$N_{pfo}$ & 0.03/0.01 & 0.21/0.29 & 0.02/0.02 & 0.01/0.02 & 0.00/0.00 & 0.00/0.00 \\ \hline
$\cos \theta_{ZT}$ & 0.05/0.05 & 0.40/0.39 & 0.06/0.06 & 0.12/0.04 & 0.03/0.06 & 0.06/0.07 \\
$T$ & 0.03/0.00 & 0.18/0.17 & 0.19/0.19 & 0.17/0.24& 0.00/0.00 & 0.00/0.00 \\
$O$ & 0.02/0.00 & 0.10/0.19 & 0.17/0.17 & 0.16/0.21 & 0.00/0.00 & 0.00/0.00 \\
$y_{32}$ & 0.03/0.01 & 0.10/0.13 & 0.14/0.14 & 0.07/0.15 & 0.00/0.01 & 0.00/0.00 \\
$y_{43}$ & 0.01/0.00 & 0.10/0.05 & 0.08/0.08 & 0.06/0.08 & 0.00/0.00 & 0.00/0.00 \\ \hline
\end{tabular}
\caption{Hadron channel BDT input variables signal separation power $\langle S^2 \rangle$ for each background process and polarization case, $e_{L}^{-}e^{+}_{R}/e_{R}^{-}e_{L}^{+}$. The background samples are the DBD samples with pure beam polarization and hadronic $Z$ and $W$ decays.}
\label{tab:separation}
\end{center}
\end{table}

The hadron channel kinematics of the $Z$ candidate have a distinct signature in signal events so the kinematic parameters from the cut-based selection are included as inputs to the BDTs. Moreover the track and PFO multiplicity are also included. In addition to the $Z$ candidate transverse momentum $p_T$, the longitudinal momentum $p_z$ also provides some discrimination from backgrounds and is included as an input to the BDT.

The spatial distribution of PFOs in hadron channel events can be characterized by event shape variables. In this analysis we use thrust $T$ and oblateness $O$, defined by 

\begin{eqnarray}
T & = &  \mbox{max} \frac{\sum_{i} \vec{p_i} \cdot \hat{n}}{\sum_{i} \vert \vec{p_i} \vert} \\
O & = &  \mbox{max} \frac{\sum_{i} \vec{p_i} \cdot \hat{m}}{\sum_{i} \vert \vec{p_i} \vert} - \mbox{min} \frac{\sum_{i} \vec{p_i} \cdot \hat{m}}{\sum_{i} \vert \vec{p_i} \vert} 
\end{eqnarray}

\noindent where $i$ indexes the PFOs and for $T$ the maximum is taken over variations over unit vectors $\hat{n}$. The thrust axis $\hat{n}_{max}$ maximizes $T$, and for $O$ the minimum and maximum are taken over variations over unit vectors $\hat{m}$ where $\hat{m} \cdot \hat{n}_{max}=0$. Thrust $T$ and oblateness $O$ are calculated with the LCD package Physics Tools. We also use the $y_{cut}$ parameter used in the Durham jetfinder. This parameter is a distance measure which determines when two PFO groupings can be considered distinct jets. The $y_{cut}$ value required to force events from four to three jets ($y_{43}$) and from three to two jets ($y_{32}$) are included as inputs.

Finally, the angular separation between the thrust axis calculated in the $Z$ candidate frame and the $Z$ candidate momentum is found to provide additional separation and is included as an input to the BDT. The BDT inputs are therefore as follows for the hadron channel:

\begin{itemize}

\item Parameters from the cut-based analysis: track and PFO multiplicity $N_{trk}$ and $N_{pfo}$, jet pair separation $\cos \theta_{jj}$, $Z$ candidate transverse momentum $p_{T}^{vis}$, $Z$ candidate mass $m_{vis}$, recoil mass $m_{rec}$.

\item $Z$ candidate longitudinal momentum $p_{z}^{vis}$ along the beamline.

\item Event shape variables thrust $T$ and oblateness $O$ calculated in the laboratory frame.

\item Cosine of angle between the $Z$ candidate momentum in the laboratory frame and the thrust calculated in the frame of the $Z$ candidate, $\cos \theta_{ZT}$.

\item Durham $y_{cut}$ parameters necessary for forcing the event from four to three jets and from three to two jets, $y_{43}$ and $y_{32}$.

\end{itemize}

\noindent See Table \ref{tab:separation} for the signal and background separation $\langle S^2 \rangle$ of each input variable $y$, defined by 

\begin{eqnarray}
\langle S^2 \rangle & = & \frac{1}{2} \sum_{i} \frac{(\hat{y}_{S}(y_i)-\hat{y}_{B}(y_i))^2}{\hat{y}_{S}(y_i)+\hat{y}_{B}(y_i)}
\end{eqnarray}

\noindent For identical signal and background distributions, $\langle S^2 \rangle=0$, whereas for disjoint distributions $\langle S^2 \rangle=1$. Note that $\langle S^2 \rangle$ is independent of classification method.

\begin{figure}[p]
\begin{center}
\framebox{\textbf{Electron Channel}}

\includegraphics[width=6in]{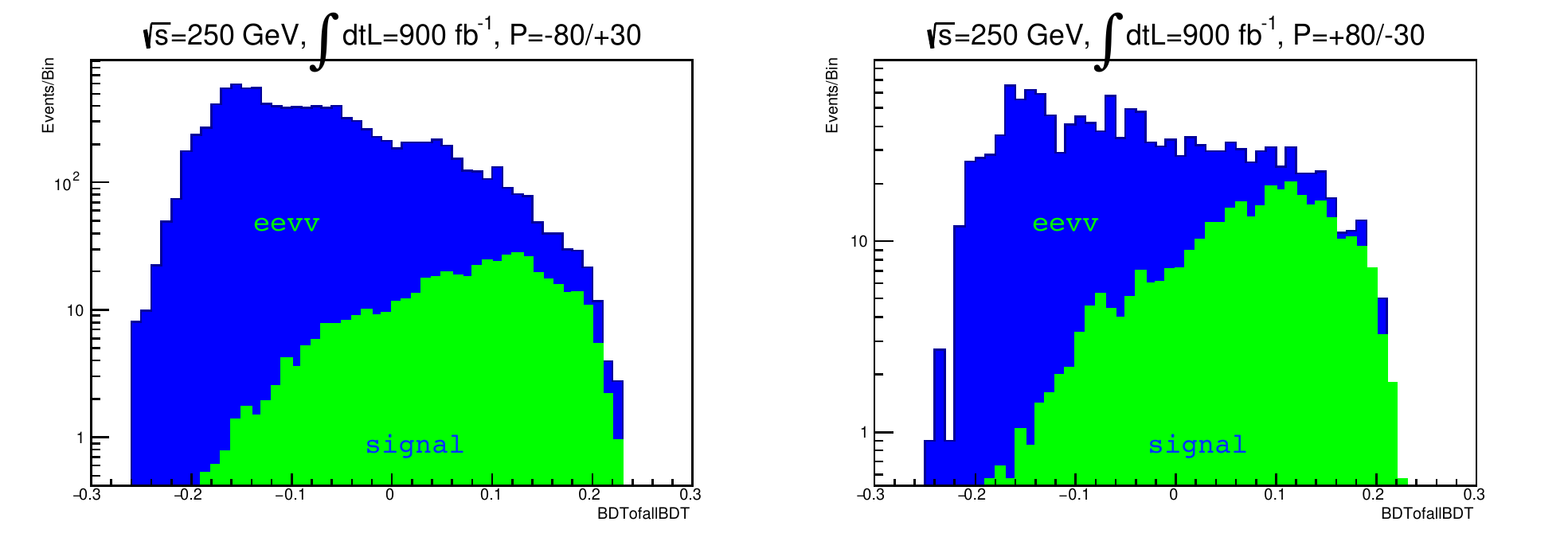}

\framebox{\textbf{Muon Channel}}

\includegraphics[width=6in]{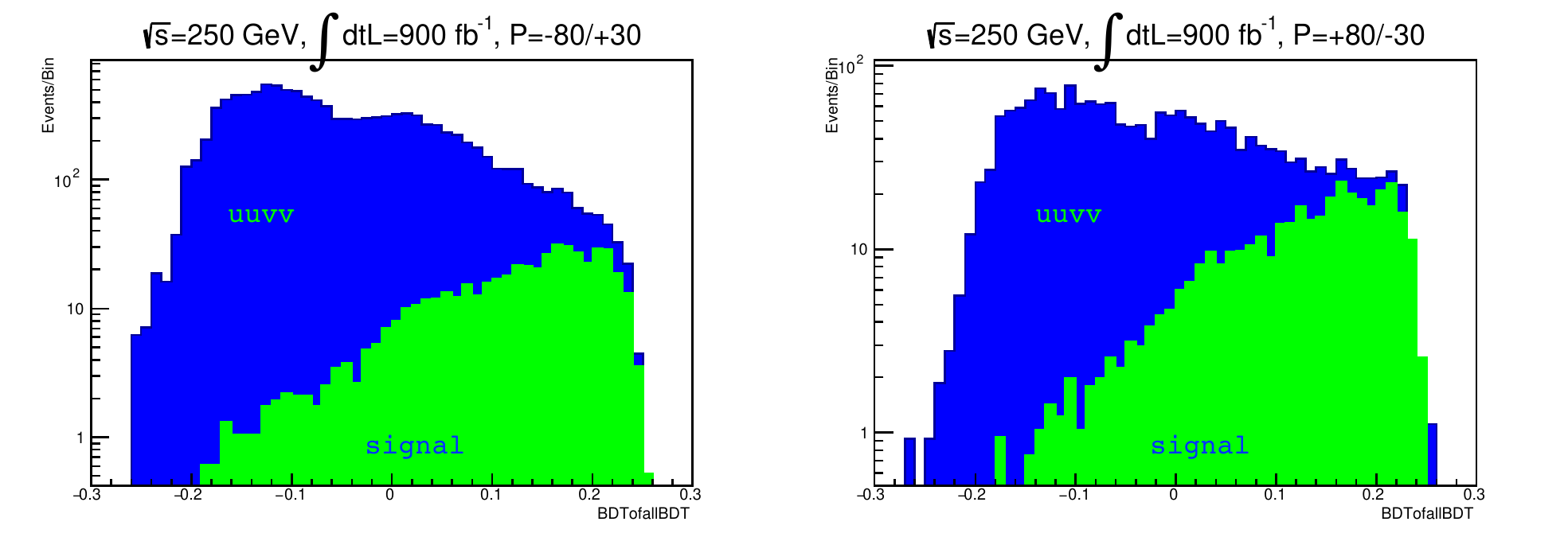}

\framebox{\textbf{Hadron Channel}}

\includegraphics[width=6in]{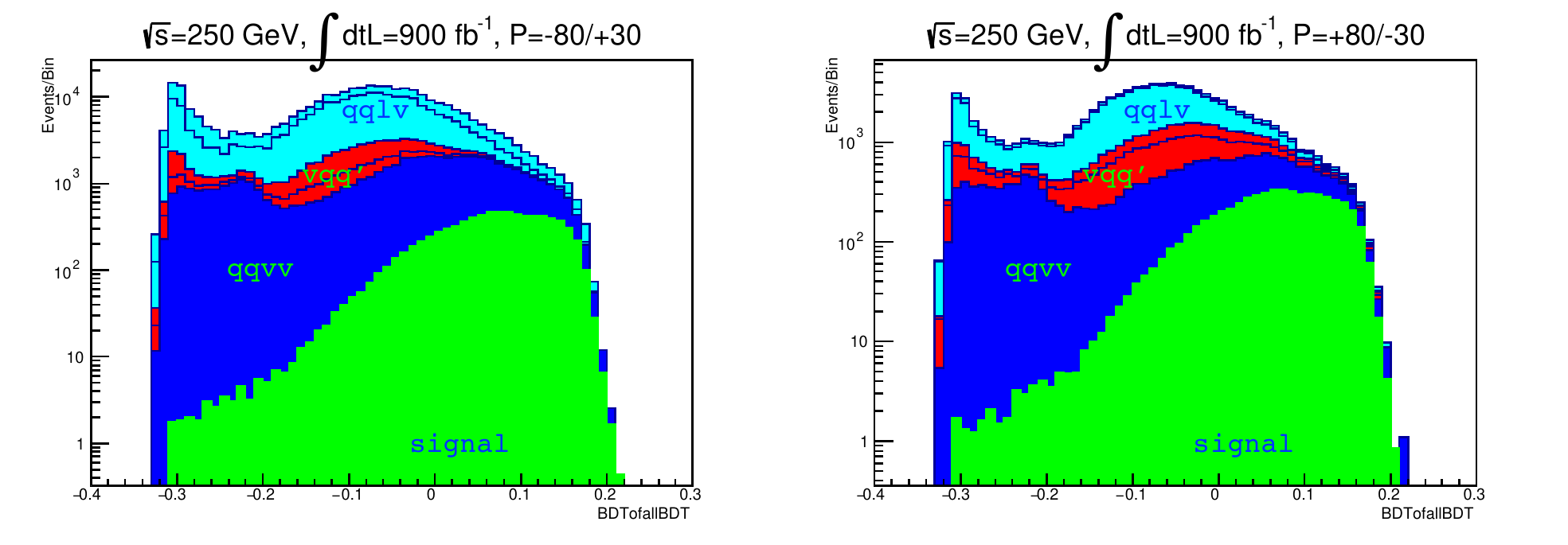}
\end{center}
\caption{Combined BDT outputs for the electron, muon, and hadron channels after all requirements in the cut-based selection are imposed. Backgrounds $ee\nu \nu, \mu \mu \nu \nu, qq \nu \nu$ (blue) are stacked on top of signal (green). Additional backgrounds in the hadron channel $\nu q q^{\prime}$ (red) and $qql\nu$ (cyan) are also stacked. The signal branching ratio is assumed to be 10\%. The integrated luminosity is 900 fb$^{-1}$ for each polarization case. These distributions are used to evaluate the expected signal upper limits.}
\label{fig:bdt}
\end{figure}

After BDT training (see below), the improvement in signal sensitivity is loosely (tightly) optimized by requiring the BDT output to be larger than 0 ($X_{opt}$). See Figure \ref{fig:bdt} for the BDT output distributions in the hadron channel after all cut-based requirements are imposed. See the final rows in Table \ref{tab:yields} for the impact on signal and background yields and sensitivities in the hadron channel.

\subsection{BDT Training}

A decision tree is a binary tree constructed iteratively to optimally separate signal from background events at each binary branching from a node to two new nodes. The root node contains all events, both signal and background. For any node with $N_{sig}$ signal events and $N_{bkg}$ background events, the signal purity $p=N_{sig}/(N_{sig}+N_{bkg})$ quantifies the separation.

Various measures based on purity can quantify the separation, but for any such measure the input distribution and cut value on that distribution are chosen which maximally increase the separation measure.  Branching to new nodes stops when a minimum number of events is reached in a node or a maximum number of layers is reached. Then these leaf nodes are labeled signal ($p > 0.5$) or background ($p < 0.5$) based on the purity of the node. The process of boosting a decision tree repeats the construction of the tree many times but with events slightly reweighted. The resulting forest of similar trees is optimally combined into a single output, which is the majority vote of the forest, and this boosted decision tree is then robust against statistical fluctuations in signal and background.

The BDTs were trained using the Root Toolkit for Multivariate Analysis (\texttt{TMVA}) \footnote{\url{https://root.cern/manual/tmva/}}. The separation measure is $p(1-p)$, the Gini index, which is optimized over twenty possible cuts on each input distribution. The minimum node size is five and the maximum number of layers is ten. The boost produces a forest of one thousand trees. For each background process identified in the \texttt{all\_SM\_background} sample, dedicated samples with higher statistics produced during the DBD are used to train independent background BDTs against signal. 

See Appendix B for some BDT evaluation plots. These plots demonstrate that, with the possible exception of the 2f samples, the training and test sample distributions match reasonably well and therefore that the BDTs have not been overtrained.

\section{Systematic Uncertainties}

\subsection{Beam Parameters}

For the cross sections used in this study, Whizard 2.6.4 is required to iterate until the theoretical uncertainty is well below the percent level. We conservatively estimate all of these uncertainties at 0.5\%. The ILC beam parameters which yield uncertainties are $\sqrt{s}$, $e^+$ and $e^-$ polarization fractions, ISR and beamstrahlung. The expected experimental precision on $\sqrt{s}$ and polarization at the ILC are 0.01\% and 0.25\% respectively \cite{kafer2008executive,fujii2018role}, yielding cross section uncertainties well below 0.5\%. We therefore treat these as negligible. 

Remaining uncertainties are due to the theoretical treatment of ISR and beamstrahlung in Whizard 2.6.4, as well as uncertainty on the choice of beam parameters which determine the beamstrahlung. In Whizard 2.6.4 ISR is estimated using the Equivalent Photon Approximation (EPA). To put a conservative upper bound on the cross section uncertainty due to ISR, we compare cross sections with ISR turned off with those with ISR turned on for each important signal and background process. The differences are typically of order 5\%, so we estimate a conservative 2\% uncertainty on all processes due to ISR treatment.

In addition to the uncertainty due to the theoretical treatment of beamstrahlung in Whizard 2.6.4, there is an uncertainty due to the choice of beam parameters. In this study we used the staged ILC250 beam parameters \cite{Evans:2017rvt} as input, although the parameters at runtime will certainly be different. Therefore we estimate the effect of beamstrahlung on cross sections by turning bremstrahlung on and off. The differences are typically of order 2\%, so we estimate a conservative 1\% uncertainty on all processes due to beamstrahlung.

See Table \ref{tab:whiz} for a summary of systematic uncertainties due to beam parameters and their impact on expected upper limits.

\subsection{Lepton Identification Efficiency}

The electron identification efficiency established in the ILC TDR for SiD achieved 90\% ($>95\%$) for electrons (muons) with $E=10$~GeV and 98\% ($>98\%$) for electrons (muons) with $E=100$~GeV \cite{Behnke:2013lya}. 

The particle flow algorithm PandoraPFA identifies muons from tracks extrapolated to the SiD muon detector and electrons from tracks extrapolated to the SiD ECal. The efficiency for identifying these leptons depends critically on the PandoraPFA algorithm matching parameters and the kinematic parameter space of the leptons.  The efficiency can be estimated from a generator sample with exactly two leptons which has been simulated and reconstructed with two tracks and two correctly identified lepton PFOs. 

From the background  $e^+ e^- \rightarrow e^+e^- \nu \bar{\nu}$ sample the electron identification efficiency is estimated to be 87\%, and from the background $e^+ e^- \rightarrow \mu^+ \mu^- \nu \bar{\nu}$  sample the muon identification efficiency is estimated to be 90\%. The binomial uncertainties on these are 0.24\% and 0.16\%, respectively, or approximately 0.2\% for each. With a dedicated effort to tune PandoraPFA parameters for optimal lepton identification, these efficiencies are expected to reach the SiD design goals.

\begin{table}[t]
\begin{center}
\begin{tabular}{|c|c|c|c|c|c|c|} \hline
Parameter & $\delta \sigma/\sigma$ & Elec. $\delta$UL/UL &  Muon $\delta$UL/UL  & Had. $\delta$UL/UL  \\ \hline \hline
Generator & $\pm 0.5\%$  & $0.5\%/0.4\%$ & $0.4\%/0.3\%$  &  $0.3\%/0.3\%$ \\
Beamstrahlung &$\pm 1\%$ & $0.9\%/0.7\%$ & $0.9\%/0.7\%$ & $0.6\%/0.5\%$\\ 
ISR &$\pm 2\%$ & $1.8\%/1.4\%$ & $1.7\%/1.3\%$  & $1.0\%/0.9\%$ \\ \hline
\end{tabular}
\caption{Whizard 2.6.4 systematic uncertainties and their impact on the cross section uncertainty $\delta \sigma/\sigma$ for each polarization case $e_{L}^{-}e_{R}^{+}/e_{R}^{-}e_{L}^{+}$. The uncertainty on the upper limit is estimated by varying signal and background process cross sections within their uncertainties over many trials. }
\label{tab:whiz}
\end{center}
\end{table}

\subsection{Tracker Momentum Resolution}

The momentum resolution established in the ILC TDR for SiD \cite{Behnke:2013lya} gave $\delta (1/p_T)=2\times 10^{-5}$/GeV. This uncertainty in track momentum determines a mass uncertainty in the dilepton masses of the $Z \rightarrow e^+ e^-$ and $Z \rightarrow \mu^+ \mu^-$ candidates in the electron and muon channels, respectively. It also partly determines the djet mass uncertainty of the $Z \rightarrow q\bar{q}$ in the hadron channel to the extent that the particle flow algorithm correctly identifies charged particle constituents of the jets.

The visible mass $m_{vis}$ and recoil mass $m_{rec}$, calculated from the lepton pair energy and $\sqrt{s}$, provide the greatest separation power $\langle S^2 \rangle$ in the lepton channels, though they are highly correlated.  Since the $\sqrt{s}$ uncertainty is expected to be negligible compared to the lepton pair energy measurement, we estimate the tracking momentum uncertainty based on reweighting the lepton pair mass distribution. 

Fitting the lepton pair mass after full selection in the electron and muon channels yields an uncertainty of order the natural $Z$ width, 2.7\%, reflecting the high precision expected from the SiD Tracker. We estimate the systematic uncertainty on the selection yields due to $\delta p/p$ by reweighting the lepton pair mass distribution in the signal sample such that the pair width is increased by 1\% above the natural $Z$ width, to 3.7\%, and reoptimizing the resulting reweighted lepton channel BDT distributions.

The overall uncertainty on the dijet masses of the $Z \rightarrow q\bar{q}$ candidates depends critically on the performance of the particle flow algorithm. In the limit of perfect track matching to calorimeter clusters, roughly 2/3 of jet constituents are charged and will have their momentum measured in the Tracker, and 1/3 are neutral and will have their energy measured in the ECal ($\pi^0$) or HCal ($K_L$ etc). The track multiplicity in the hadron channel (Figure \ref{fig:trkmult}), with a mean near 20, is dominated by charged pions. So we use the muon pair mass uncertainty impact on $\delta$UL/UL, added in quadrature for each track pair in the hadron channel, to estimate the impact of Tracker momentum uncertainty on the hadron channel upper limit.

\begin{table}[t]
\begin{center}
\begin{tabular}{|c|c|c|c|c|c|c|} \hline
Parameter $\delta$ & $\delta$ & Elec. $\delta$UL/UL & Muon $\delta$UL/UL & Had. $\delta$UL/UL \\ \hline \hline
Lepton ID $\delta \epsilon$ & $\pm 0.2\%$ & $0.3\%/0.2\%$ & $0.3\%/0.2\%$ & 0/0 \\ 
Tracker $\delta m/m$  & $+1\%$ & $4.9\%/0.6\%$ & $1.9\%/0.2\%$ & $5.7\%/0.6\%$\\
ECal/HCal $\delta m/m$  & $+2\%$   & 0/0 & 0/0  & $8.2\%/7.7\%$\\ \hline
\end{tabular}
\caption{Lepton identification uncertainty $\delta \epsilon$ and tracking and calorimetry $Z$ mass uncertainty $\delta m/m$ and the corresponding uncertainties on the expected upper limit for each polarization case $e_{L}^{-}e_{R}^{+}/e_{R}^{+}e_{L}^{+}$. The latter is estimated by reweighting the $\delta m/m$ distributions in signal and reoptimizing for best expected upper limit.}
\label{tab:sid}
\end{center}
\end{table}

\subsection{Calorimeter Energy Resolution}

The energy resolution in the ECal and HCal established in the ILC TDR for SiD \cite{Behnke:2013lya} are parametrized as follows:

\begin{eqnarray}
\frac{\delta E}{E} & = & 0.01 \oplus \frac{0.17}{\sqrt{E}} \\
\frac{\delta E}{E} & = & 0.094 \oplus \frac{0.56}{\sqrt{E}}
\end{eqnarray}

\noindent The energy of the $Z$ boson in Higgstrahlung events $e^+e^- \rightarrow ZH$ at $\sqrt{s}=250$~GeV is $E_{Z}\approx 110$~GeV. Assuming invisible Higgs decay, the energy uncertainty $\delta E/E$ is 2\% (11\%) in the limit of total energy deposition in the ECal (HCal). In this total deposition limit the visible mass uncertainties are approximately 3\% (16\%) for the ECal (HCal). Adding the natural $Z$ width in quadrature yields 4\% (16\%) for the ECal (HCal).

Fitting the jet pair mass after full selection in the hadron channel yields $\delta m/m=5\%$. We estimate the systematic uncertainty on the selection yields due to $\delta E/E$ and the particle flow algorithm performance by reweighting the dijet mass distribution in the signal sample such that the pair width is increased by 2\% above the nominal 5\%, to 7\%, and reoptimizing the resulting reweighted hadron channel BDT distribution. 

See Table \ref{tab:sid} for the systematic uncertainties due to lepton identification efficiency, tracker resolution, and calorimeter resolution and their impact on expected upper limits.

\section{Results and Conclusion}

We calculate the expected 68\% (95\%) confidence level upper limits on the invisible Higgs branching ratio with the \texttt{TLimit} \footnote{\url{https://root.cern/doc/master/classTLimit.html}} class in Root using the combined BDT output distributions (Figure \ref{fig:bdt}). 

\texttt{TLimit} employs the CL$_{s+b}$ technique with a Bayesian approach. We calculate the expected limits for each channel separately for 900 fb$^{-1}$ integrated luminosity for each polarization case. Then the channels and polarization case samples are combined. The combined expected upper limit for all channels and polarization cases is 0.16\% at 95\% confidence level. See Table \ref{tab:results}. 

The SM invisible branching ratio for $H \rightarrow ZZ^{\star} \rightarrow \nu \bar{\nu} \nu \bar{\nu}$ is approximately 0.10\%, which lies just below the expected upper limits at 95\% confidence level For enhanced branching ratios in BSM models, the yields in Table \ref{tab:yields} can be easily scaled down from 10\% to arbitrary levels. Evidence (discovery) is expected for invisible Higgs branching ratios of 0.50\% (0.83\%) or higher with this dataset.

We do not include systematic uncertainties in calculating these limits. The impacts on the expected upper limits for these has been estimated in the previous section and are summarized in Tables \ref{tab:whiz} and \ref{tab:sid}. We neglect these uncertainties because they are provisional and expected to reduce significantly as ISR and beamstrahlung are better understood and detector reconstruction is tuned to optimal performance. These limits should be regarded as realistic and achievable \emph{with the current SiD design}.

Moreover the sensitivity is expected to improve with improved detector performance. The SiD design is being carefully reconsidered in light of recent advancements in subdetector design \cite{breidenbach2021updating}. For the Tracker and ECal the MAPS technology promises significant improvement in measurement precision \cite{ballin2009digital}. The ECal and HCal energy resolution is expected to improve significantly when machine learning techniques are employed to recover calorimeter energy leakage \cite{Braun:2020eme}.

In conclusion, we expect that the SiD detector at the ILC will allow a precision measurement of the invisible Higgs branching ratio. The expected limit is 0.16\% at 95\% confidence level with data samples of 900fb$^{-1}$ at $\sqrt{s}=250$~GeV for each polarization case. The SM invisible branching ratio lies just below this expected limit. For BSM enhanced invisible Higgs, evidence or discovery is expected above the half-percent level.

\begin{table}[t]
\begin{center}
\begin{tabular}{|c|c|c|c|} \hline
Channel & 80\% $e^{-}_{L}e^{+}_{R}$ 30\% & 80\% $e^{-}_{R}e^{+}_{L}$ 30\% & Combined \\ \hline \hline
Electron & 1.12\% & 0.35\% & 0.33\% \\ 
Muon  & 0.77\% & 0.29\% & 0.27\% \\
Hadron & 0.42\% & 0.31\% & 0.25\% \\ \hline
Combined  & 0.35\% & 0.18\% & 0.16\% \\ \hline
\end{tabular}
\caption{Expected upper limits at 95\% confidence level on the invisible Higgs branching ratio. The integrated luminosity is 900 fb$^{-1}$ for each polarization case. Systematic uncertainties are not included, but their impact is estimated in Tables \ref{tab:whiz} and \ref{tab:sid}.}
\label{tab:results}
\end{center}
\end{table}

\begin{center}\textbf{Acknowledgments}\end{center}

We thank our colleagues on ILD for useful discussions of backgrounds to this important channel, the ILC International Development Team (IDT) for organizing the global effort toward realization of the ILC, and the organizers of the DPF Snowmass 2021/2022 community planning exercise for coordinating this important community effort. We thankfully acknowledge support from the US Department of Energy grant DE-SC0017996.

\bibliography{paper}

\newpage

\section*{Appendix A: Monte Carlo Generator Samples \label{app:samples}}

\rule{0.7\paperwidth}{0.4pt}

\begin{itemize}

\item \textbf{SiD MC20/21}\footnote{\url{https://pages.uoregon.edu/ctp/SiD_private.html}} (80\% $e^-$, 30\% $e^+$ polarization). Signal used in training BDTs, signal and background used in evaluating signal sensitivity.

\begin{itemize}

\item (sidmc20)\_ilc250\_eLpR\_2f1hinv.*.whizard\_2\_6\_4.stdhep
\item (sidmc20)\_ilc250\_eRpL\_2f1hinv.*.whizard\_2\_6\_4.stdhep

\item sidmc21a\_ilc250\_eLpR\_ap3f*.stdhep, sidmc21a\_ilc250\_eLpR\_ap3f*.stdhep
\item sidmc21a\_ilc250\_eLpR\_ea3f*.stdhep, sidmc21a\_ilc250\_eLpR\_ea3f*.stdhep

\item sidmc21a\_ilc250\_eLpR\_eevv*.stdhep, sidmc21a\_ilc250\_eRpL\_eevv*.stdhep
\item sidmc21a\_ilc250\_eLpR\_mumuvv*.stdhep, sidmc21a\_ilc250\_eRpL\_mumuvv*.stdhep
\item sidmc21a\_ilc250\_eLpR\_qqvv*.stdhep, sidmc21a\_ilc250\_eRpL\_qqvv*.stdhep
\item sidmc21a\_ilc250\_eLpR\_qqev*.stdhep, sidmc21a\_ilc250\_eRpL\_qqev*.stdhep
\item sidmc21a\_ilc250\_eLpR\_qqlv*.stdhep, sidmc21a\_ilc250\_eRpL\_qqlv*.stdhep
\end{itemize}

\item \textbf{Barklow SiD DBD Mixed Samples} (80\% $e^-$, 30\% $e^+$ polarization). Used only in estimating background yields and distributions at all stages in the analysis chain. Not used in training BDTs or the final sensitivity estimate.

\begin{itemize}
\item all\_SM\_background\_-80e-\_+30e+\_*.stdhep
\item all\_SM\_background\_+80e-\_-30e+\_*.stdhep
\end{itemize}

\item \textbf{DBD Two-Fermion} Samples (100\% $e^-$, 100\% $e^+$ polarization). Used only in training BDTs.

\begin{itemize}
\item E250-TDR\_ws.P2f\_z\_l.Gwhizard-1\_95.eL.pR.I106605.*.stdhep
\item E250-TDR\_ws.P2f\_z\_l.Gwhizard-1\_95.eR.pL.I106606.*.stdhep
\item E250-TDR\_ws.P2f\_z\_h.Gwhizard-1\_95.eL.pR.I106607.*.stdhep
\item E250-TDR\_ws.P2f\_z\_h.Gwhizard-1\_95.eR.pL.I106608.*.stdhep
\end{itemize}

\item \textbf{DBD Three-Fermion} Samples (100\% $e^-$, 100\% $e^+$ polarization). Used only in training BDTs.

\begin{itemize}
\item E0250-TDR\_ws.Pea\_vxy.Gwhizard-1.95.eL.pW.I37785.*.stdhep
\item E0250-TDR\_ws.Pea\_vxy.Gwhizard-1.95.eL.pB.I37786.*.stdhep
\item E0250-TDR\_ws.Pae\_vxy.Gwhizard-1.95.eW.pR.I37815.*.stdhep
\item E0250-TDR\_ws.Pae\_vxy.Gwhizard-1.95.eB.pR.I37816.*.stdhep
\end{itemize}

\item \textbf{DBD Four-Fermion} Samples (100\% $e^-$, 100\% $e^+$ polarization). Used only in training BDTs.

\begin{itemize}

\item E250-TDR\_ws.P4f\_sw\_sl.Gwhizard-1\_95.eL.pR.I106564.*.stdhep
\item E250-TDR\_ws.P4f\_sw\_sl.Gwhizard-1\_95.eR.pL.I106566.*.stdhep
\item E250-TDR\_ws.P4f\_sw\_l.Gwhizard-1\_95.eL.pR.I106586.*.stdhep
\item E250-TDR\_ws.P4f\_sw\_l.Gwhizard-1\_95.eR.pL.I106588.*.stdhep
\item E250-TDR\_ws.P4f\_szeorsw\_l.Gwhizard-1\_95.eL.pR.I106568.*.stdhep
\item E250-TDR\_ws.P4f\_sznu\_sl.Gwhizard-1\_95.eL.pR.I106571.*.stdhep
\item E250-TDR\_ws.P4f\_sznu\_sl.Gwhizard-1\_95.eR.pL.I106572.*.stdhep
\item E250-TDR\_ws.P4f\_sznu\_l.Gwhizard-1\_95.eL.pR.I106589.*.stdhep
\item E250-TDR\_ws.P4f\_zz\_sl.Gwhizard-1\_95.eL.pR.I106575.*.stdhep
\item E250-TDR\_ws.P4f\_zz\_sl.Gwhizard-1\_95.eR.pL.I106576.*.stdhep
\item E250-TDR\_ws.P4f\_ww\_sl.Gwhizard-1\_95.eL.pR.I106577.*.stdhep
\item E250-TDR\_ws.P4f\_ww\_sl.Gwhizard-1\_95.eR.pL.I106578.*.stdhep
\item E250-TDR\_ws.P4f\_ww\_l.Gwhizard-1\_95.eL.pR.I106581.*.stdhep
\item E250-TDR\_ws.P4f\_ww\_l.Gwhizard-1\_95.eR.pL.I106582.*.stdhep

\end{itemize}
\end{itemize}

\begin{center}\textbf{SiD MC20/21 Whizard 2.6.4 Generator Parameters}\end{center}

\begin{itemize}

\item Whizard 2.6.4 Beam Parameters

\begin{itemize}
\item sqrts=250 GeV
\item beams= "e-", "e+"  =$>$ isr, isr
\item beams\_pol\_fraction=80\%, 30\%
\item either beams\_pol\_density=@(-1), @(+1) or beams\_pol\_density=@(+1), @(-1)
\item epa\_q\_min=1 GeV and epa\_x\_min=0.01
\end{itemize}

\item Whizard 2.6.4 Aliases and Decays

\begin{itemize}
\item alias nu=nue:numu:nutau and alias nubar=nuebar:numubar:nutaubar
\item alias q=u:d:s:c:b and alias qbar=ubar:dbar:sbar:cbar:bbar
\item process wpdecay=Wp =$>$ q, qbar and process wmdecay=Wm =$>$ q, qbar
\item unstable Wp(wpdecay) and unstable Wm(wmdecay)
\end{itemize}

\item Whizard 2.6.4 Processes and Cuts

\begin{itemize}

\item process ilc250\_eLpR\_eevv= "e-", "e+" =$>$ "e+", "e-", nu, nubar and cuts=all Pt $>$ 0.5 GeV [charged] and all M $>$ 70. GeV ["e+","e-"] and all M $<$ 110. GeV ["e+","e-"] and all M $>$ 100. GeV [nu, nubar] and all M $<$ 160. GeV [nu, nubar]

\item process ilc250\_eLpR\_mumuvv= "e-", "e+" =$>$ "mu+", "mu-", nu, nubar and cuts=all M $>$ 60. GeV ["mu+","mu-"] and all M $<$ 120. GeV ["mu+","mu-"] and all M $>$ 90. GeV [nu, nubar] and all M $<$ 170. GeV [nu, nubar]

\item process ilc250\_eLpR\_qqvv= "e-", "e+" =$>$ q, qbar, nu, nubar and cuts=all M $>$ 60. GeV [q, qbar] and all M $<$ 120. GeV [q, qbar] and all M $>$ 90. GeV [nu, nubar] and all M $<$ 170. GeV [nu, nubar]

\item process ilc250\_eLpR\_qqev= "e-", "e+" =$>$ (Wp, "e-", nuebar) + ("e+", nue, Wm) and cuts=all M $>$ 60. GeV [q,qbar] and all M $<$ 120. GeV [q, qbar] 

\item process ilc250\_eLpR\_qqlv= "e-", "e+" =$>$ (q, qbar, "mu-", numubar) + ("mu+", numu, q, qbar) + (q, qbar, "tau-", nutaubar) + ("tau+", nutau, q, qbar) and cuts=all M $>$ 60. GeV [q,qbar] and all M $<$ 120. GeV [q, qbar]

\item process ilc250\_eLpR\_ap3f= photon, "e+" =$>$ ("e+", q, qbar)+(nuebar, q, qbar) and cuts=all M $>$ 75. GeV [q, qbar] and all M $<$ 105. GeV [q,qbar] and all Pt $>$ 20. GeV [q,qbar] and all Pt $<$ 60. GeV [q,qbar]

\item process ilc250\_eLpR\_ea3f= "e-", photon =$>$ ("e-", q, qbar)+(nue, q, qbar) and cuts=all M $>$ 75. GeV [q, qbar] and all M $<$ 105. GeV [q,qbar] and all Pt $>$ 20. GeV [q,qbar] and all Pt $<$ 60. GeV [q,qbar]

\end{itemize}

\end{itemize}

\newpage

\section*{Appendix B: BDT Evaluation Plots \label{app:bdt}}

\rule{0.7\paperwidth}{0.4pt}

\begin{figure}[H]
\begin{center}
\framebox{\textbf{Electron Channel BDT of BDTs}}

\vspace{0.15in}

\includegraphics[height=1.6in]{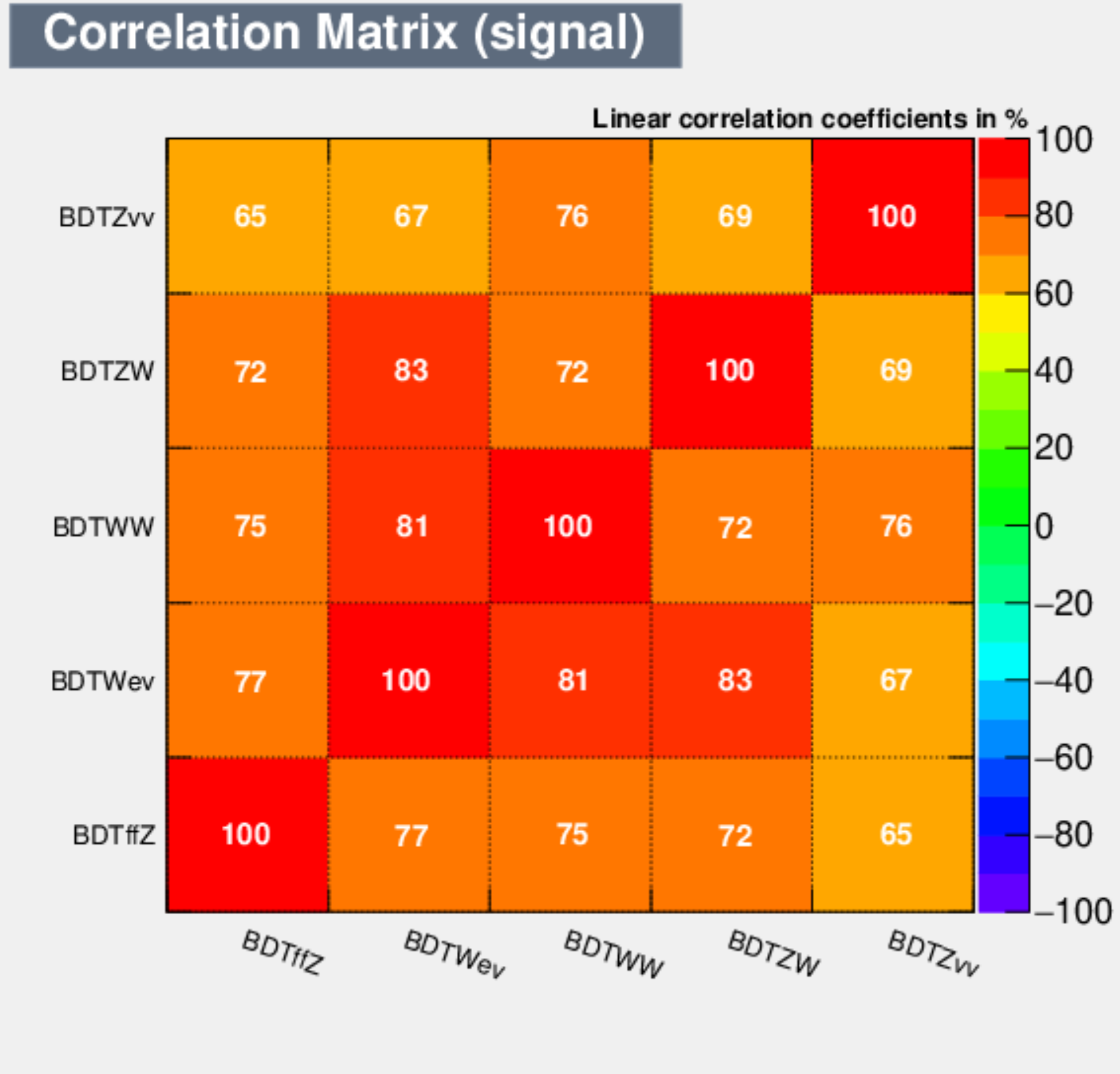}
\includegraphics[height=1.6in]{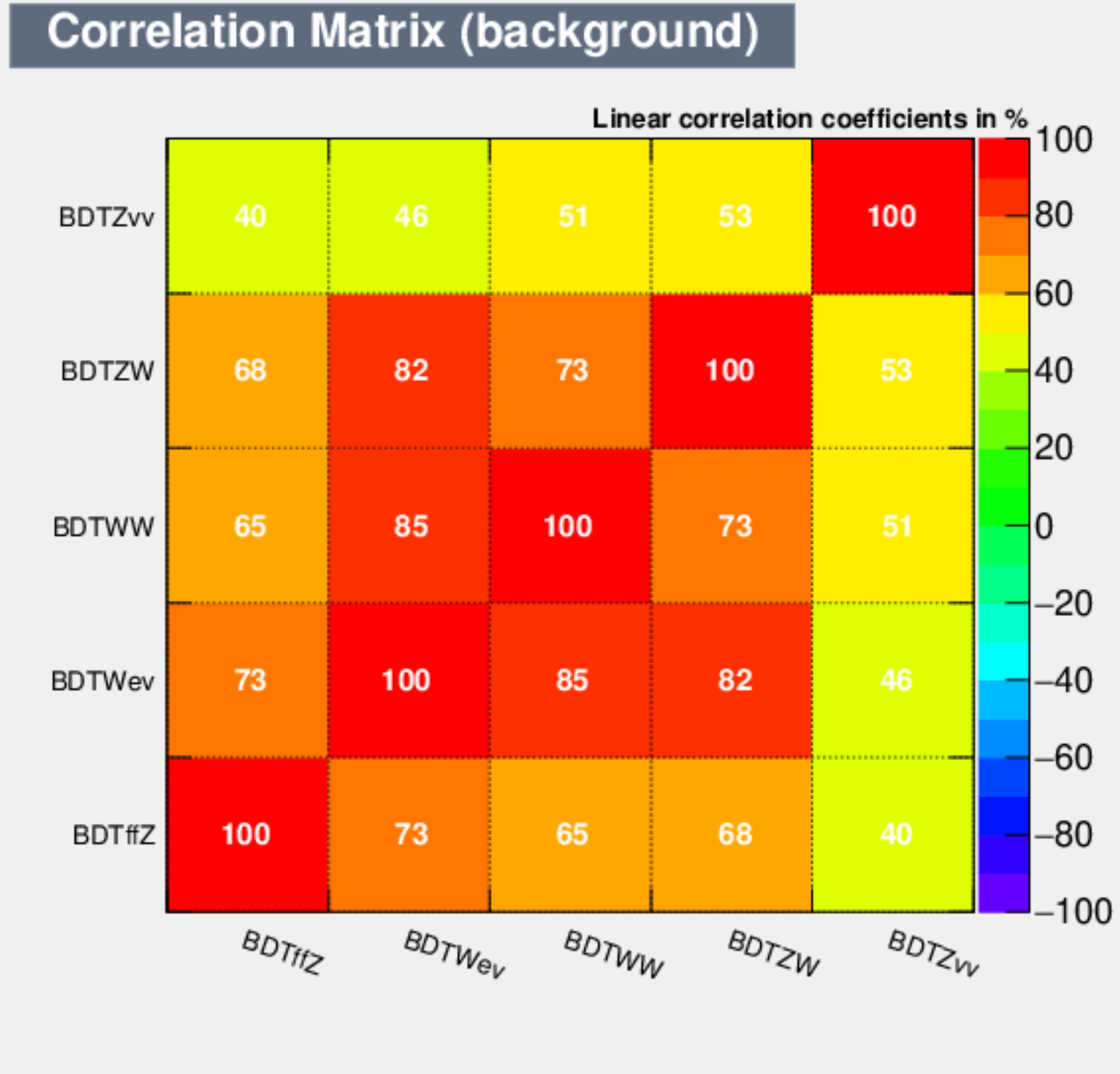}
\includegraphics[height=1.6in]{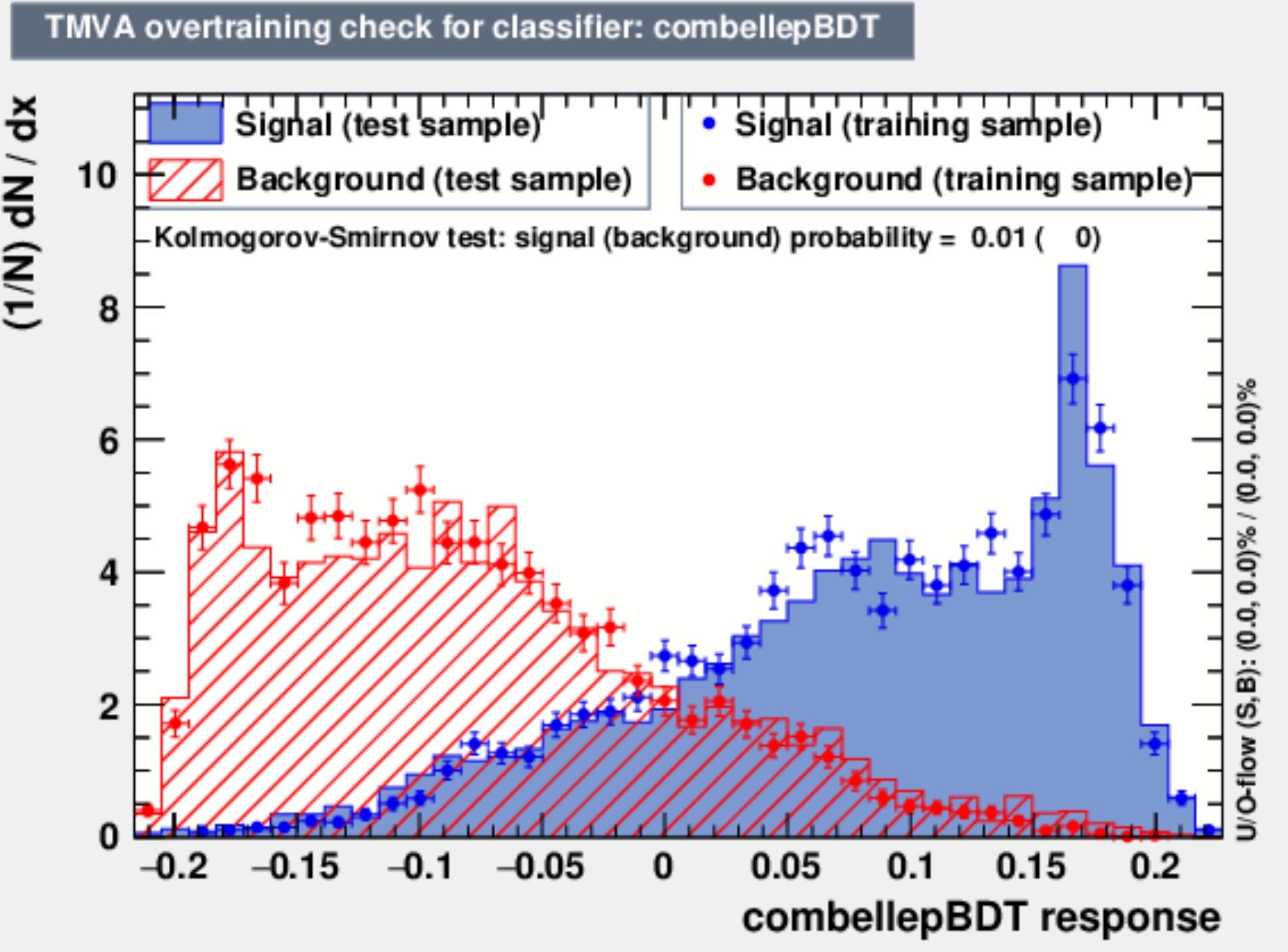}

\vspace{0.15in}

\framebox{\textbf{Muon Channel BDT of BDTs}}

\vspace{0.15in}

\includegraphics[height=1.6in]{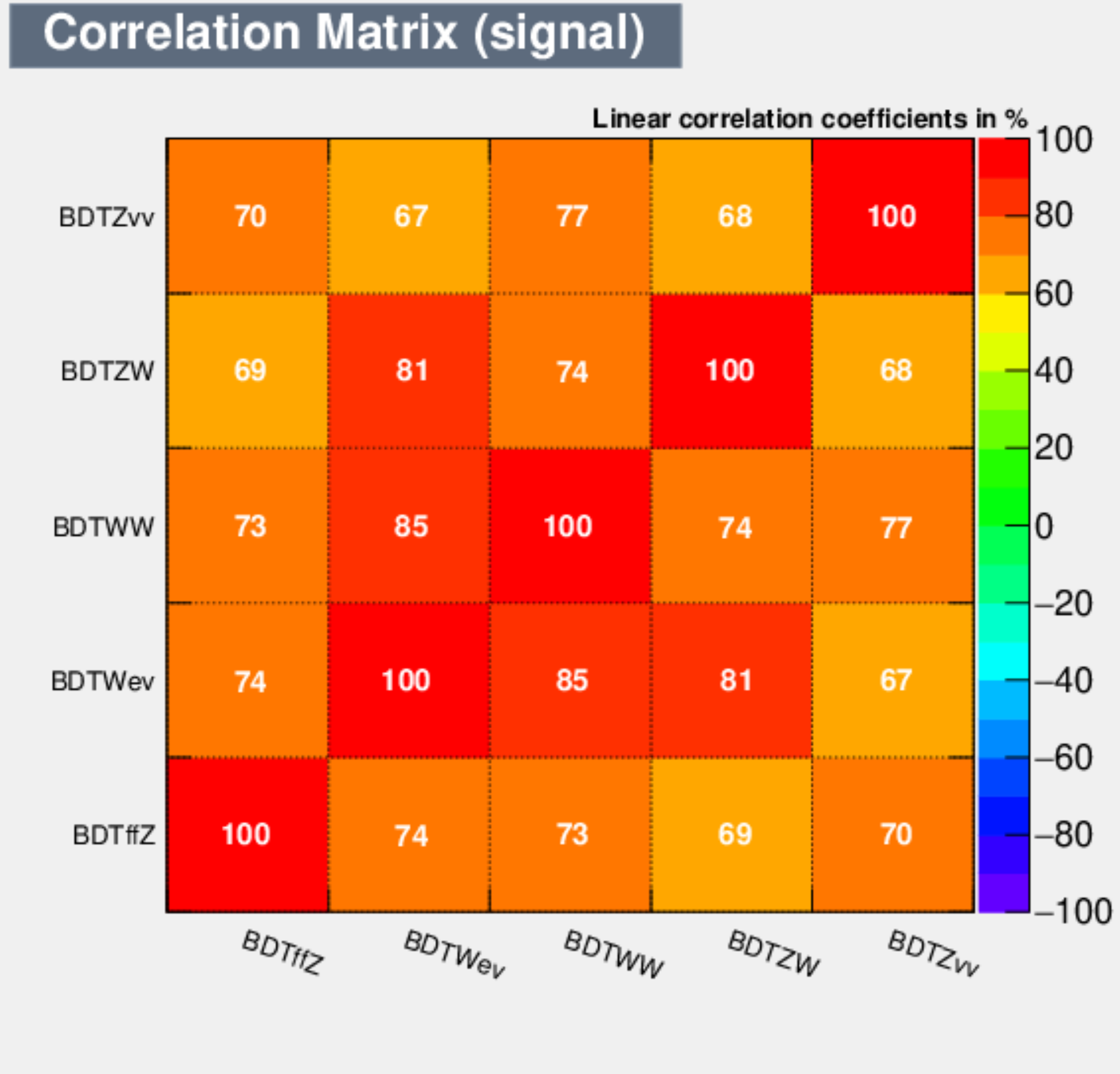}
\includegraphics[height=1.6in]{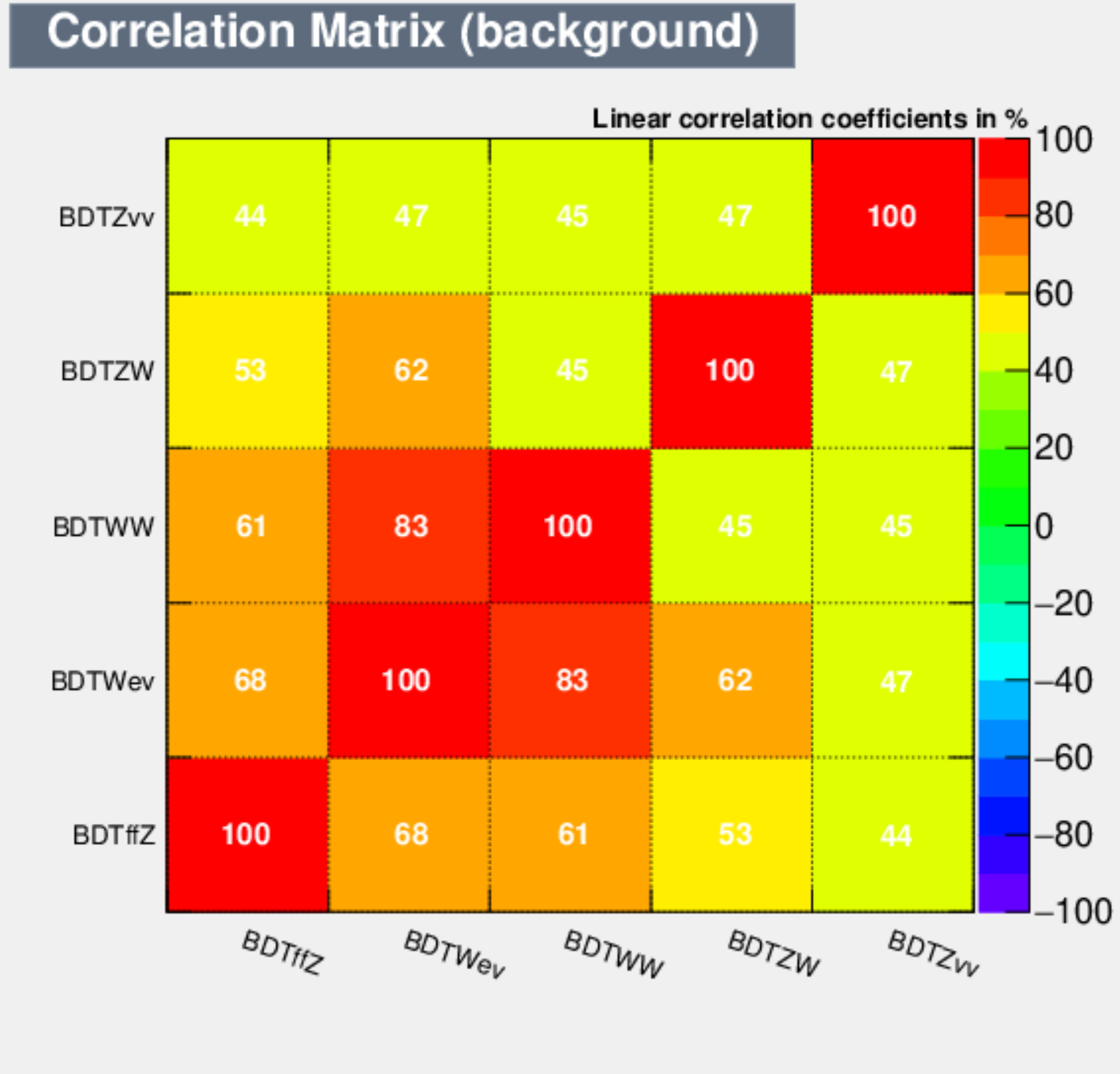}
\includegraphics[height=1.6in]{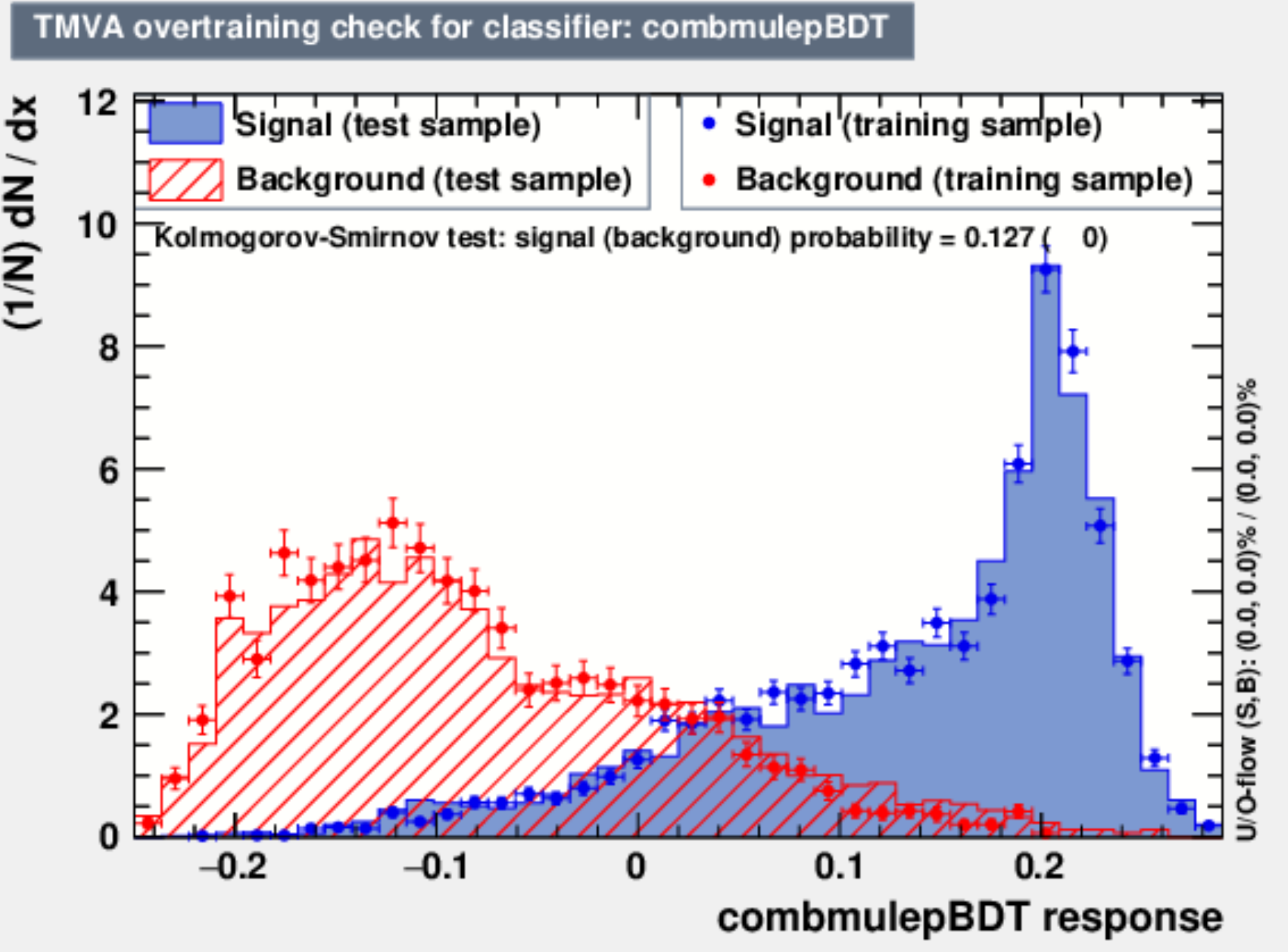}

\vspace{0.15in}

\framebox{\textbf{Hadron Channel BDT of BDTs}}

\vspace{0.15in}

\includegraphics[height=1.6in]{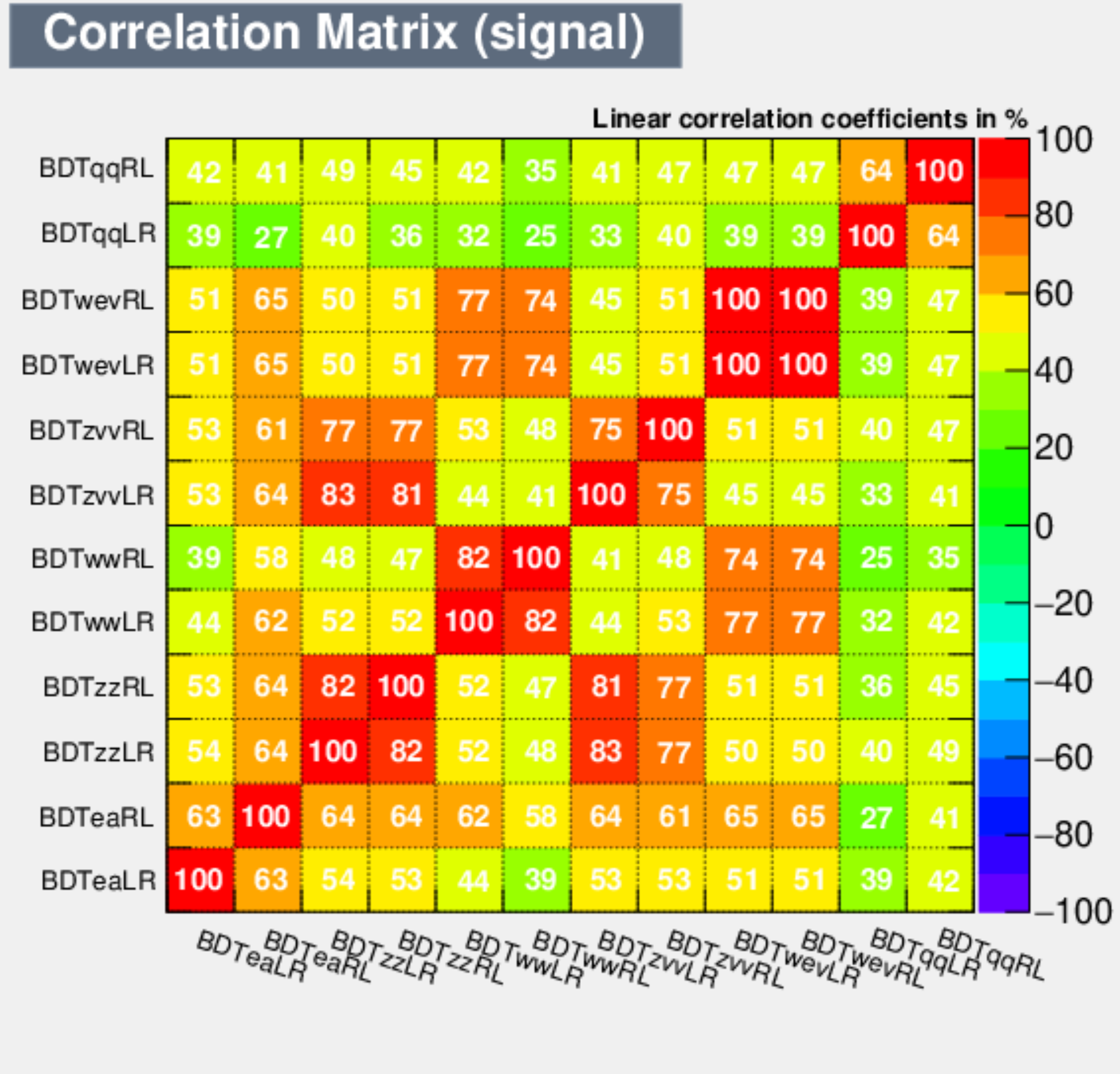}
\includegraphics[height=1.6in]{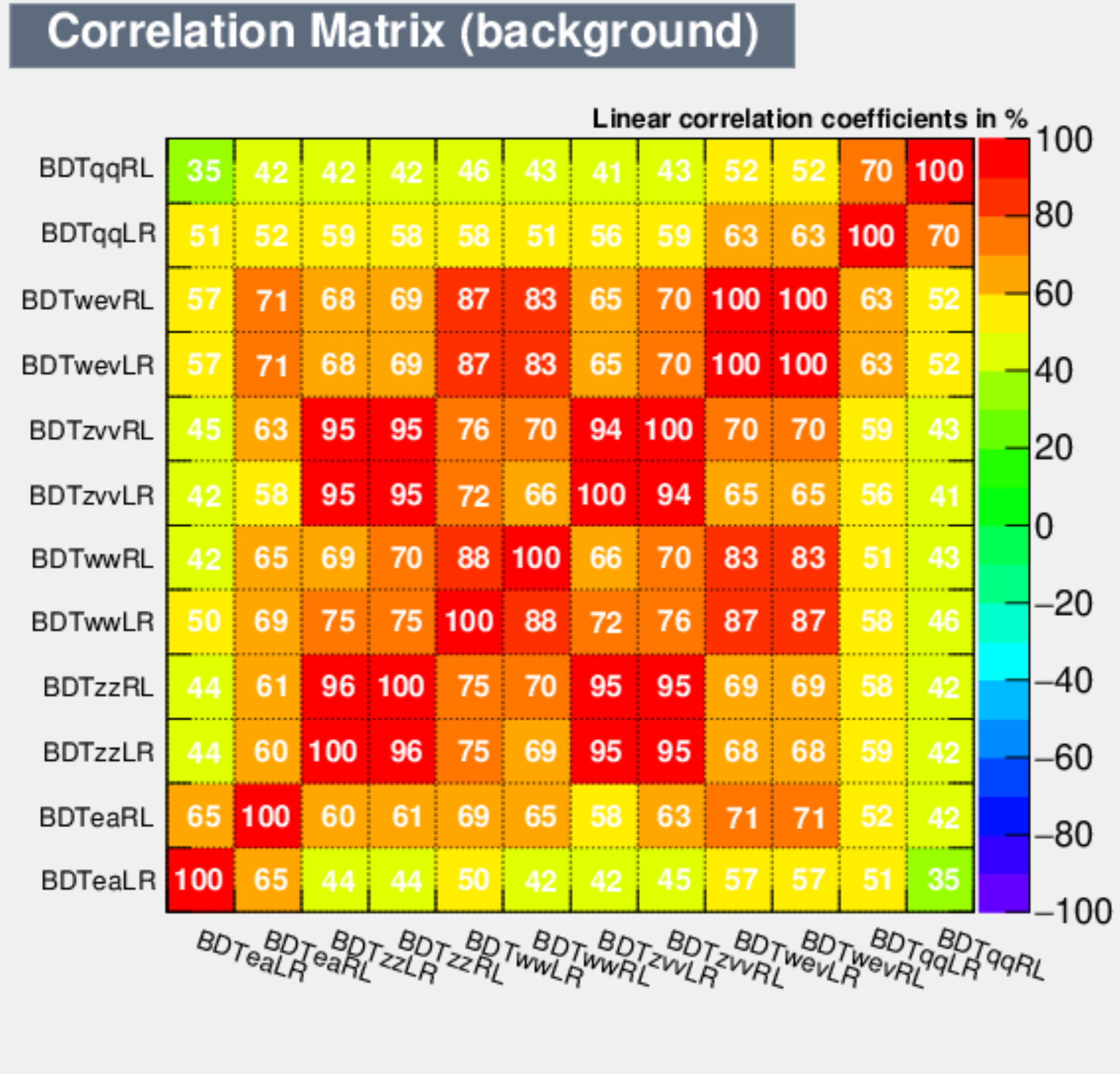}
\includegraphics[height=1.6in]{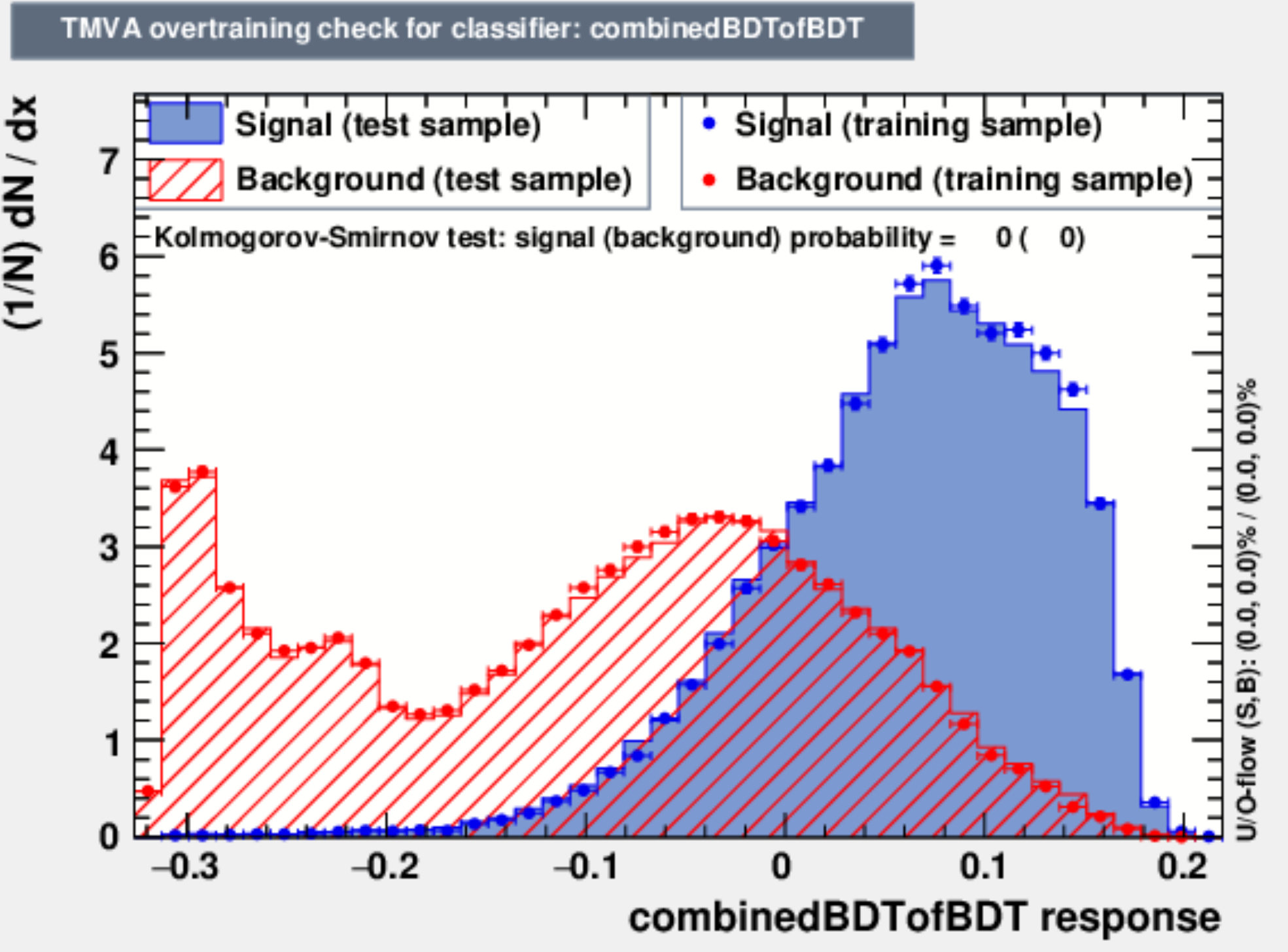}

\end{center}
\caption{Combined BDT signal (left) and background (middle) BDT inputs linear correlation coefficients and outputs (right) for the electron, muon and hadron channels. In the BDT output distributions, test samples and training samples are plotted separately and show no evidence of overtraining.}
\label{fig:bdtbdt}
\end{figure}

\begin{figure}[p]
\begin{center}
\framebox{\textbf{Electron and Muon Channel BDTs: All Backgrounds}}

\vspace{0.25in}

\includegraphics[height=1.35in]{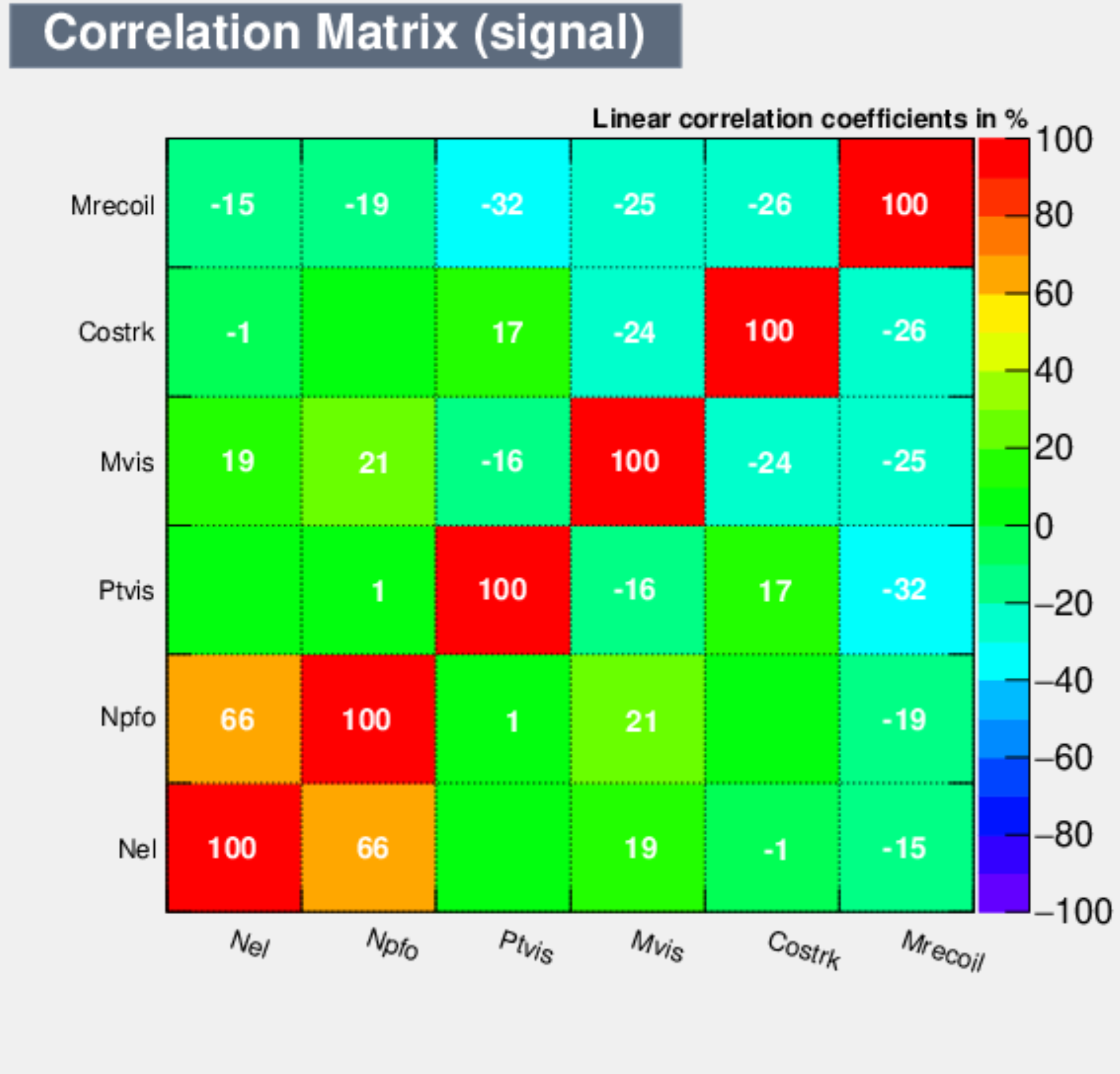}
\includegraphics[height=1.35in]{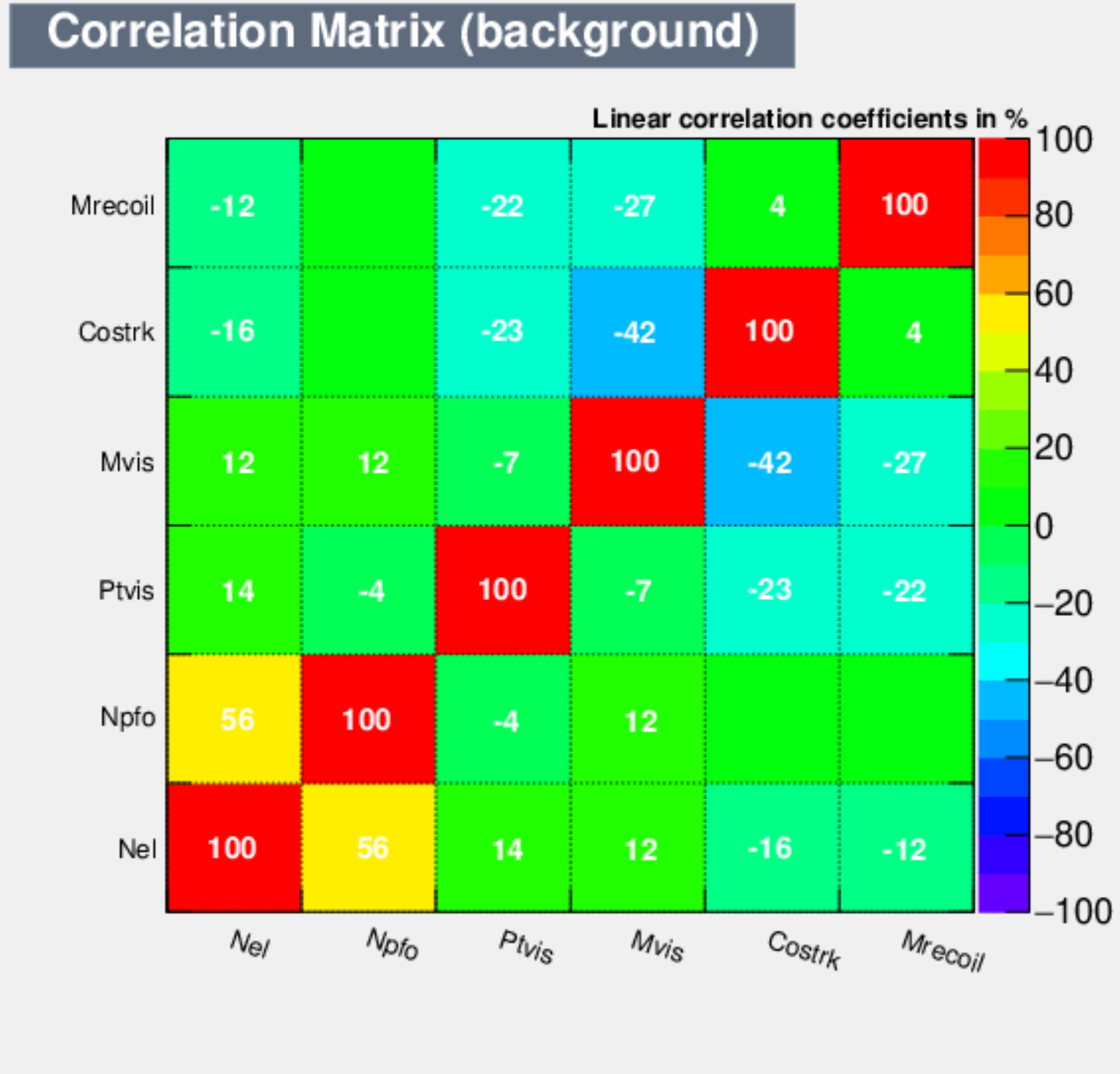}
\includegraphics[height=1.35in]{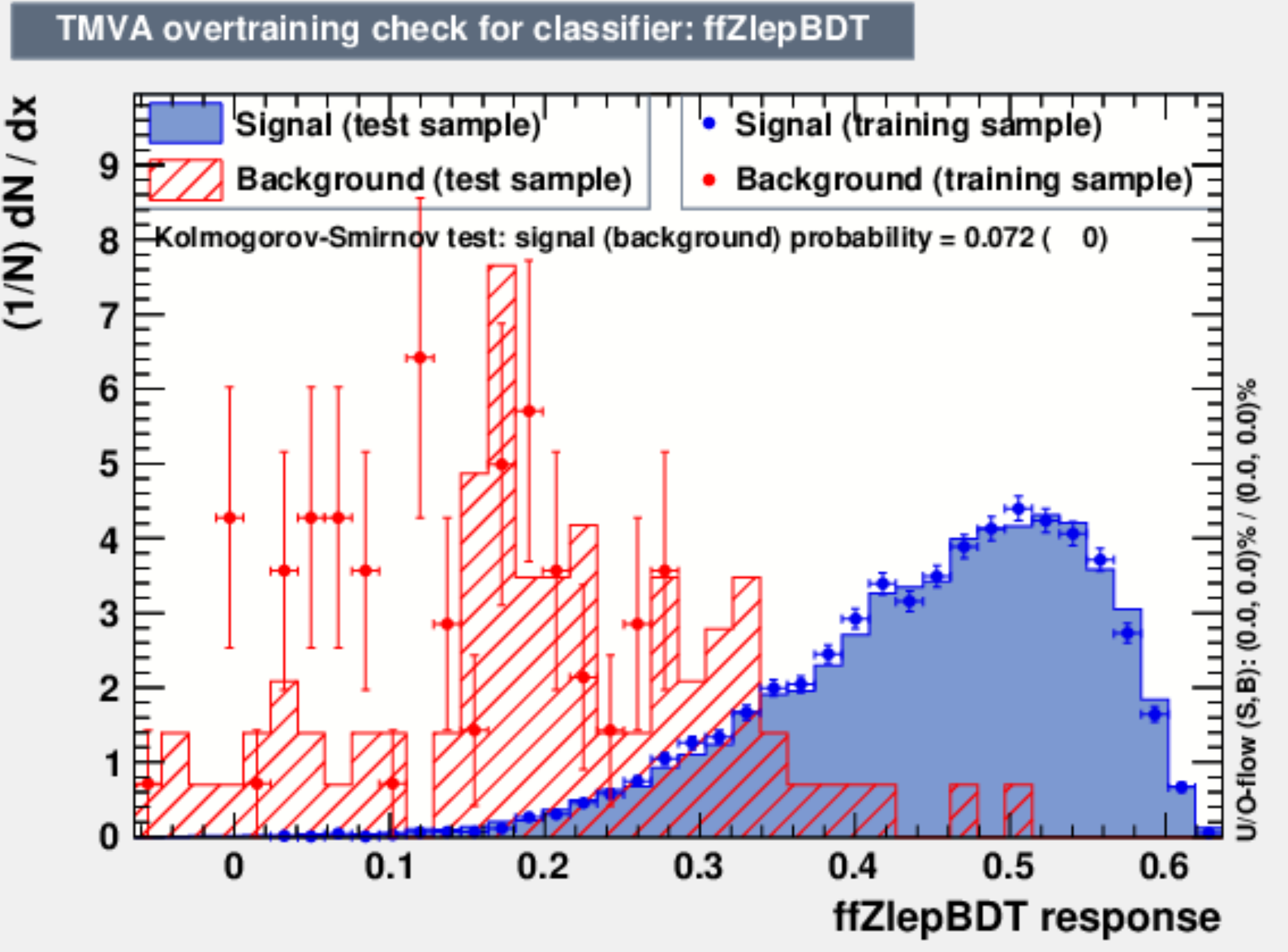}

\includegraphics[height=1.35in]{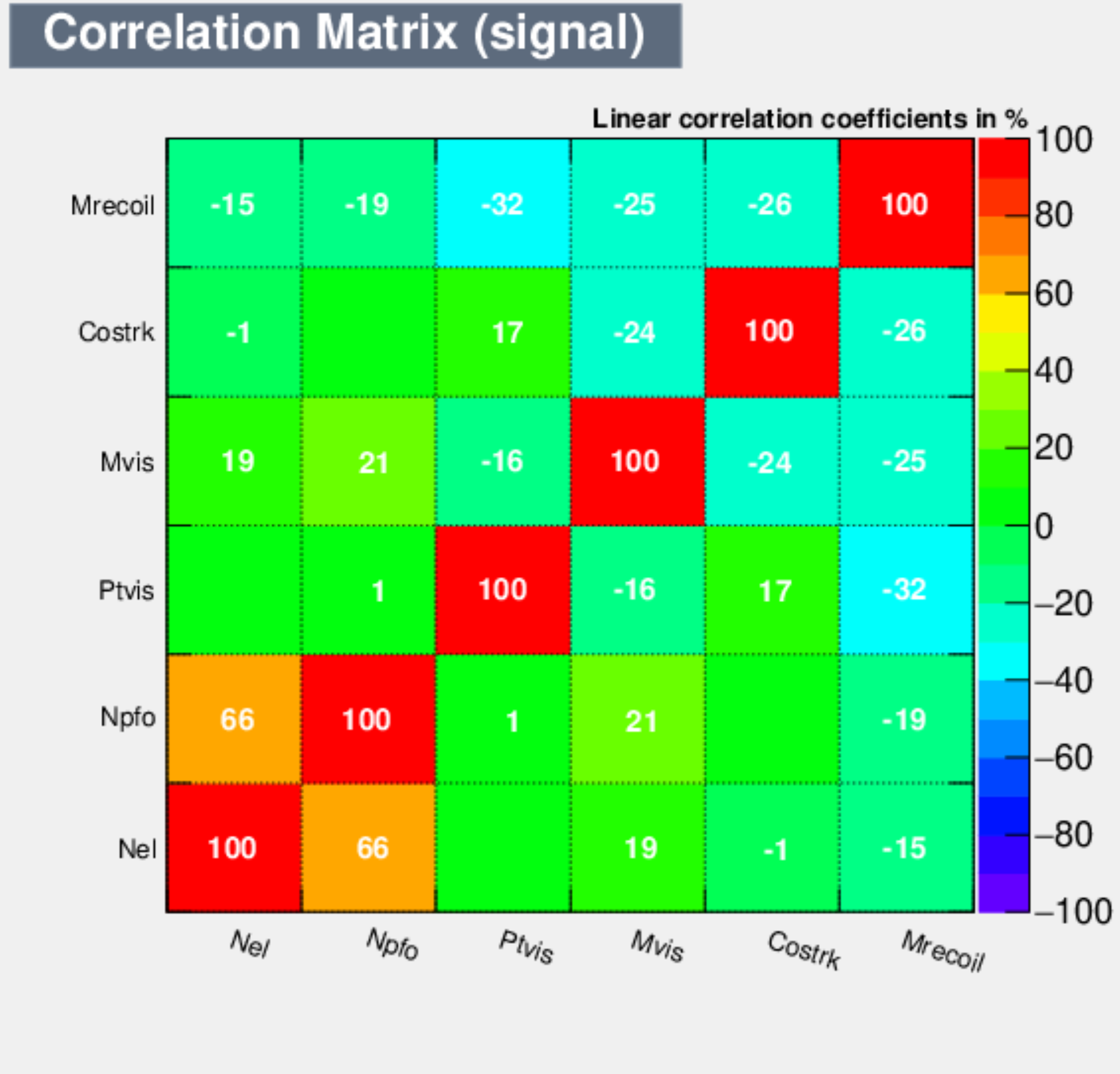}
\includegraphics[height=1.35in]{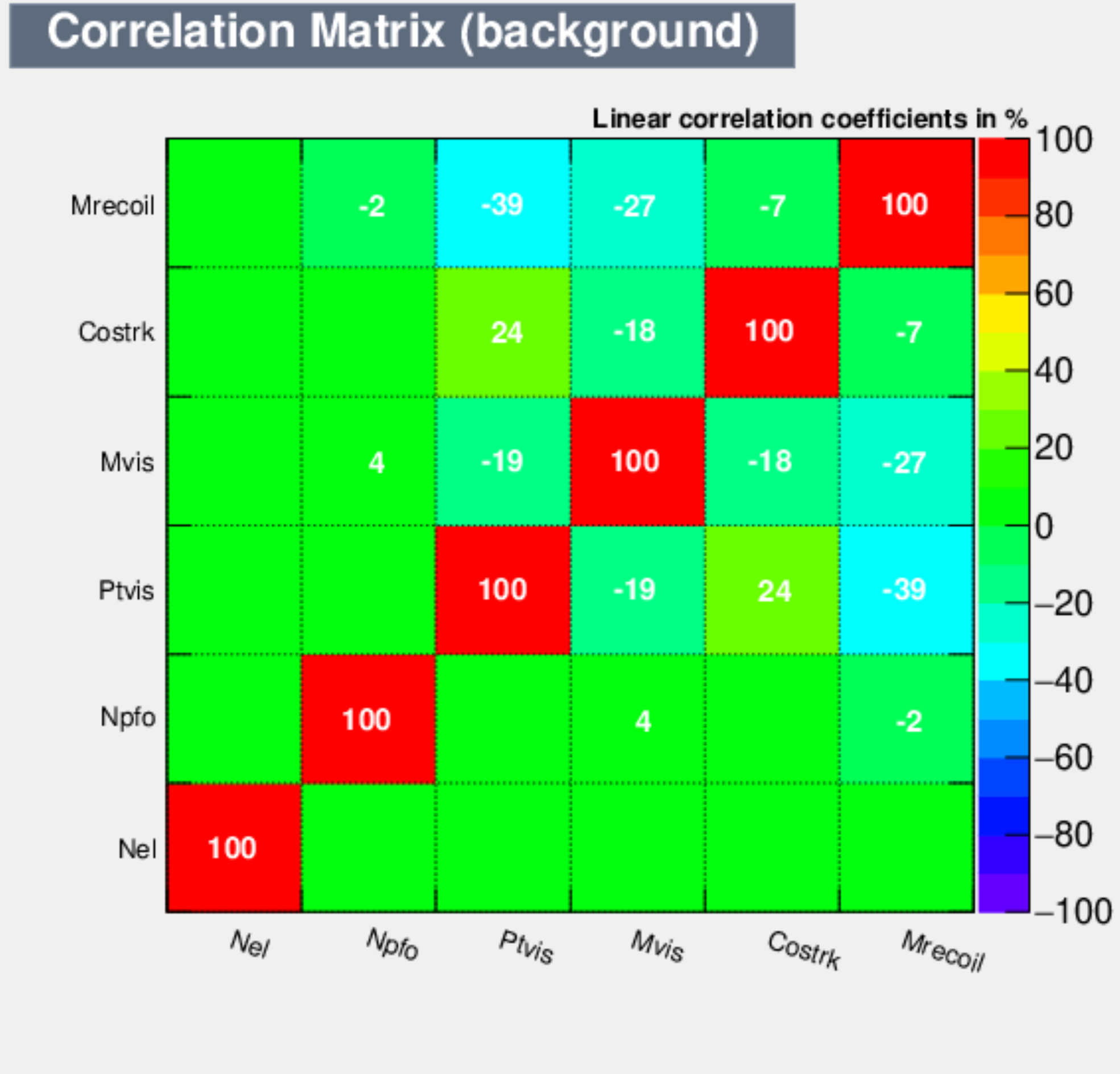}
\includegraphics[height=1.35in]{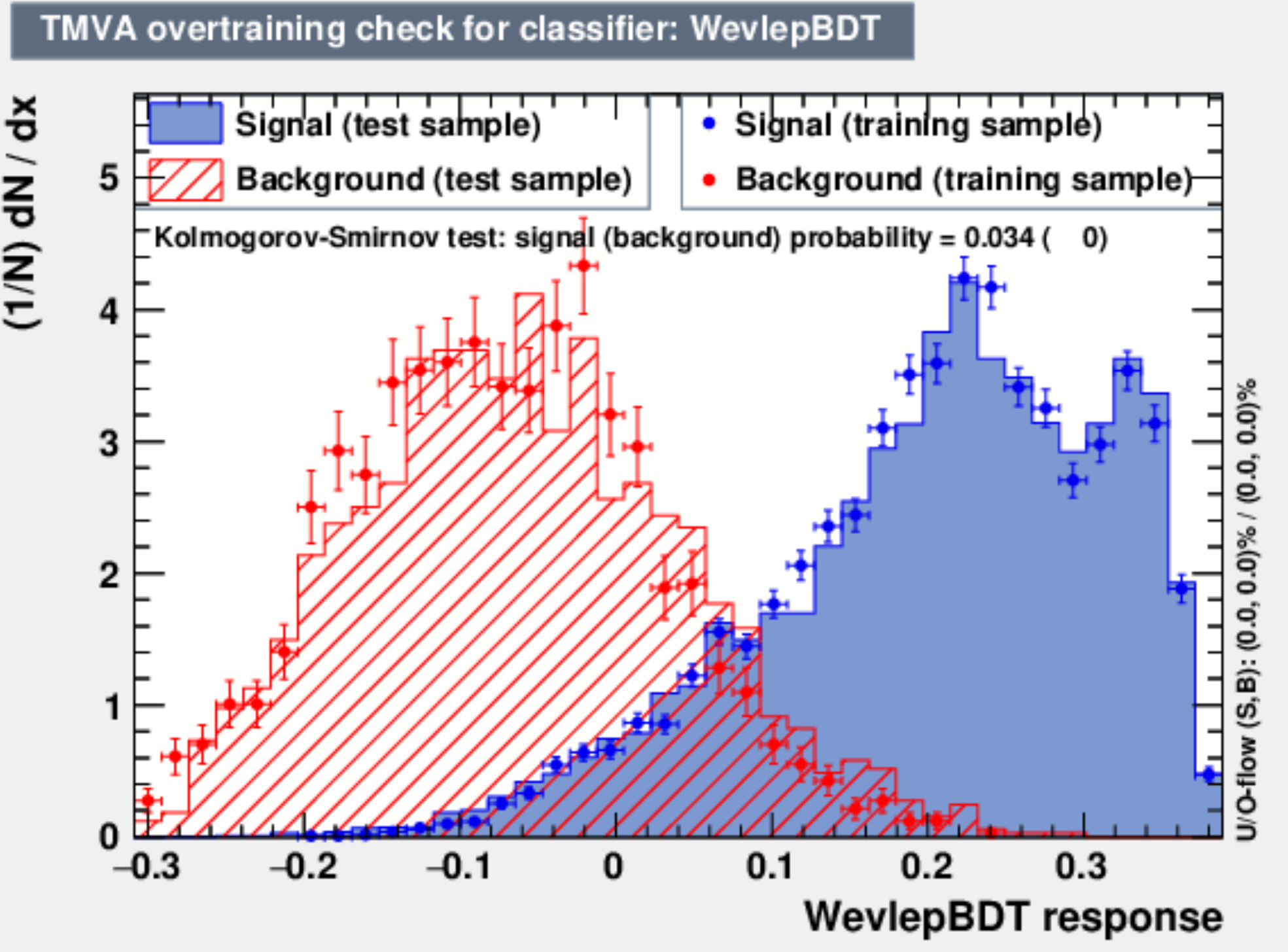}

\includegraphics[height=1.35in]{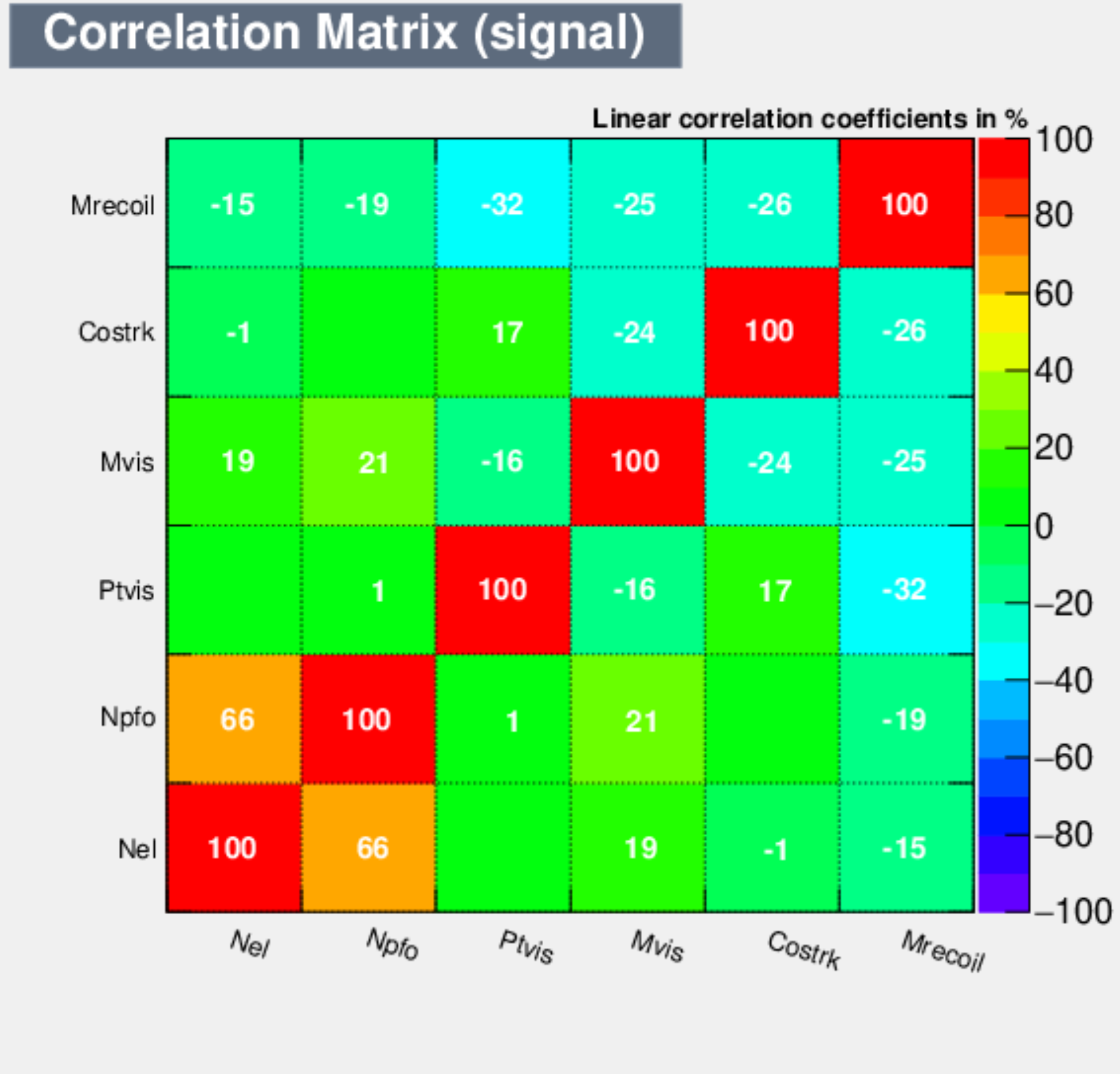}
\includegraphics[height=1.35in]{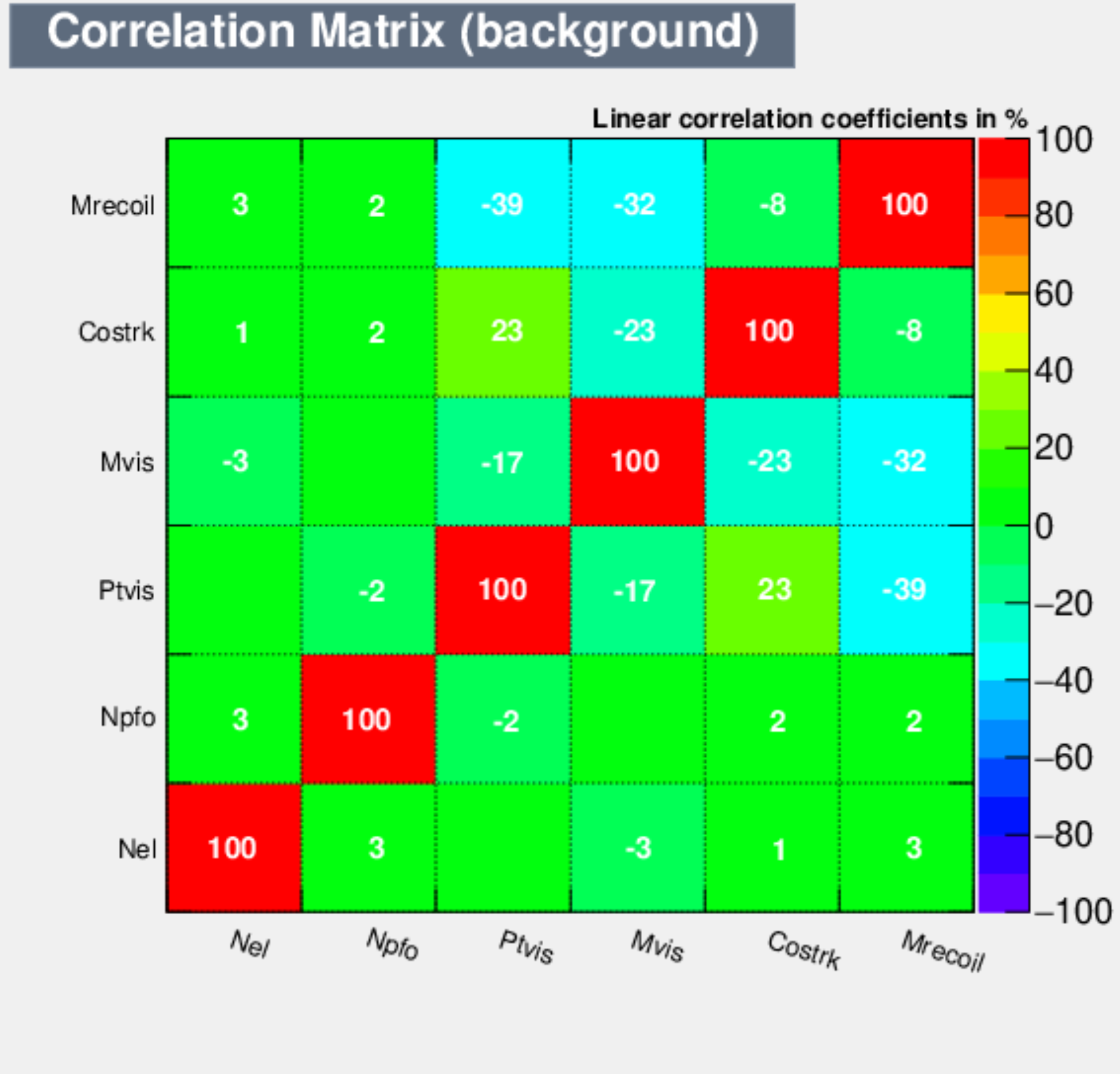}
\includegraphics[height=1.35in]{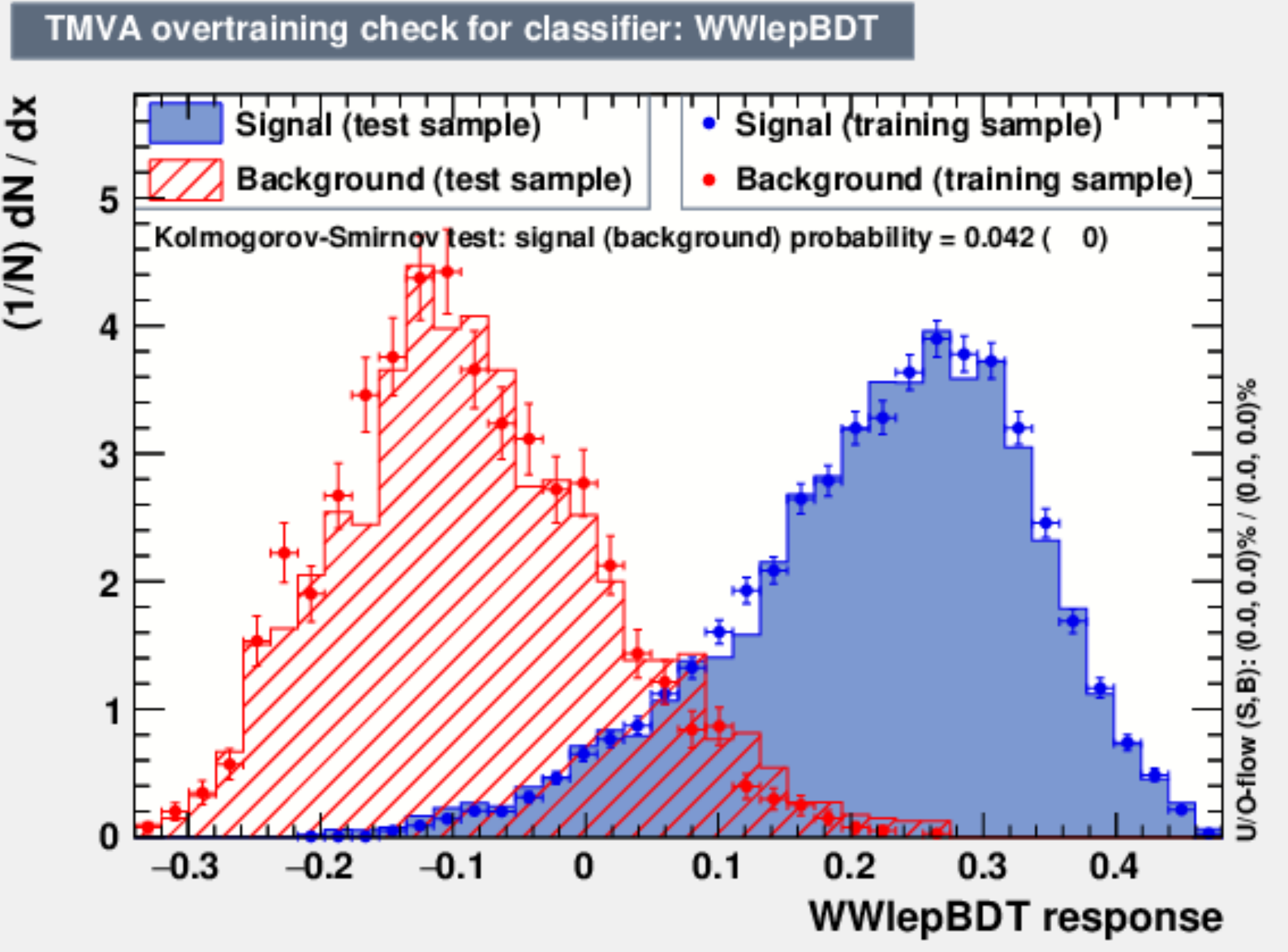}

\includegraphics[height=1.35in]{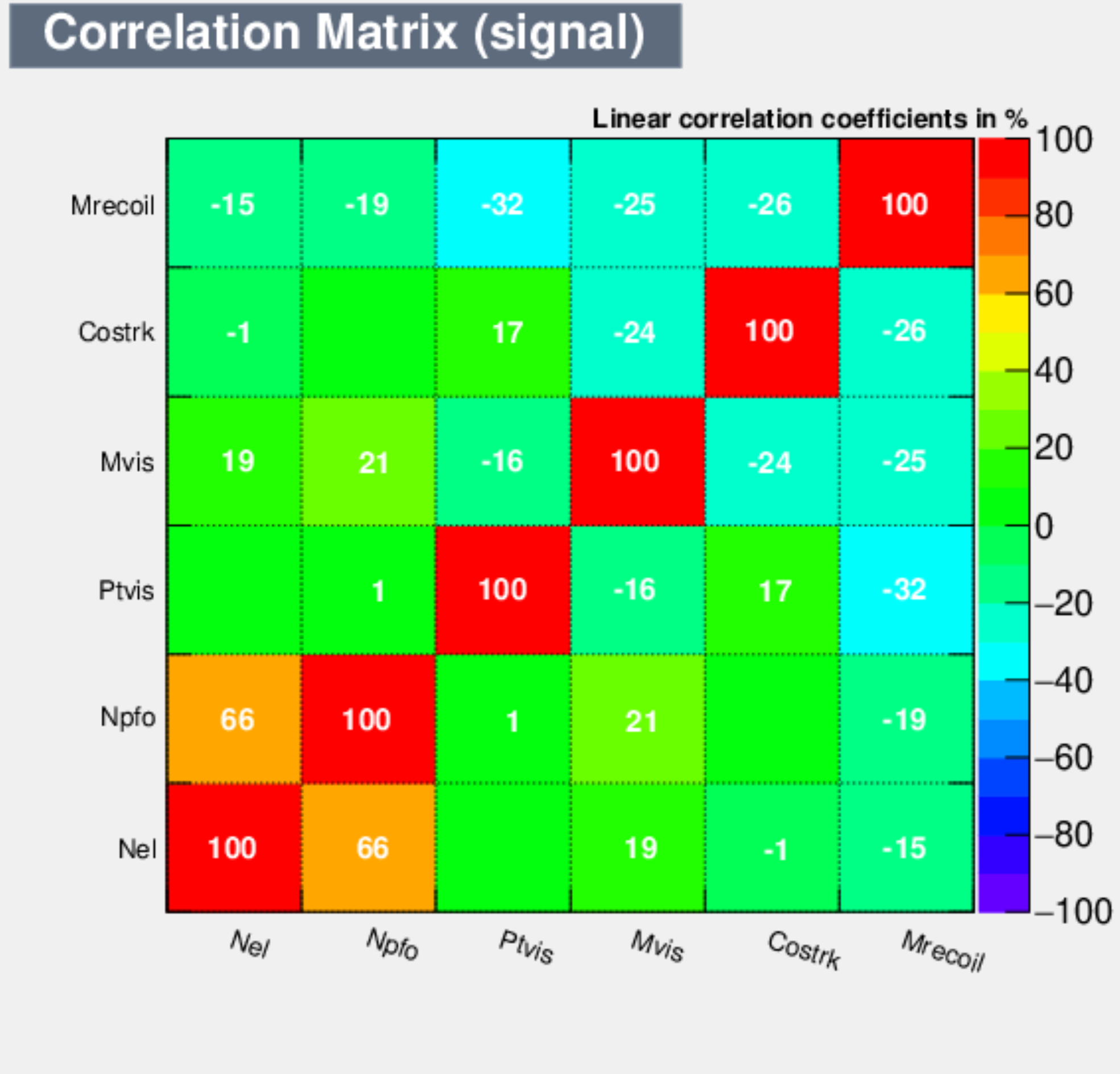}
\includegraphics[height=1.35in]{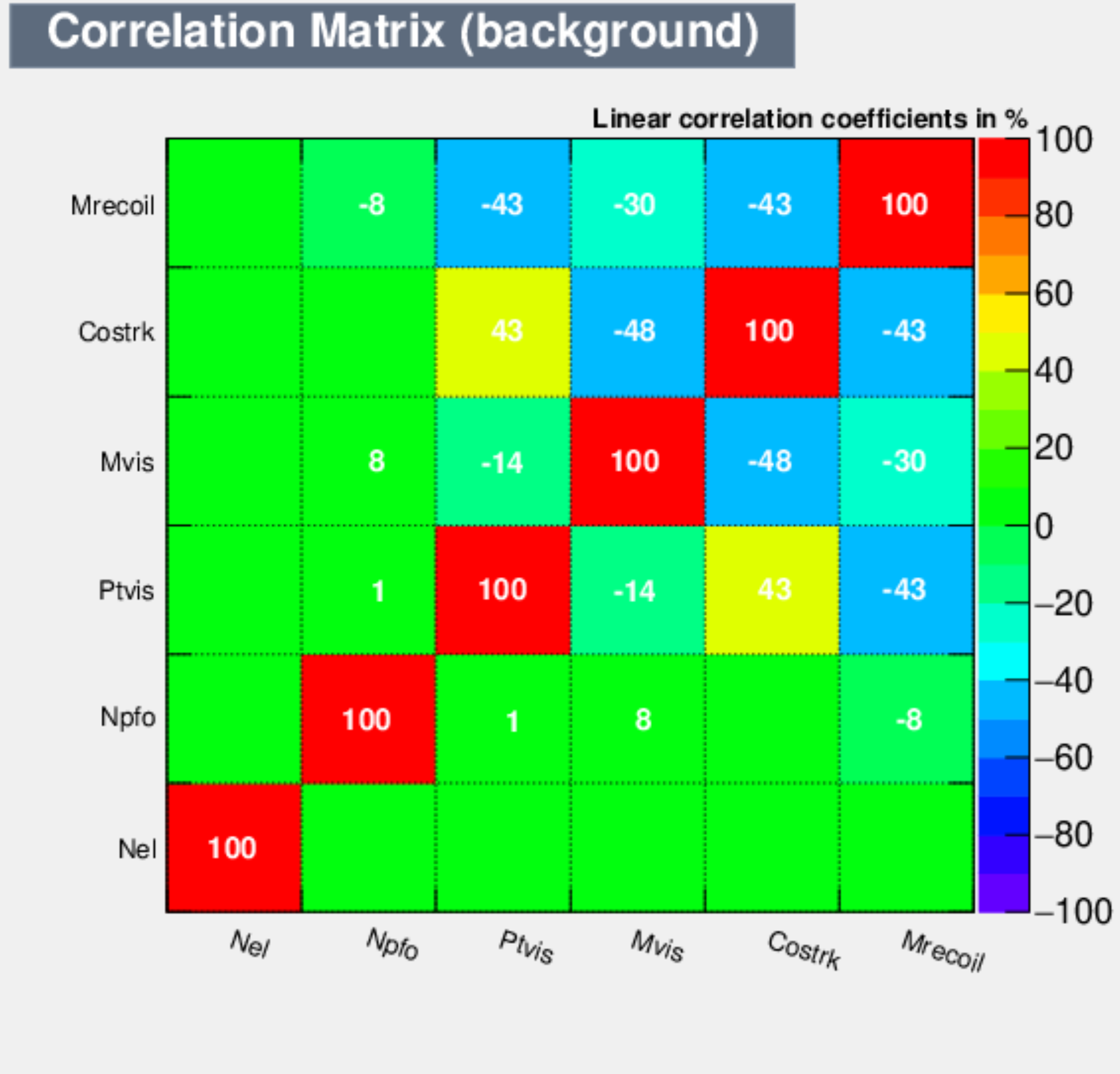}
\includegraphics[height=1.35in]{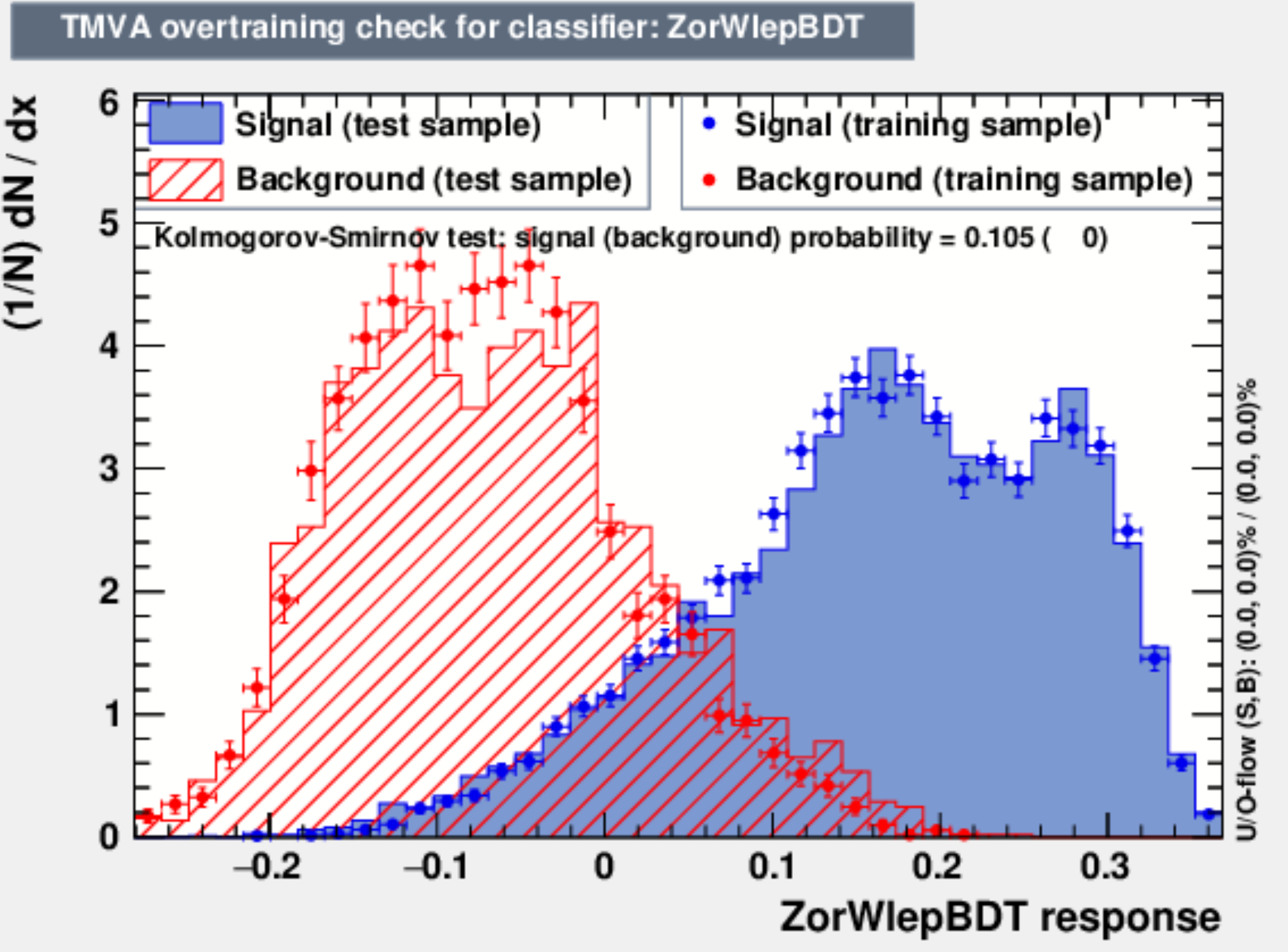}

\includegraphics[height=1.35in]{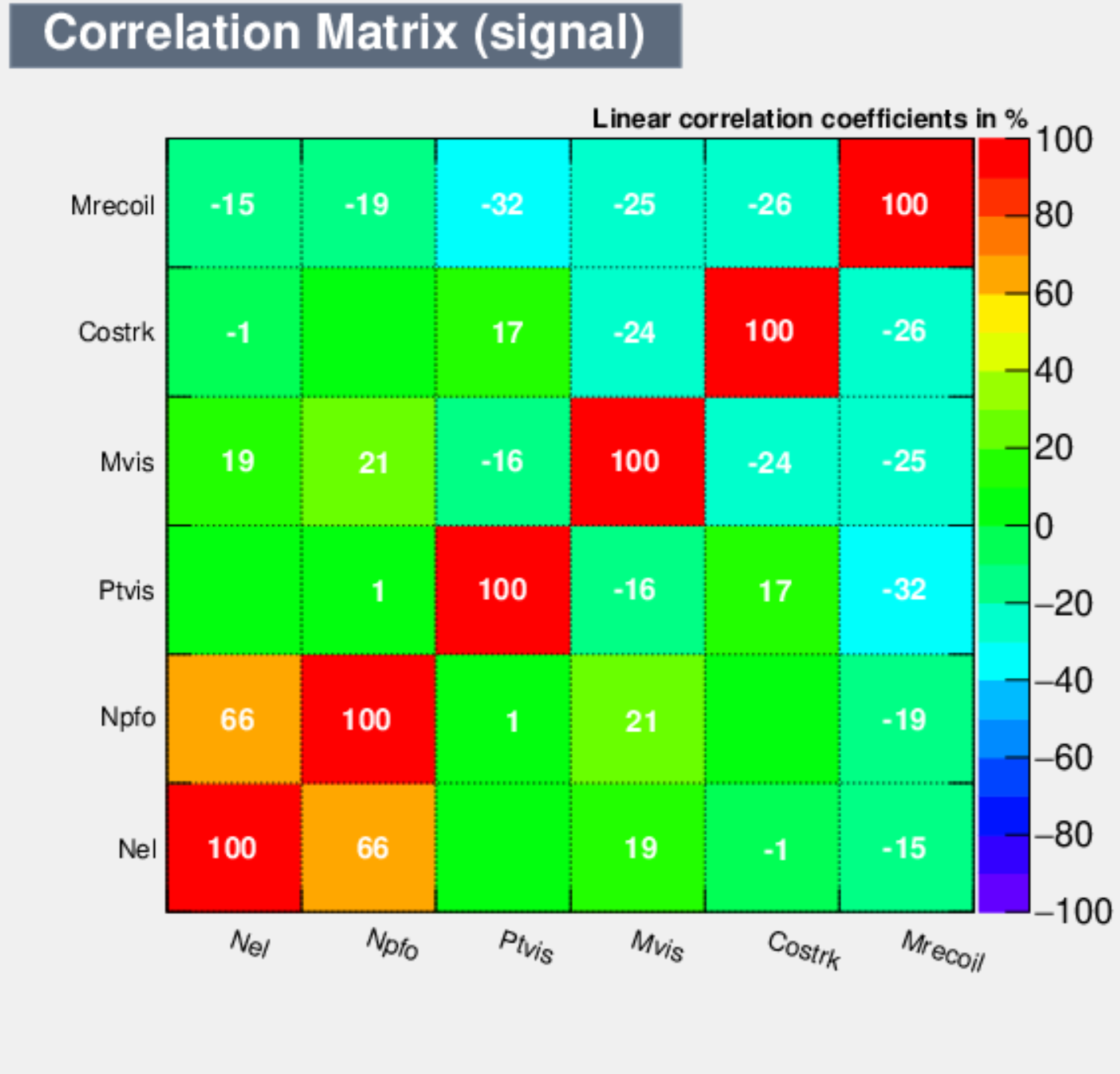}
\includegraphics[height=1.35in]{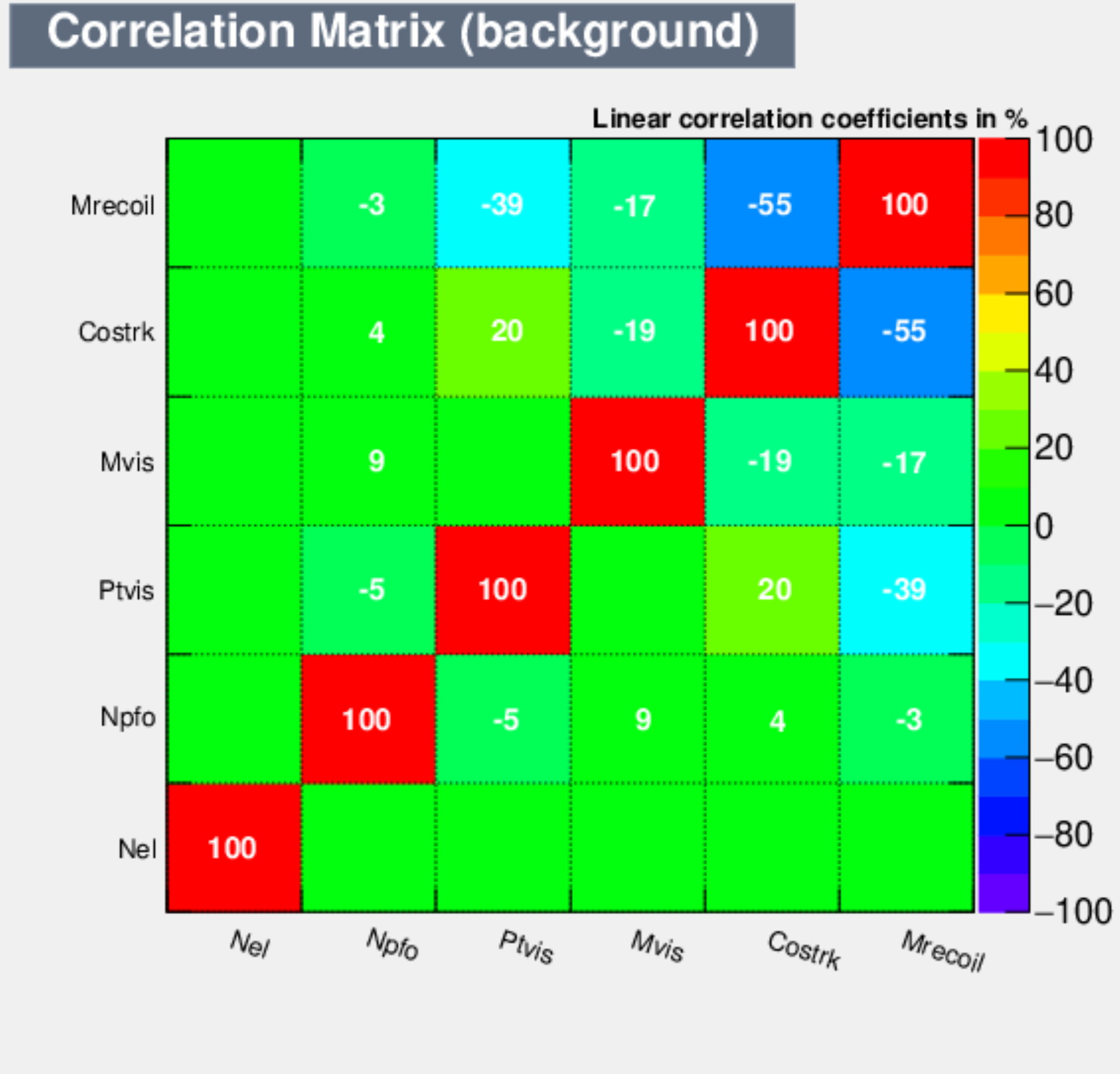}
\includegraphics[height=1.35in]{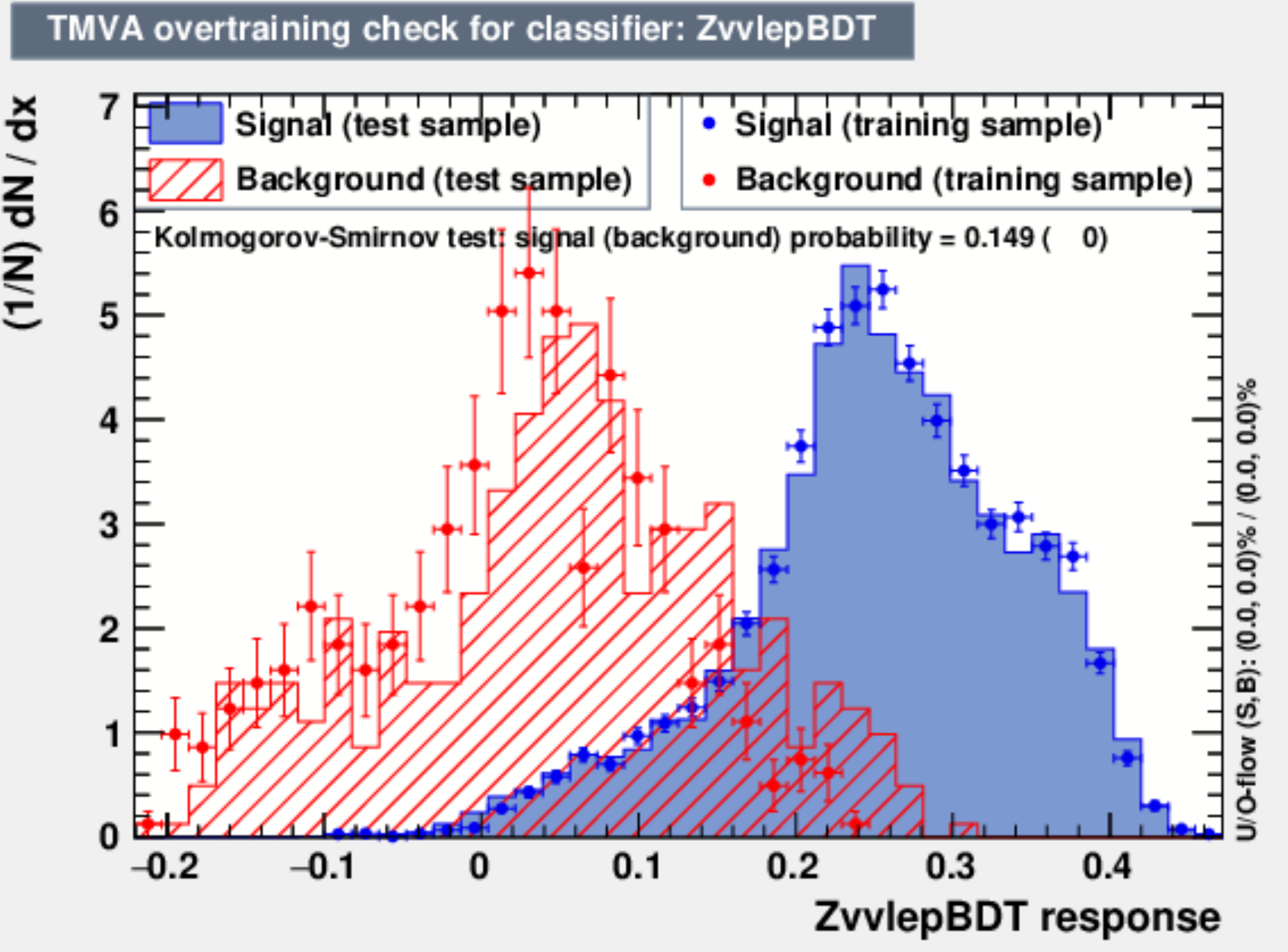}

\end{center}
\caption{BDT signal (left) and background (middle) inputs and correlations and outputs (right) for the lepton channels. From top to bottom are $e^- e^+ \rightarrow \ell^+ \ell^-$ (106605/106606), $e^- e^+ \rightarrow  e\nu W$ (106586/106588), $e^- e^+ \rightarrow  WW$ (106581/106582), single $e^- e^+ \rightarrow  Z/W$ (106568), and $e^- e^+ \rightarrow  \nu \bar{\nu}Z$ (106589) background samples. In the BDT output distributions, test samples and training samples are plotted separately and show no evidence of overtraining.}
\label{fig:lep}
\end{figure}

\begin{figure}[p]
\begin{center}
\framebox{\textbf{Hadron Channel BDTs: 2f and 3f Backgrounds}}

\vspace{0.25in}

\includegraphics[height=1.65in]{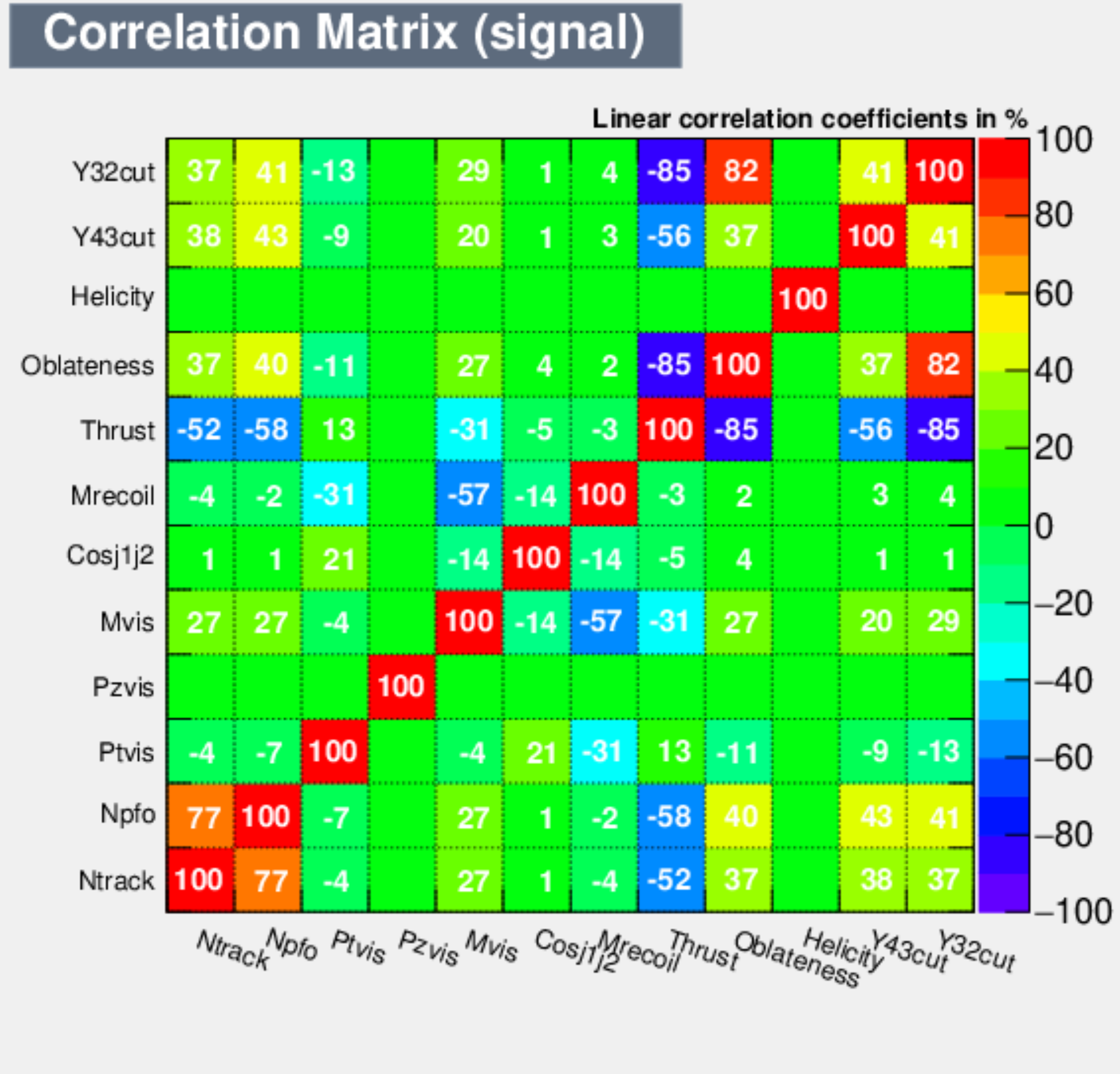}
\includegraphics[height=1.65in]{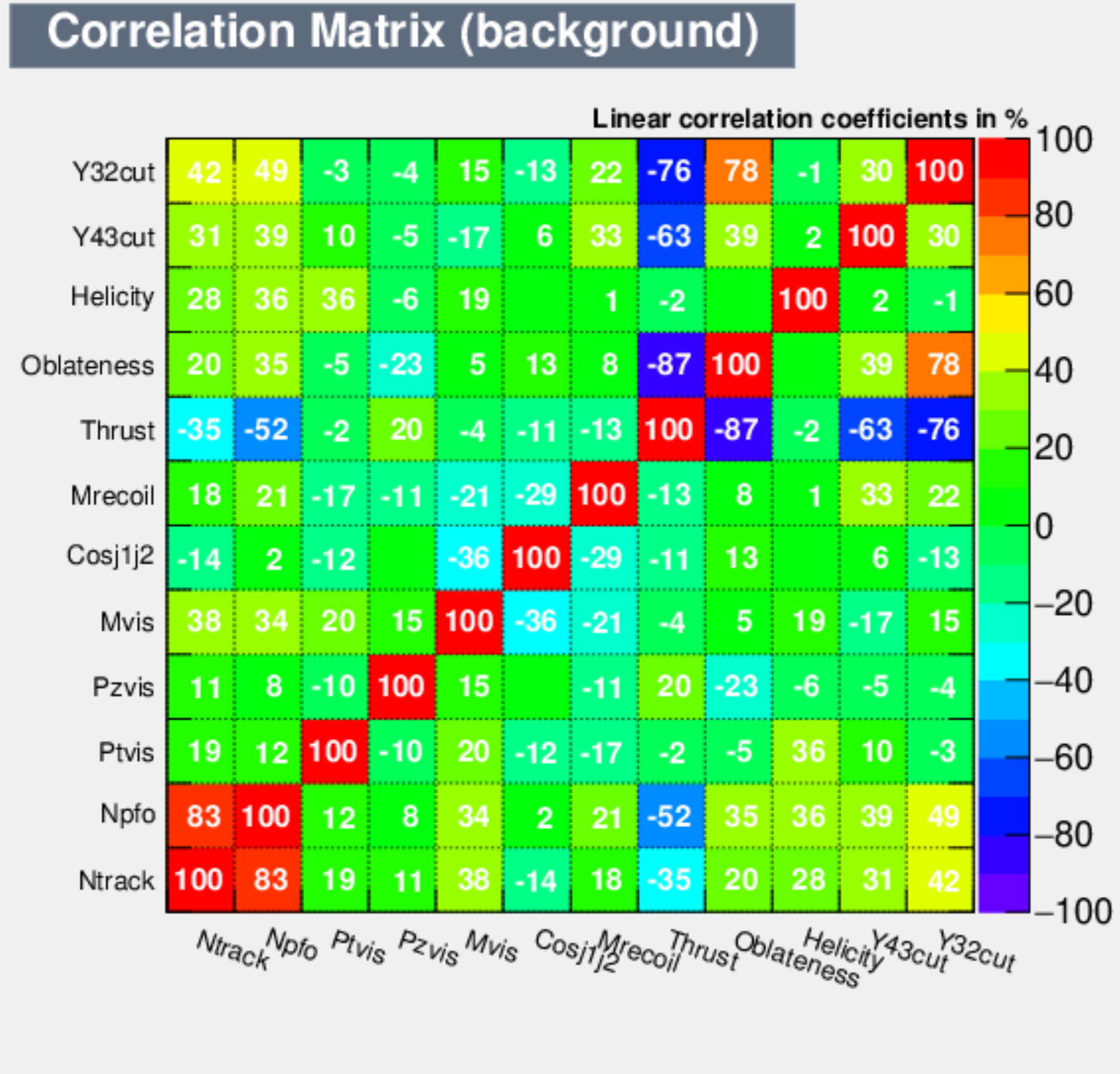}
\includegraphics[height=1.65in]{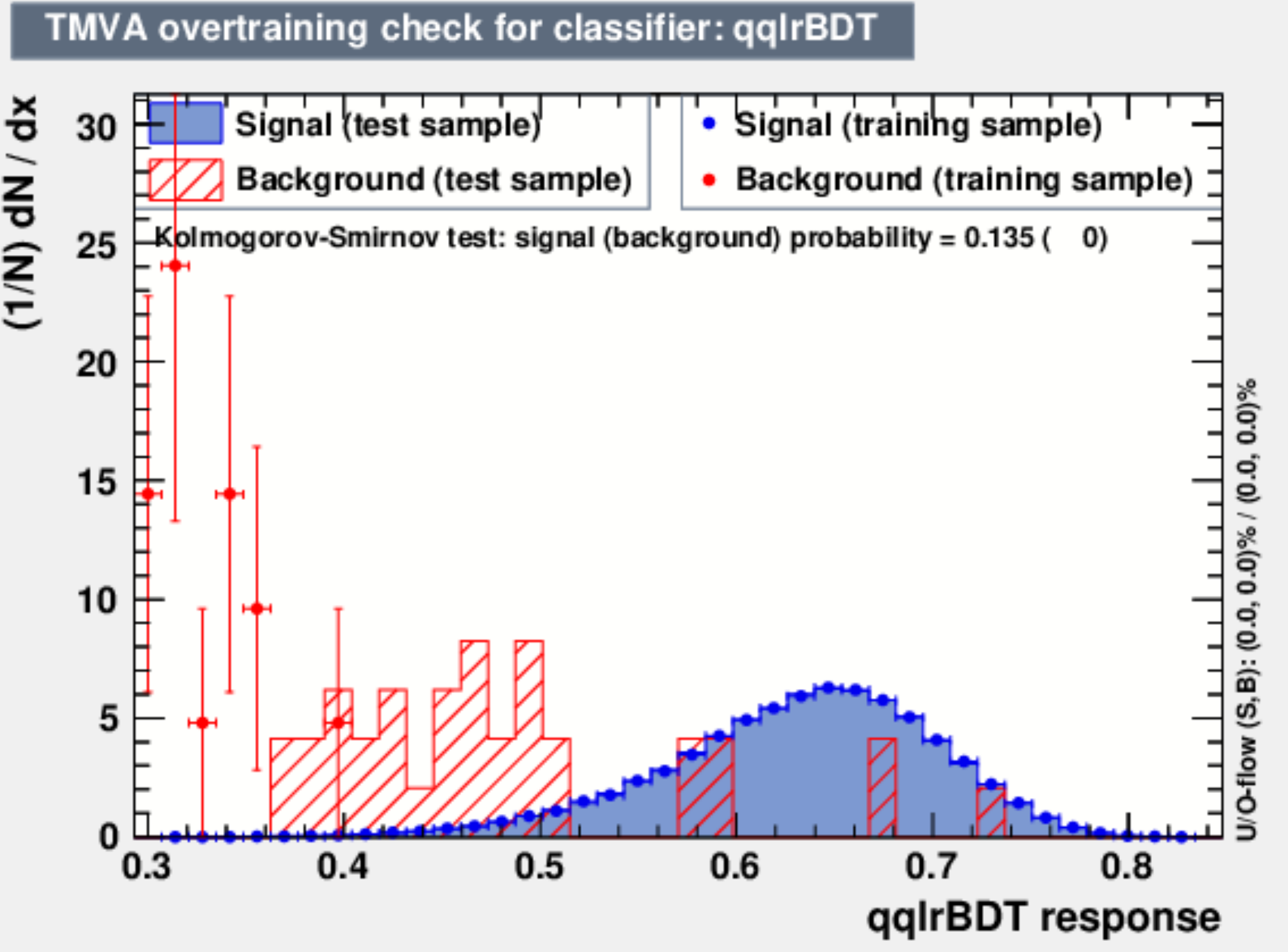}

\includegraphics[height=1.65in]{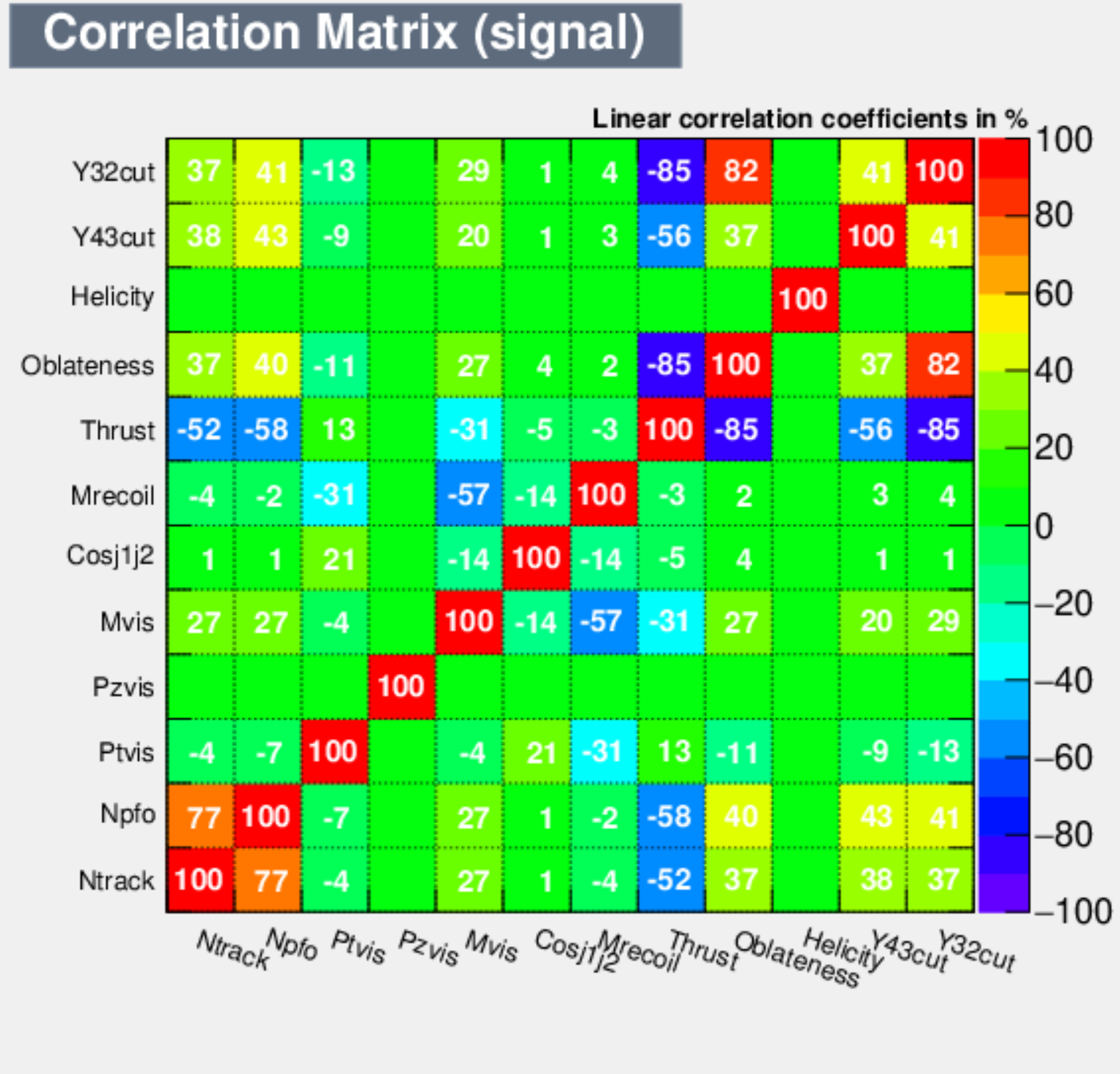}
\includegraphics[height=1.65in]{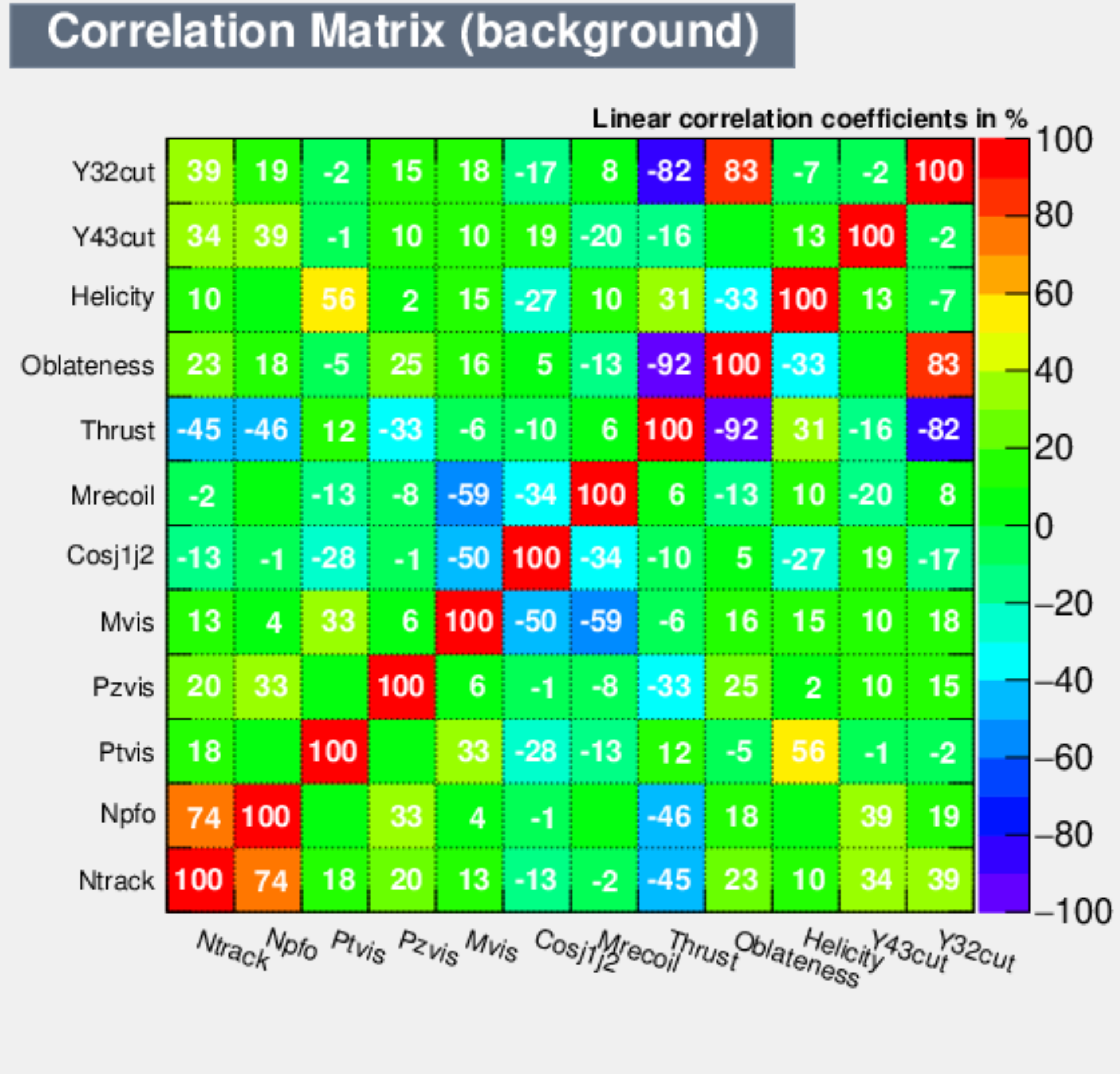}
\includegraphics[height=1.65in]{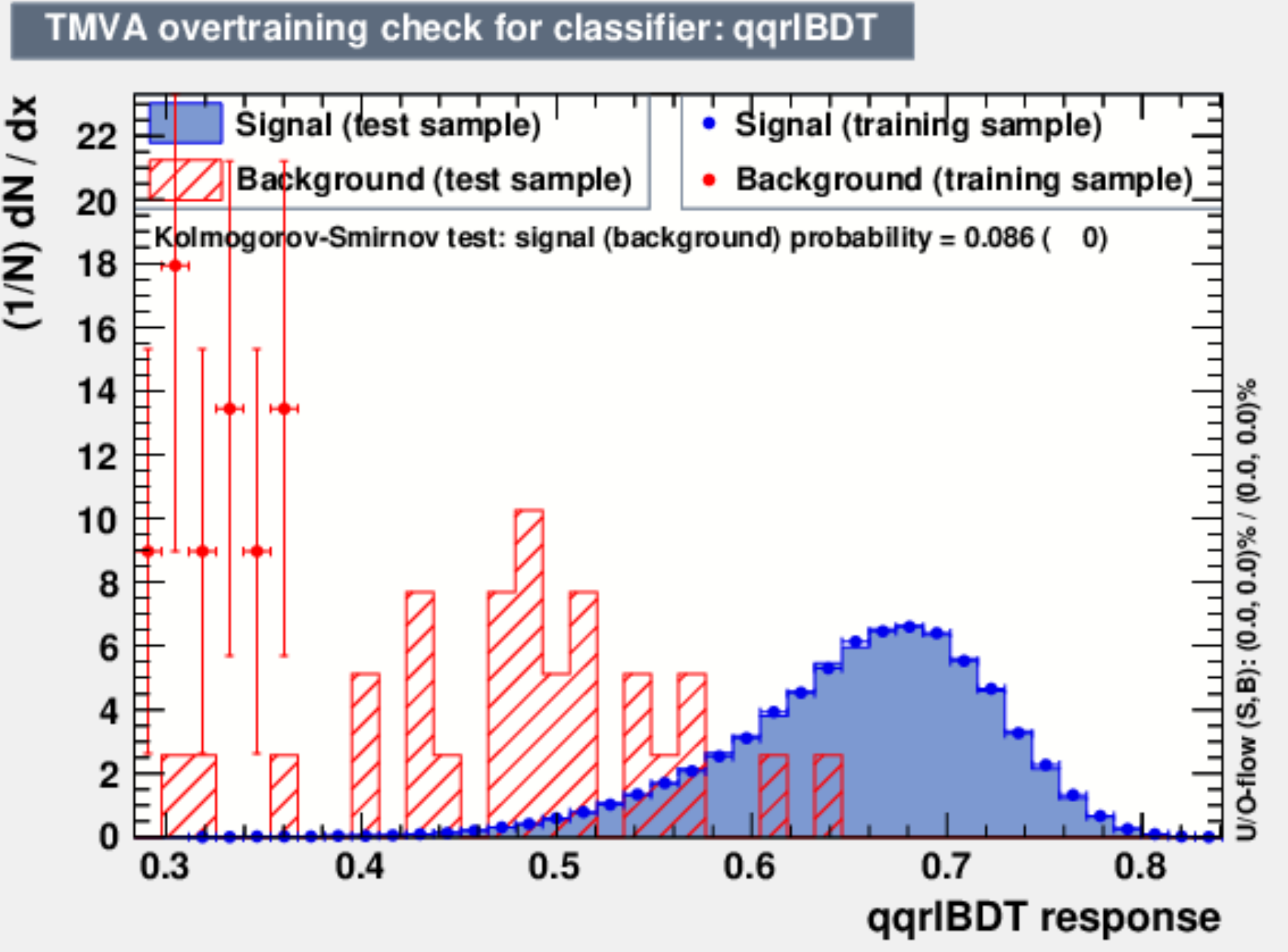}

\includegraphics[height=1.65in]{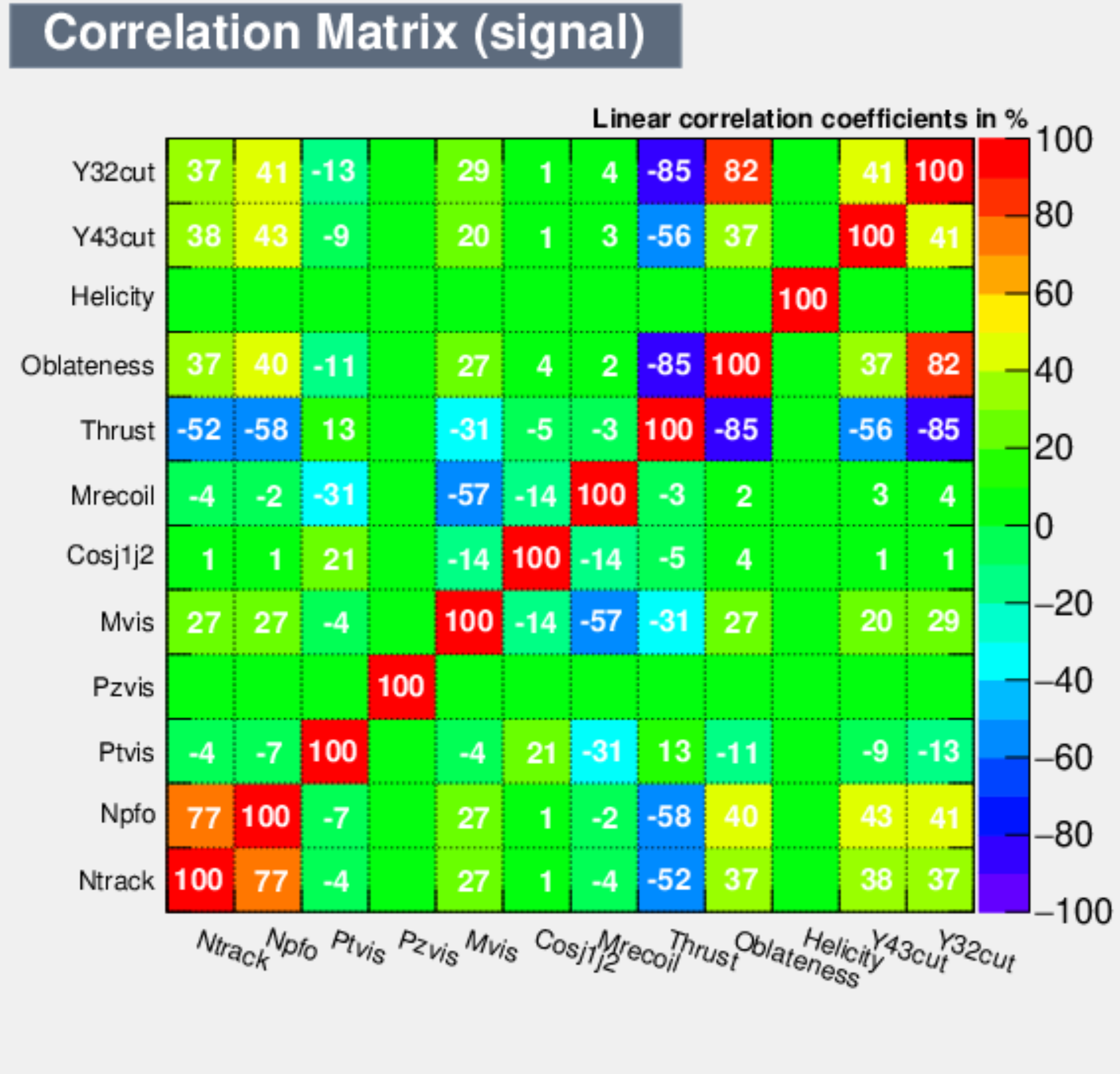}
\includegraphics[height=1.65in]{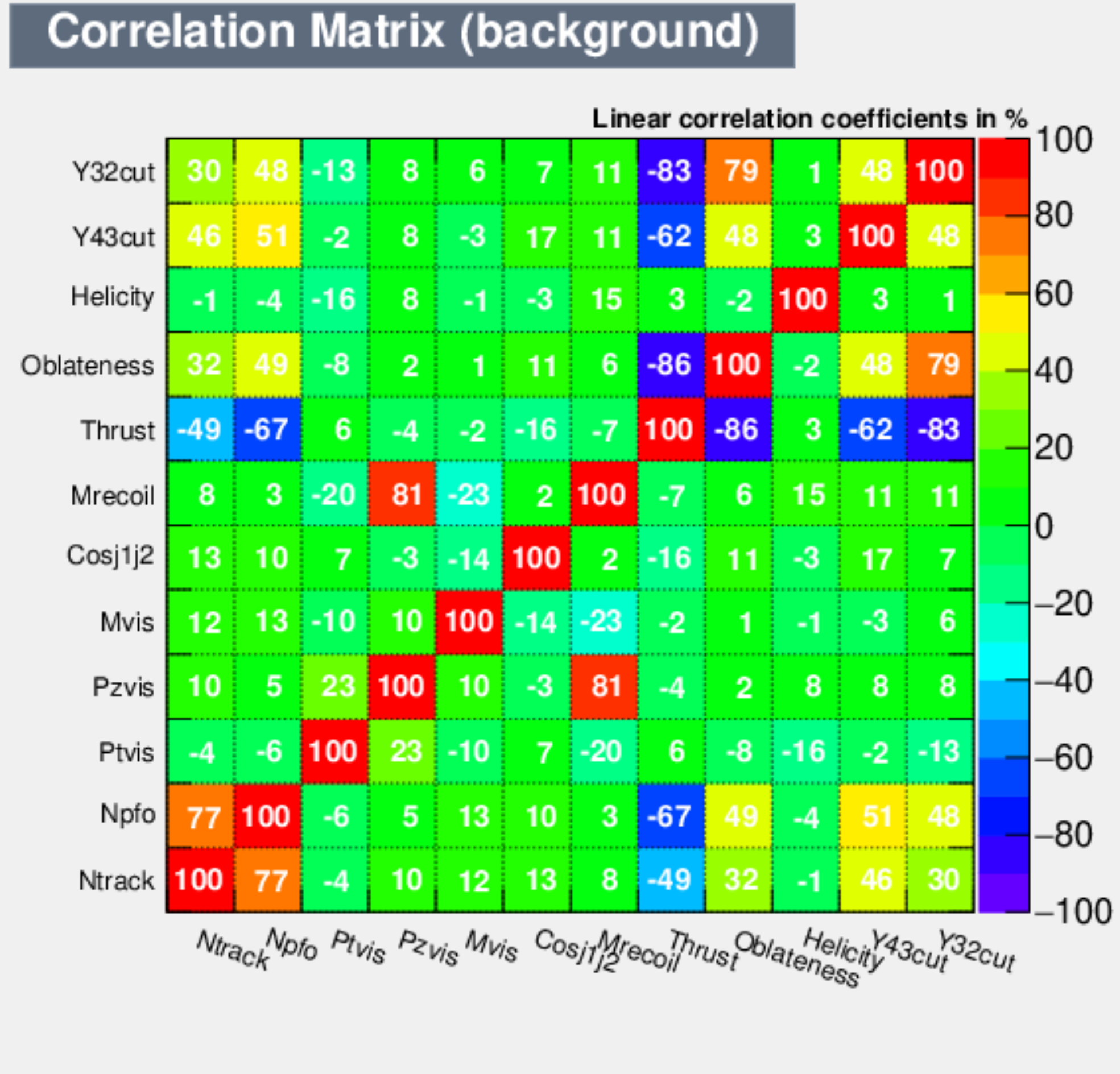}
\includegraphics[height=1.65in]{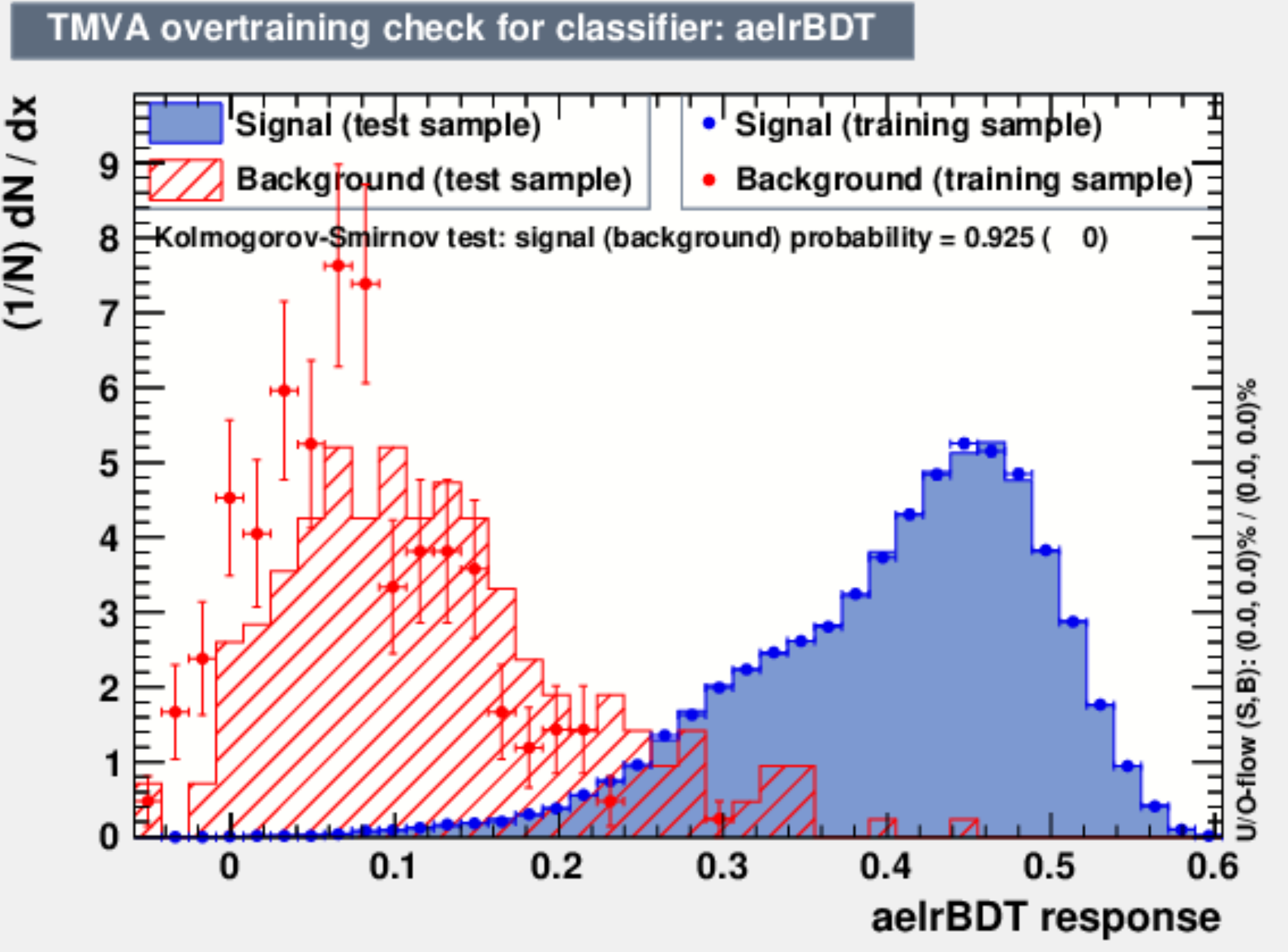}

\includegraphics[height=1.65in]{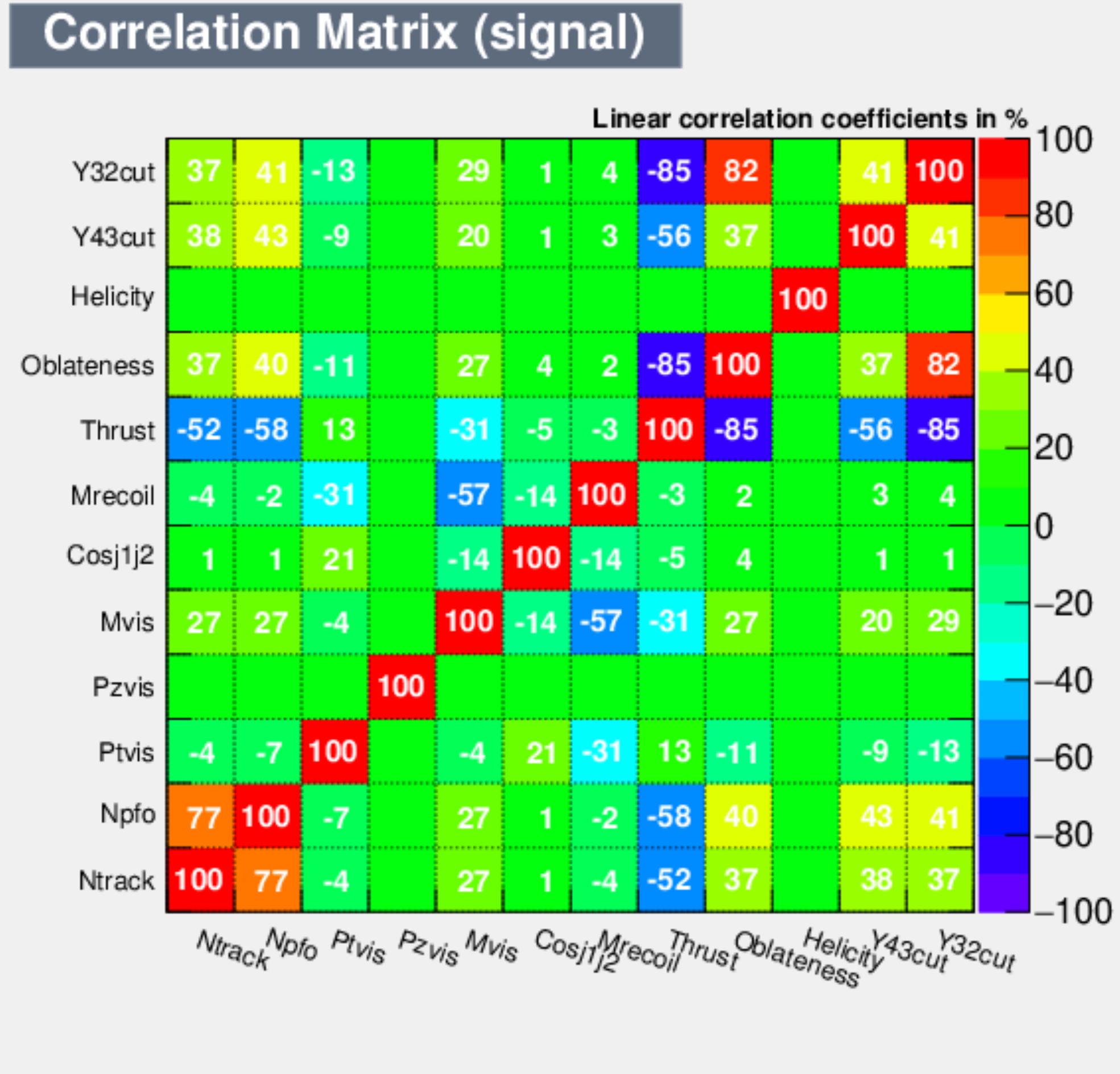}
\includegraphics[height=1.65in]{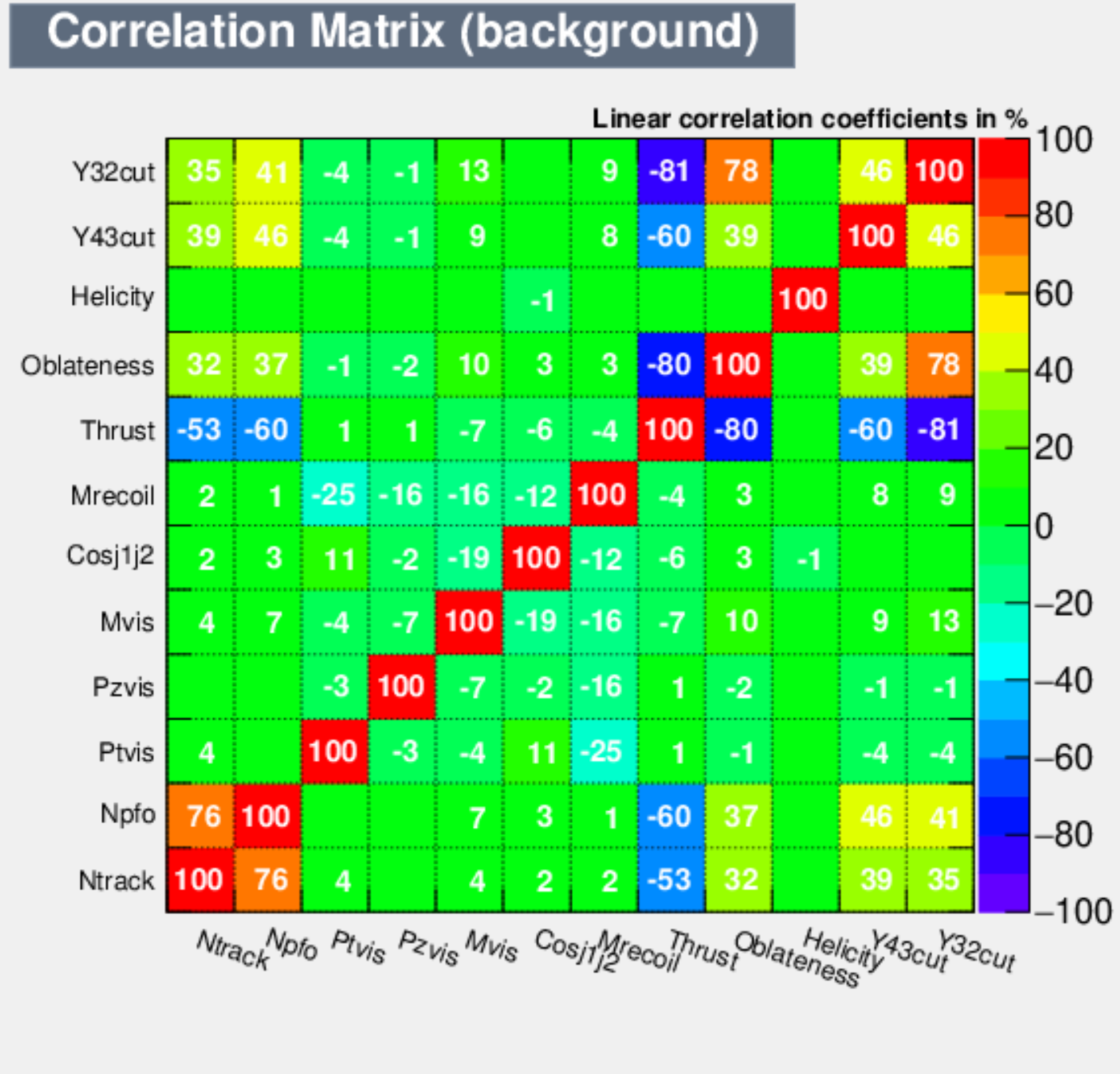}
\includegraphics[height=1.65in]{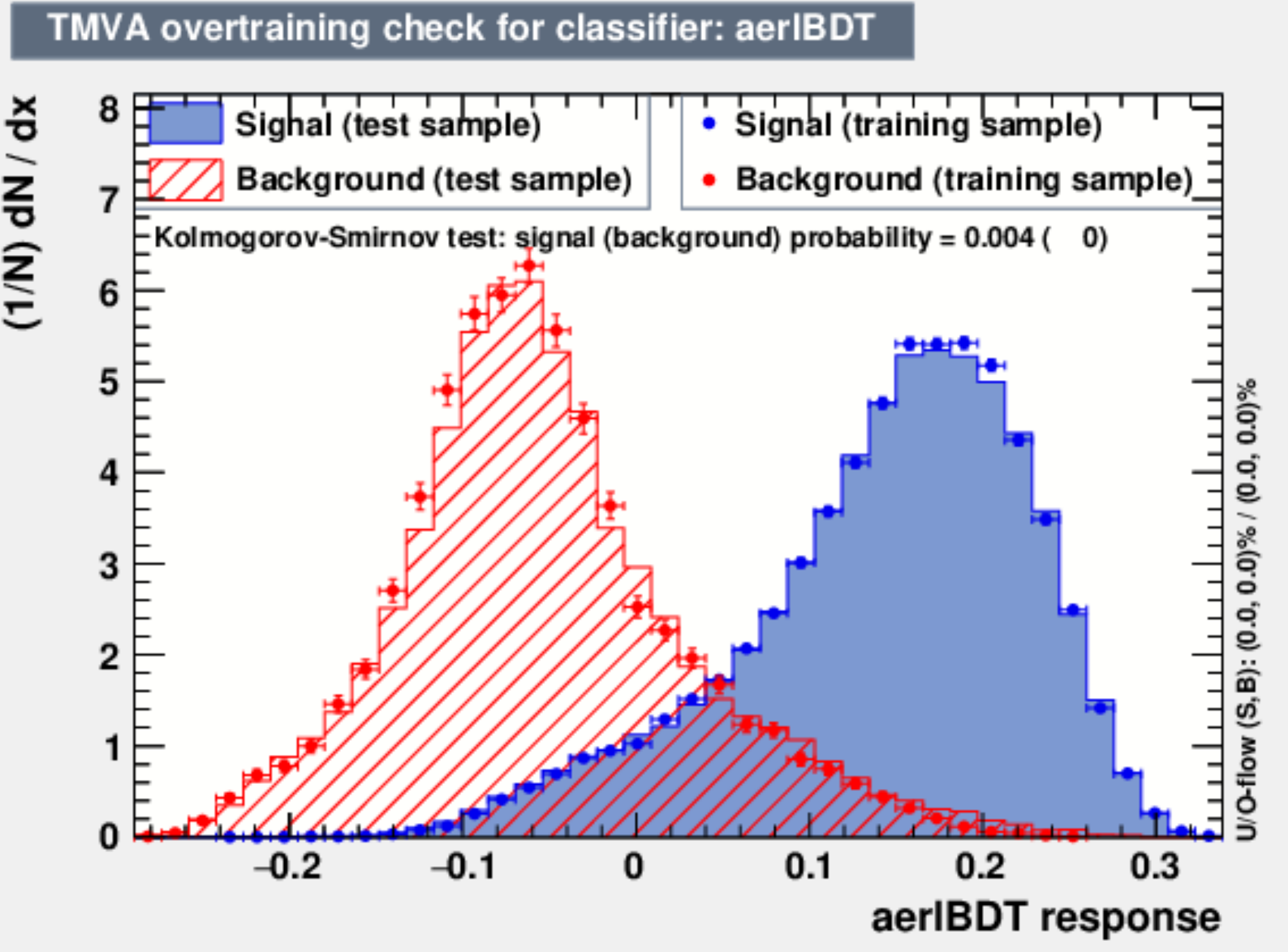}

\end{center}
\caption{BDT signal (left) and background (middle) inputs and correlations and outputs (right) for the hadron channel. From top to bottom are $e^{-}_{L}e^{+}_{R} \rightarrow q\bar{q}$ (106607),  $e^{-}_{R}e^{+}_{L} \rightarrow q\bar{q}$ (106608) , $e_{L} \gamma \rightarrow eZ,\nu W$ (37785) and $e_{R} \gamma \rightarrow eZ,\nu W$ (37786). In the BDT output distributions, test samples and training samples are plotted separately and show no evidence of overtraining.}
\label{fig:had1}
\end{figure}

\begin{figure}[p]
\begin{center}
\framebox{\textbf{Hadron Channel BDTs: 4f $e\nu W$ and $WW$ Backgrounds}}

\vspace{0.25in}

\includegraphics[height=1.65in]{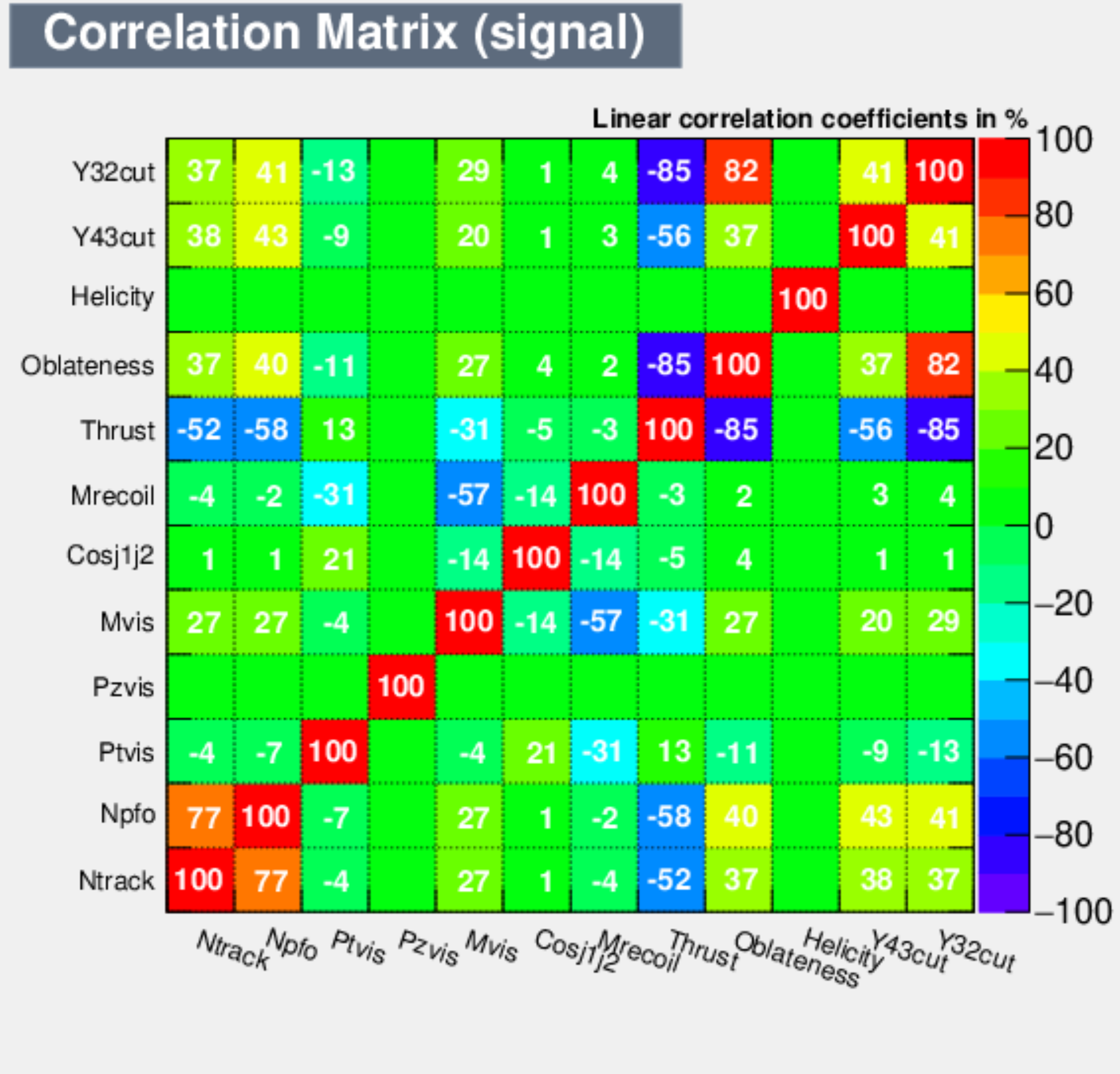}
\includegraphics[height=1.65in]{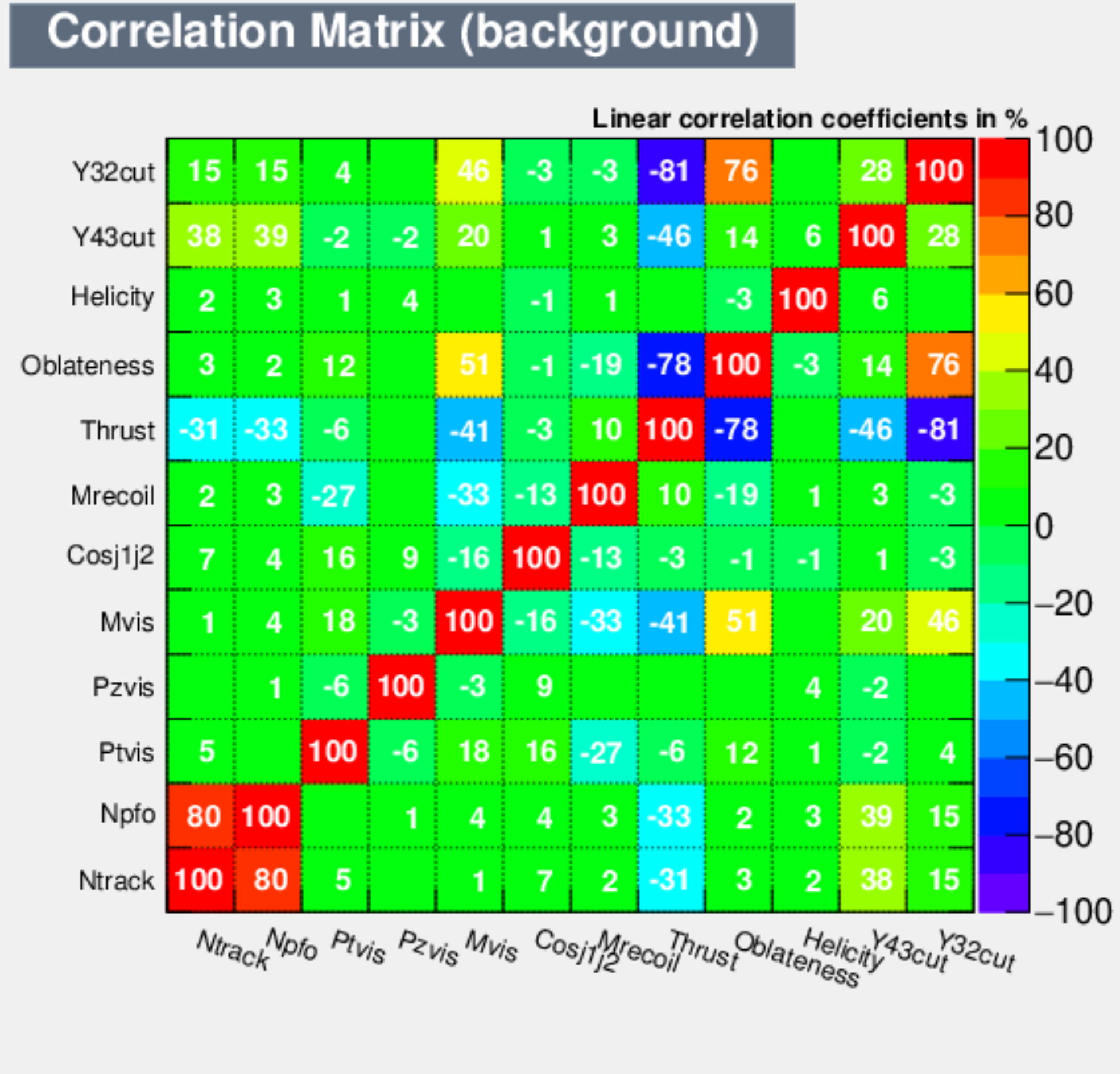}
\includegraphics[height=1.65in]{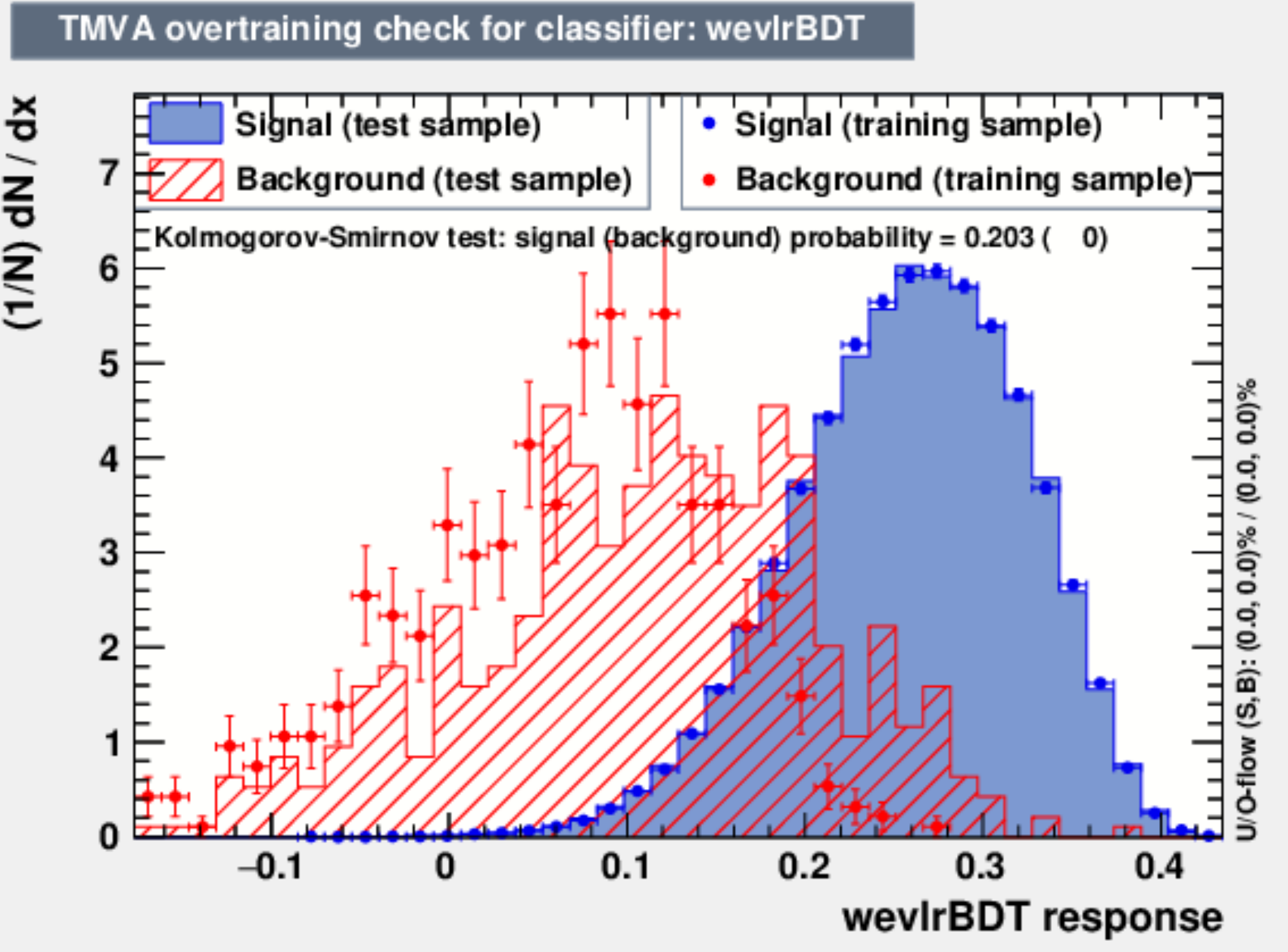}

\includegraphics[height=1.65in]{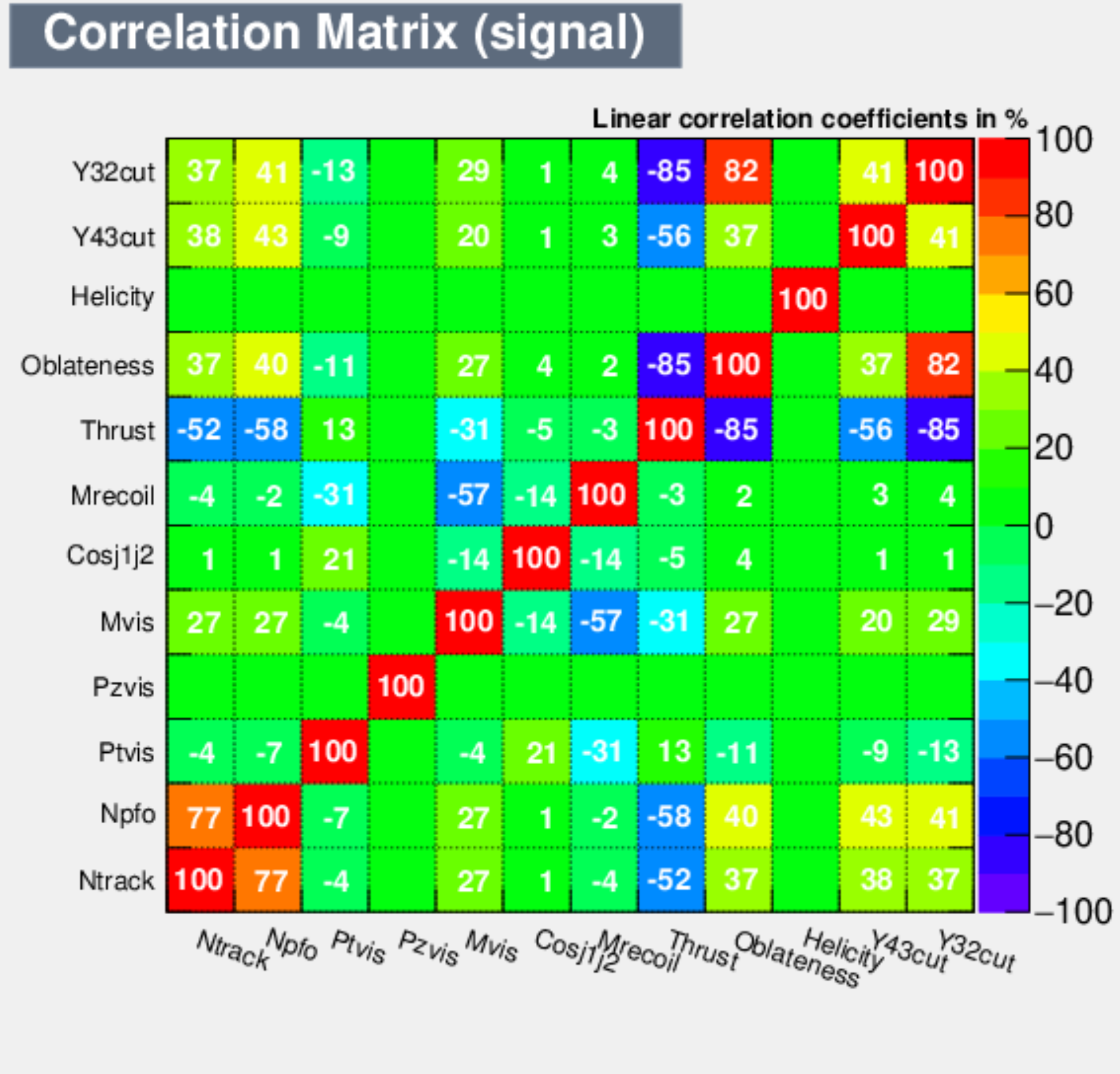}
\includegraphics[height=1.65in]{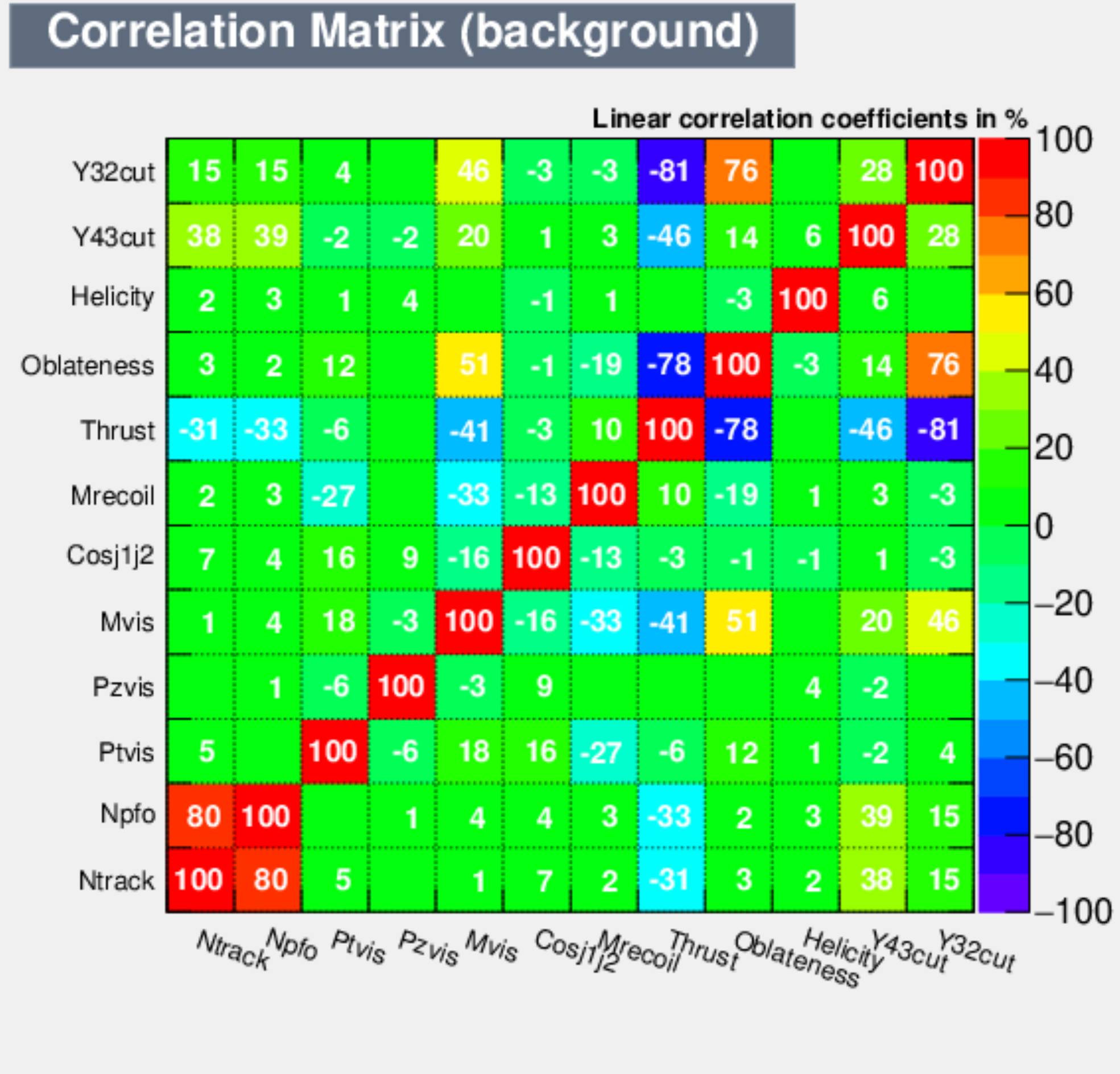}
\includegraphics[height=1.65in]{overtrain_wevlrBDT.eps.pdf}

\includegraphics[height=1.65in]{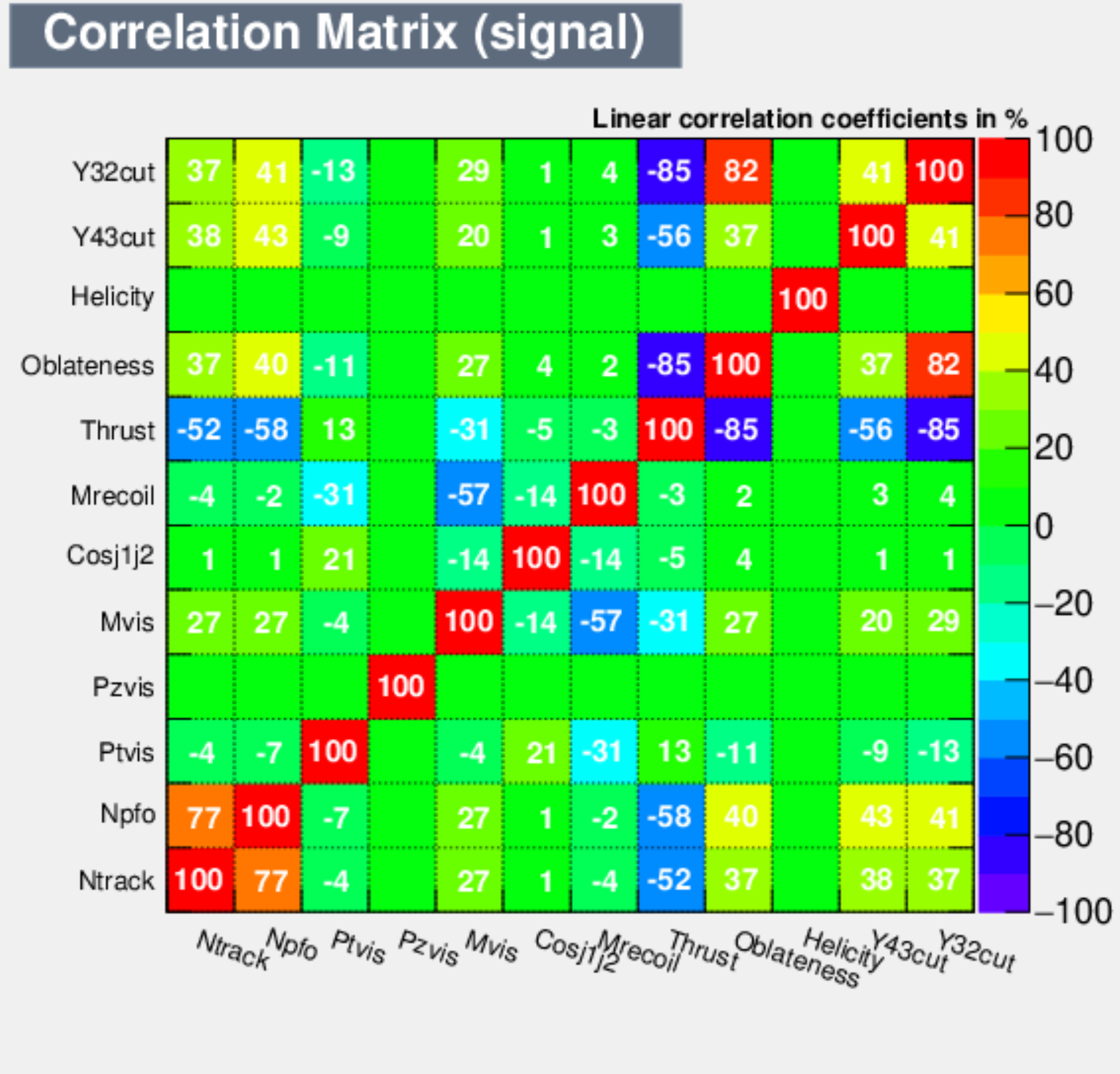}
\includegraphics[height=1.65in]{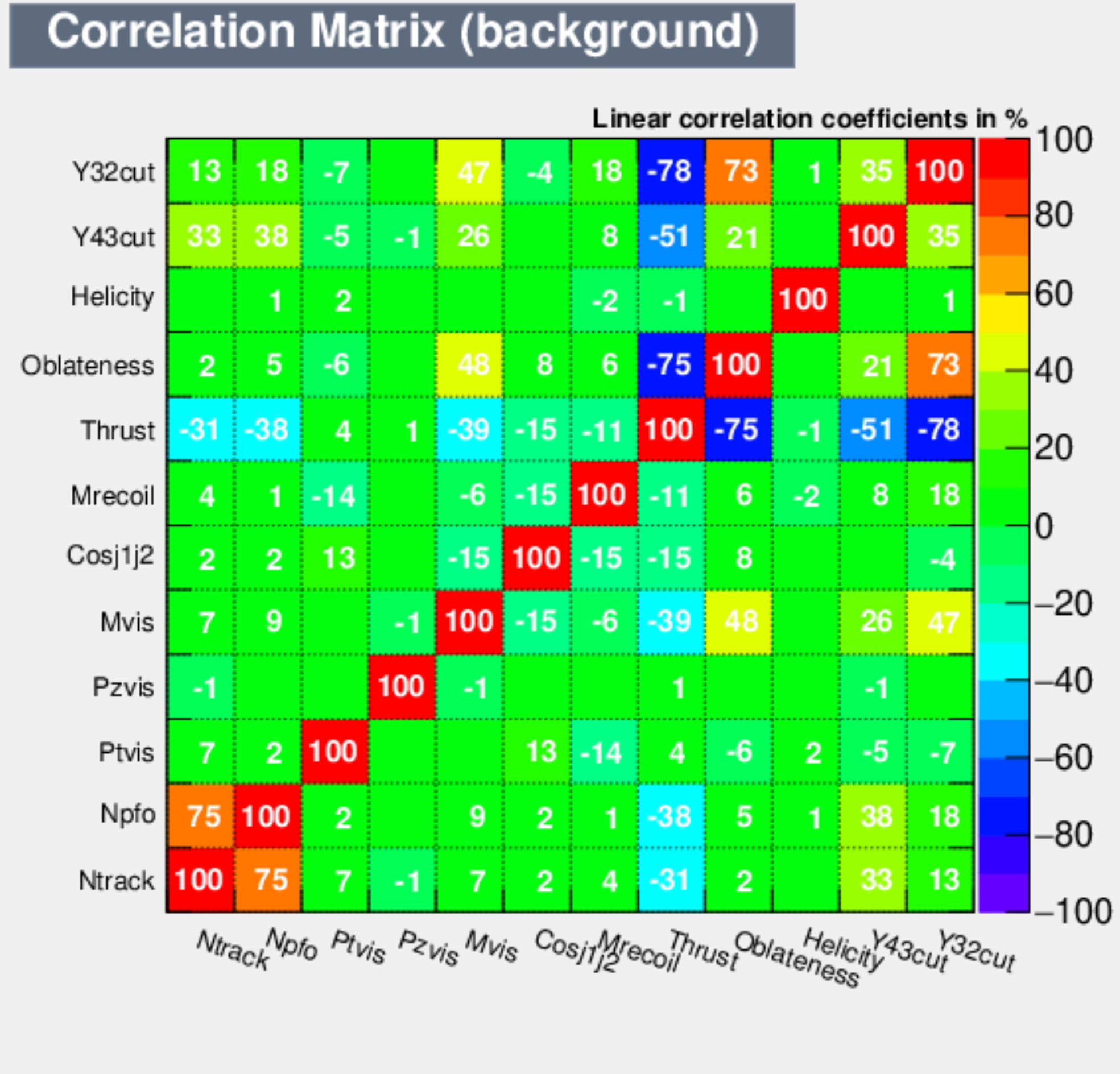}
\includegraphics[height=1.65in]{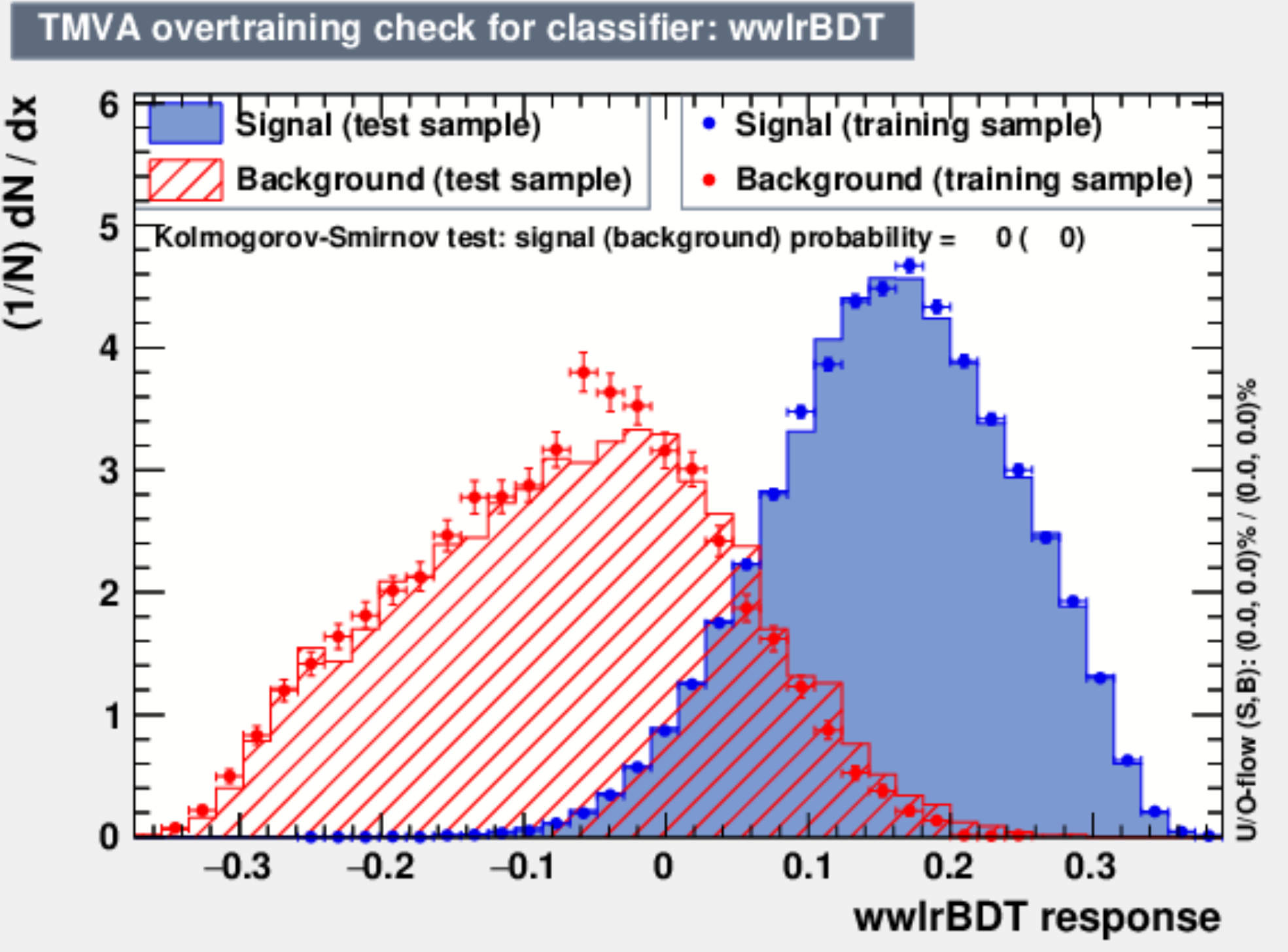}

\includegraphics[height=1.65in]{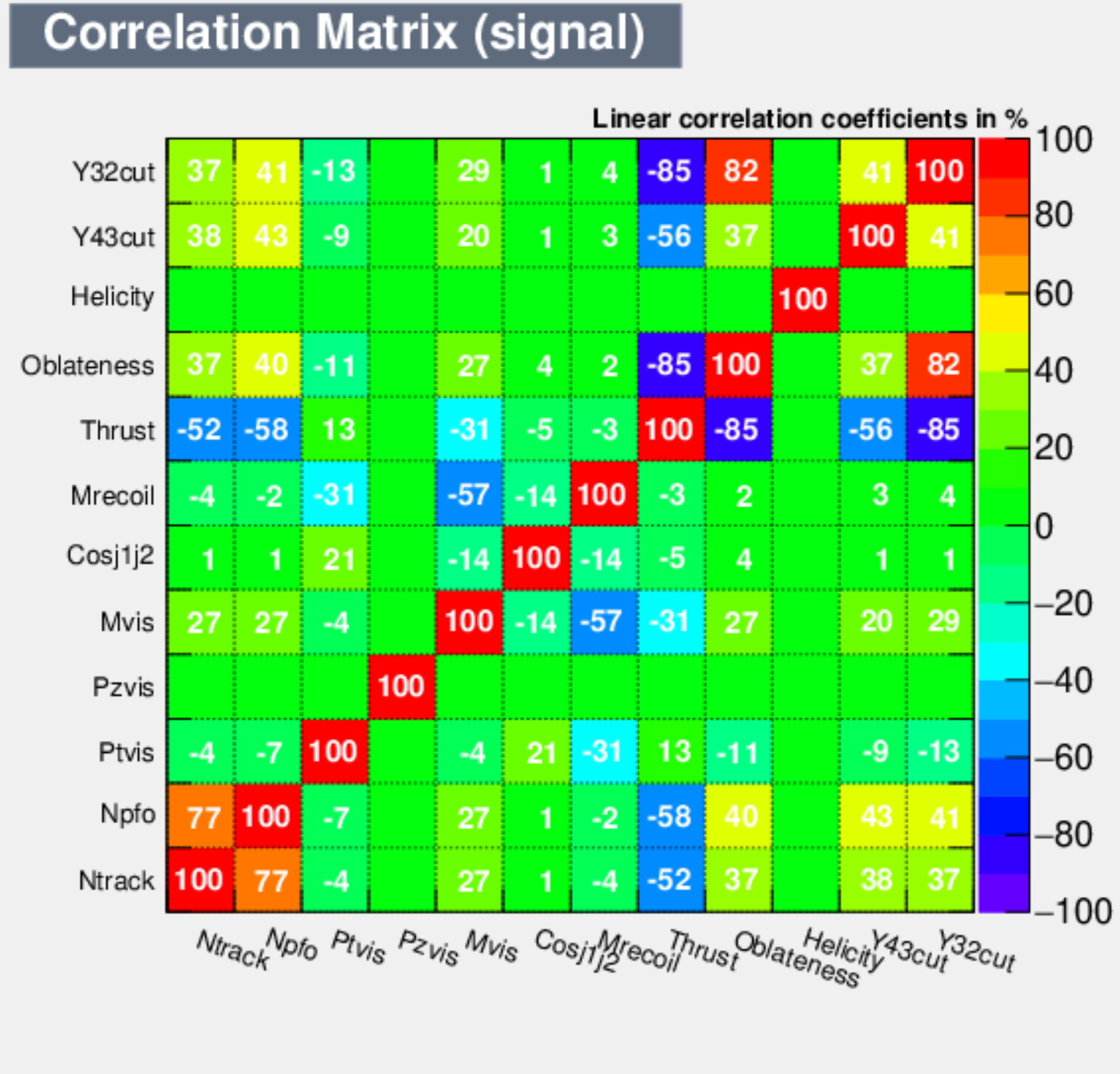}
\includegraphics[height=1.65in]{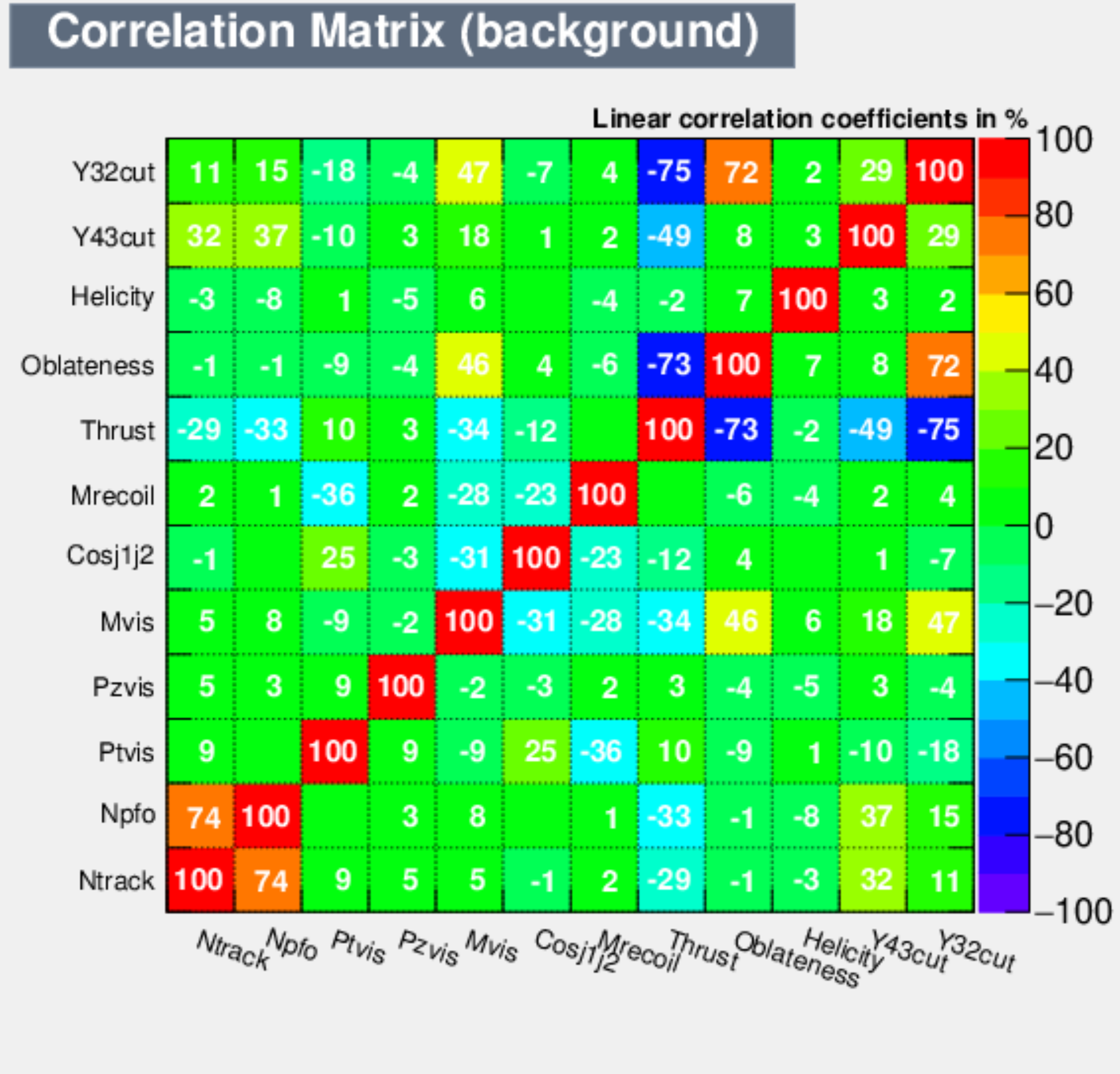}
\includegraphics[height=1.65in]{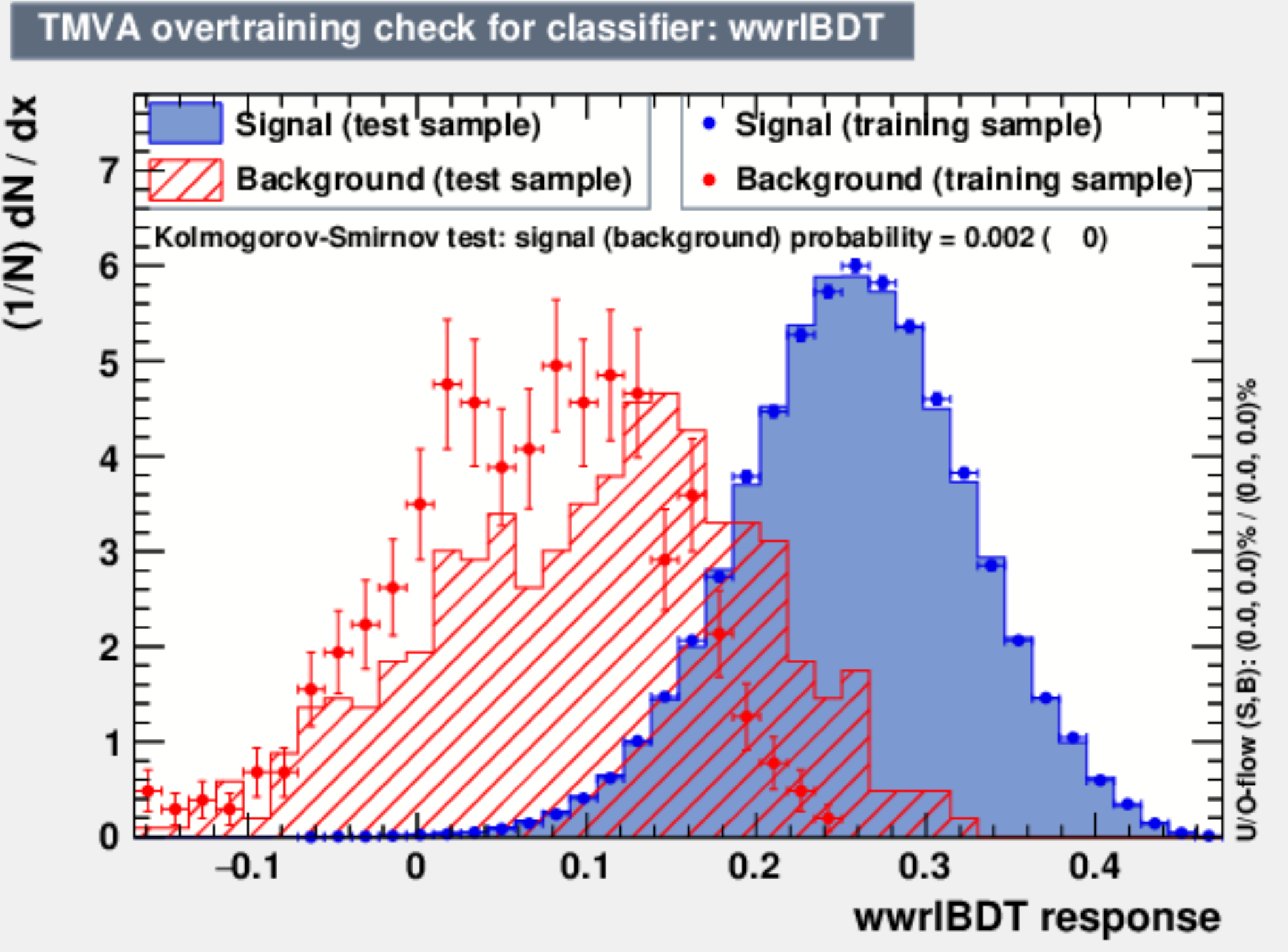}

\end{center}
\caption{BDT signal (left) and background (middle) BDT inputs and correlations and outputs (right) for the hadron channel. From top to bottom are $e^{-}_{L}e^{+}_{R} \rightarrow e \nu W$ (106564),  $e^{-}_{R}e^{+}_{L} \rightarrow e\nu W$ (106565), $e^{-}_{L}e^{+}_{R} \rightarrow WW$ (106577),  and $e^{-}_{R}e^{+}_{L} \rightarrow WW$ (106578). In the BDT output distributions, test samples and training samples are plotted separately and show no evidence of overtraining.}
\label{fig:had2}
\end{figure}

\begin{figure}[p]
\begin{center}
\framebox{\textbf{Hadron Channel BDTs: 4f $\nu \bar{\nu} Z$ and $ZZ$ Backgrounds}}

\vspace{0.25in}

\includegraphics[height=1.65in]{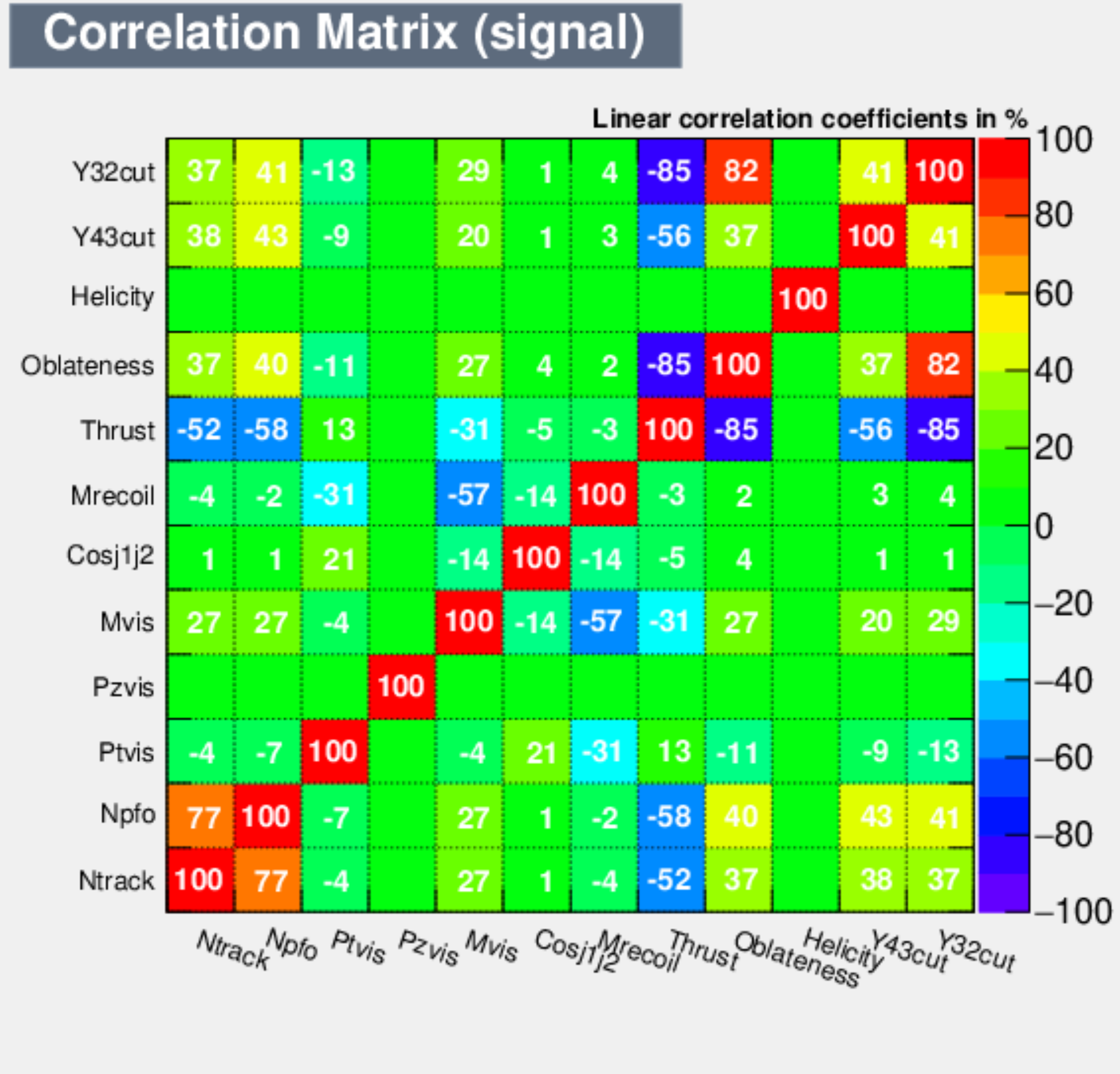}
\includegraphics[height=1.65in]{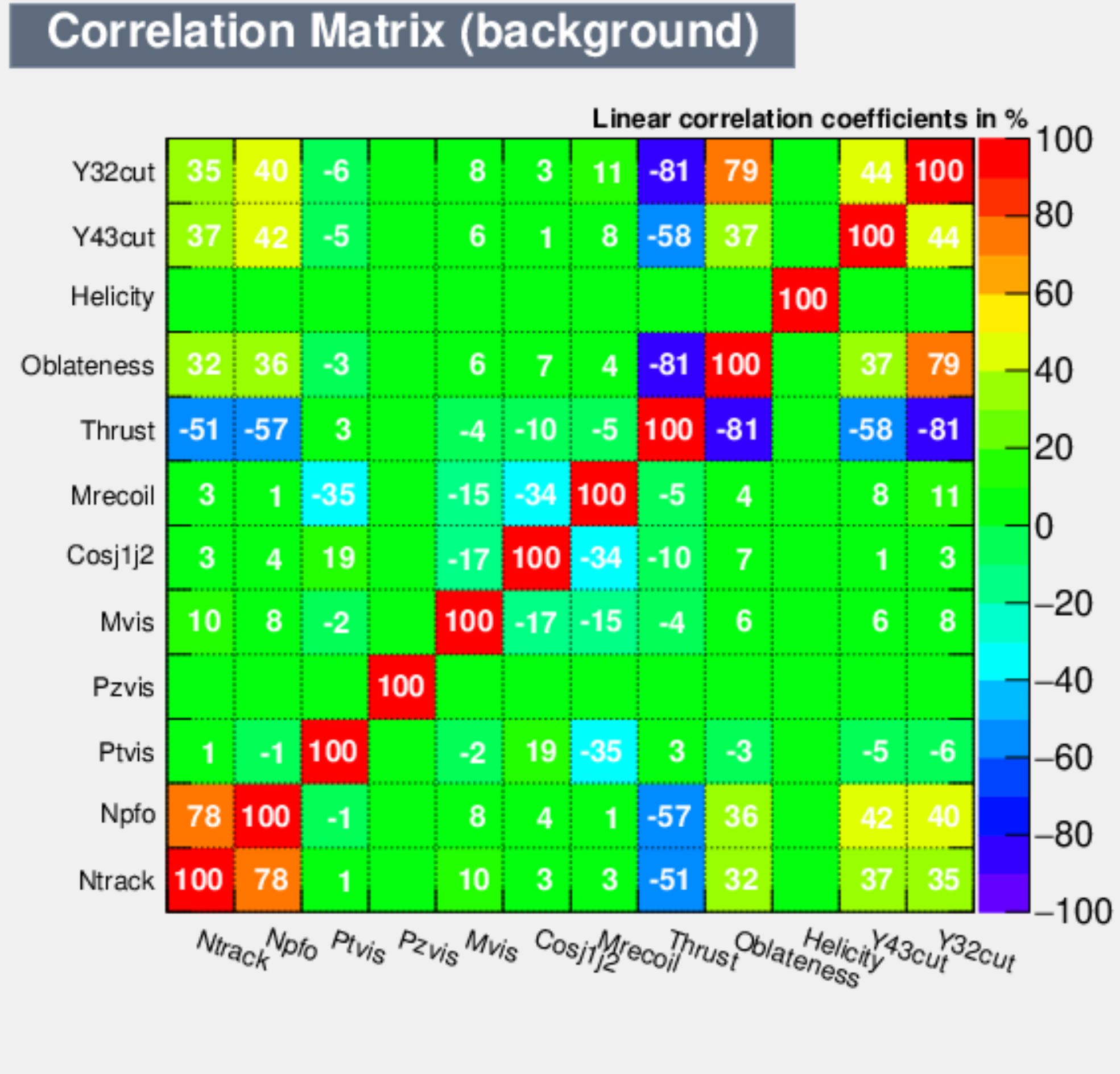}
\includegraphics[height=1.65in]{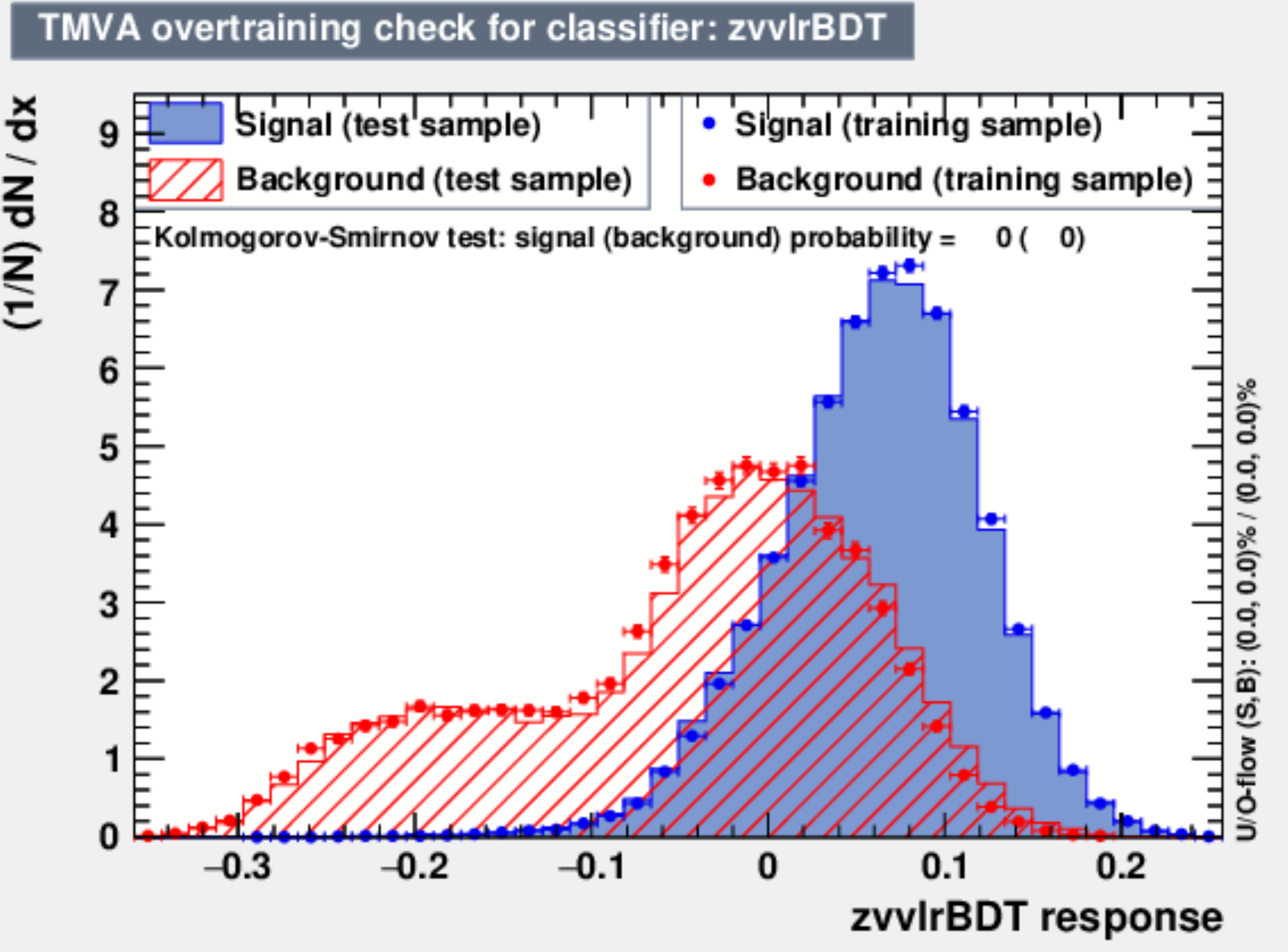}

\includegraphics[height=1.65in]{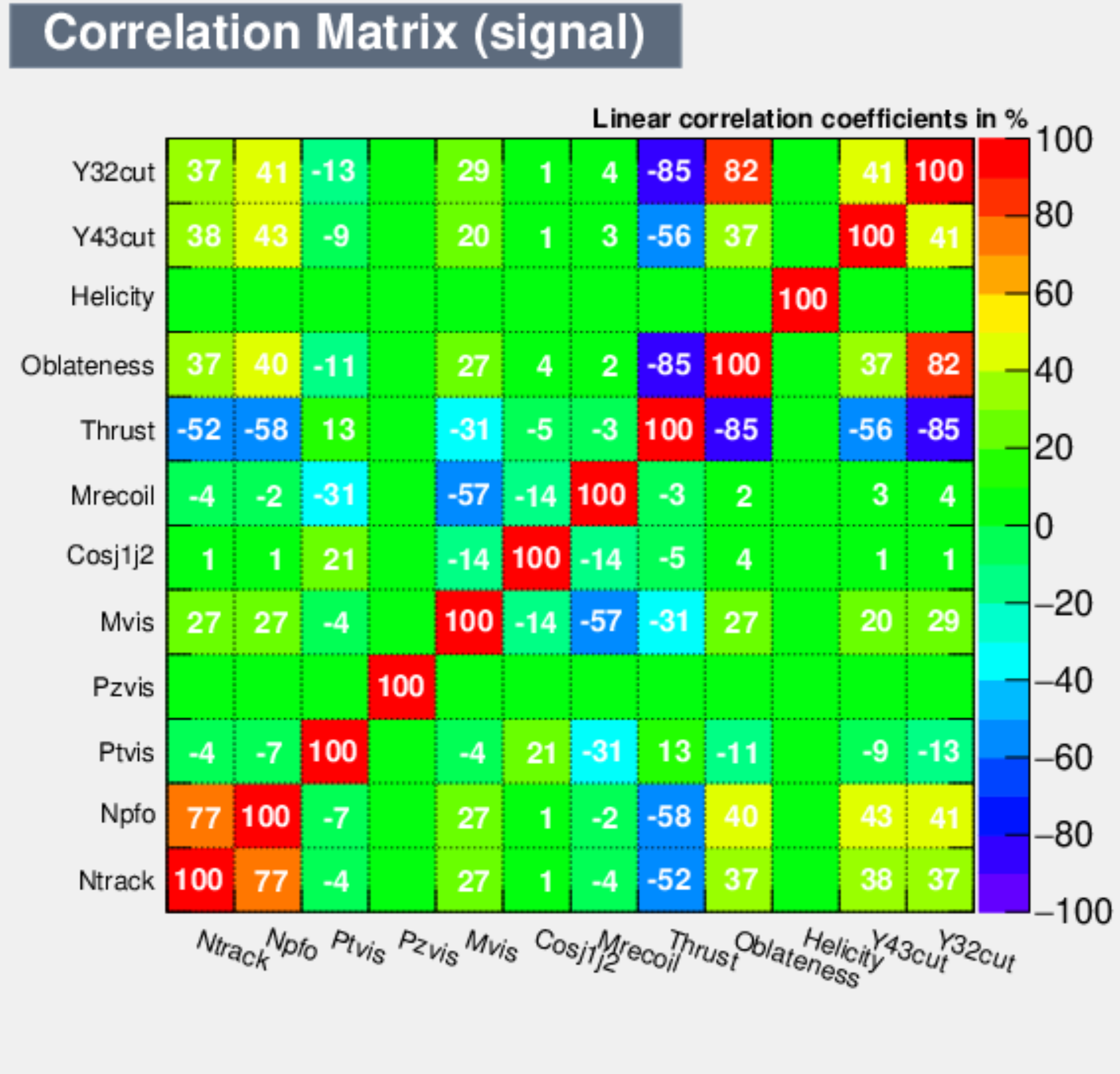}
\includegraphics[height=1.65in]{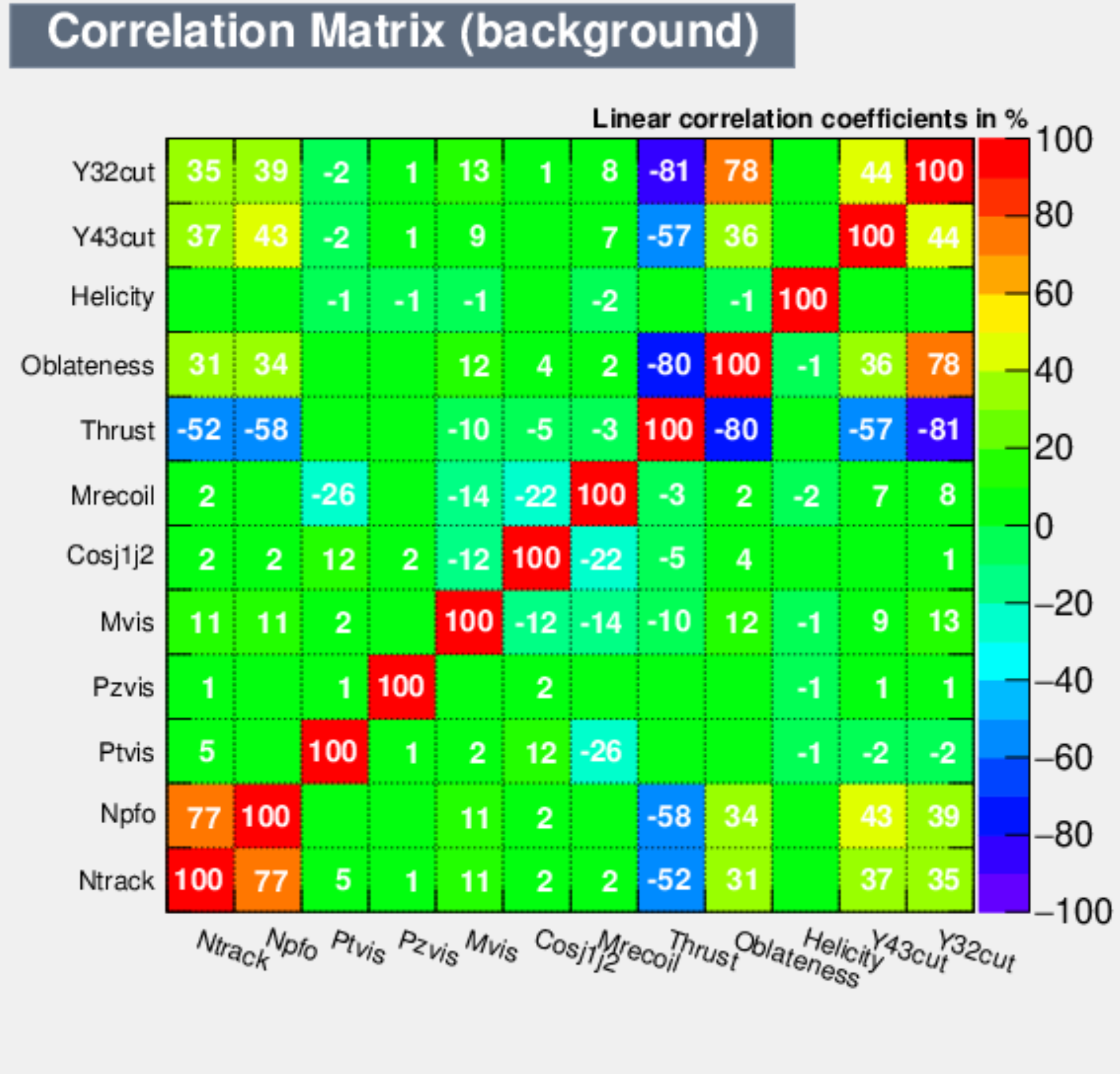}
\includegraphics[height=1.65in]{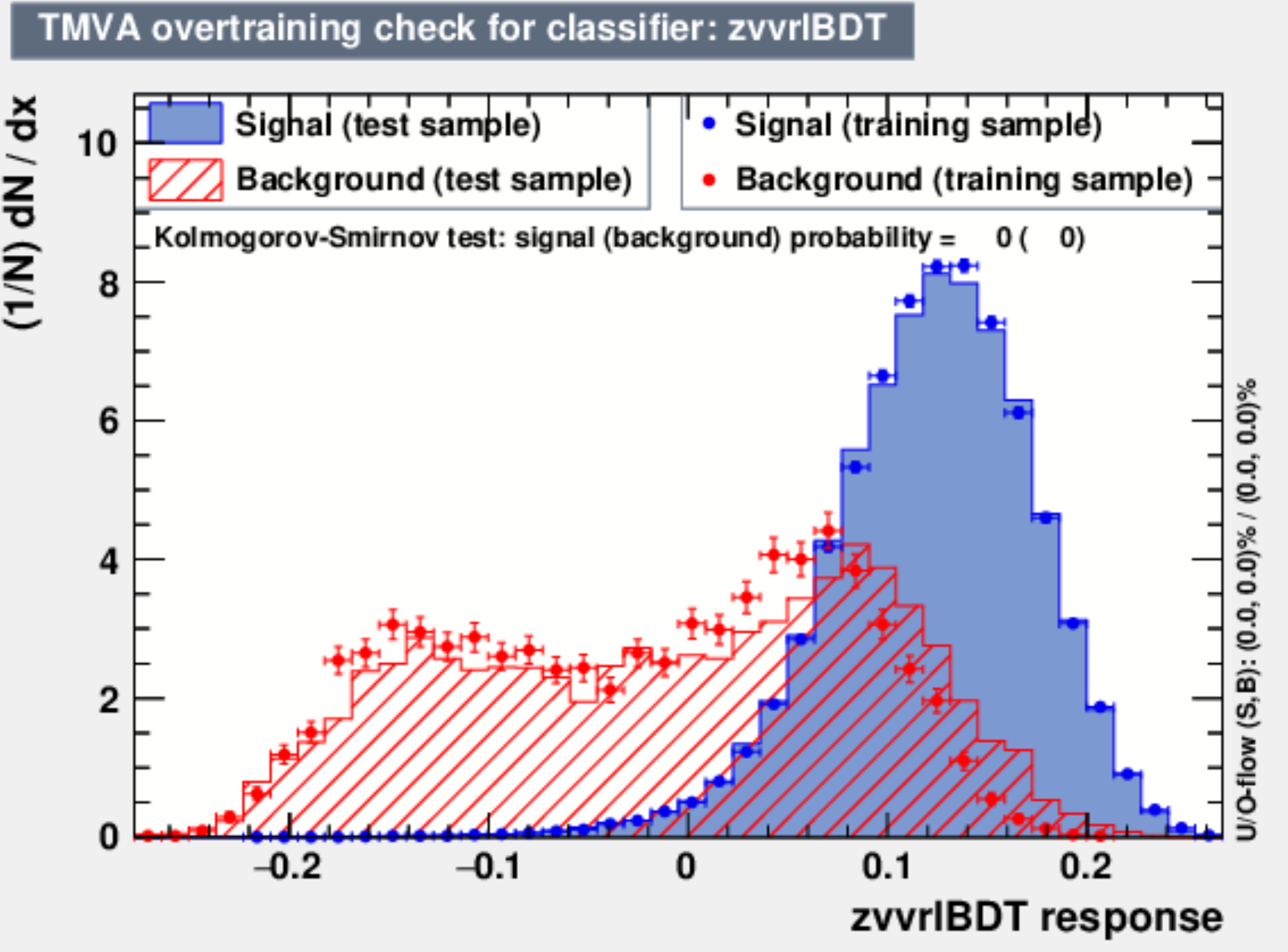}

\includegraphics[height=1.65in]{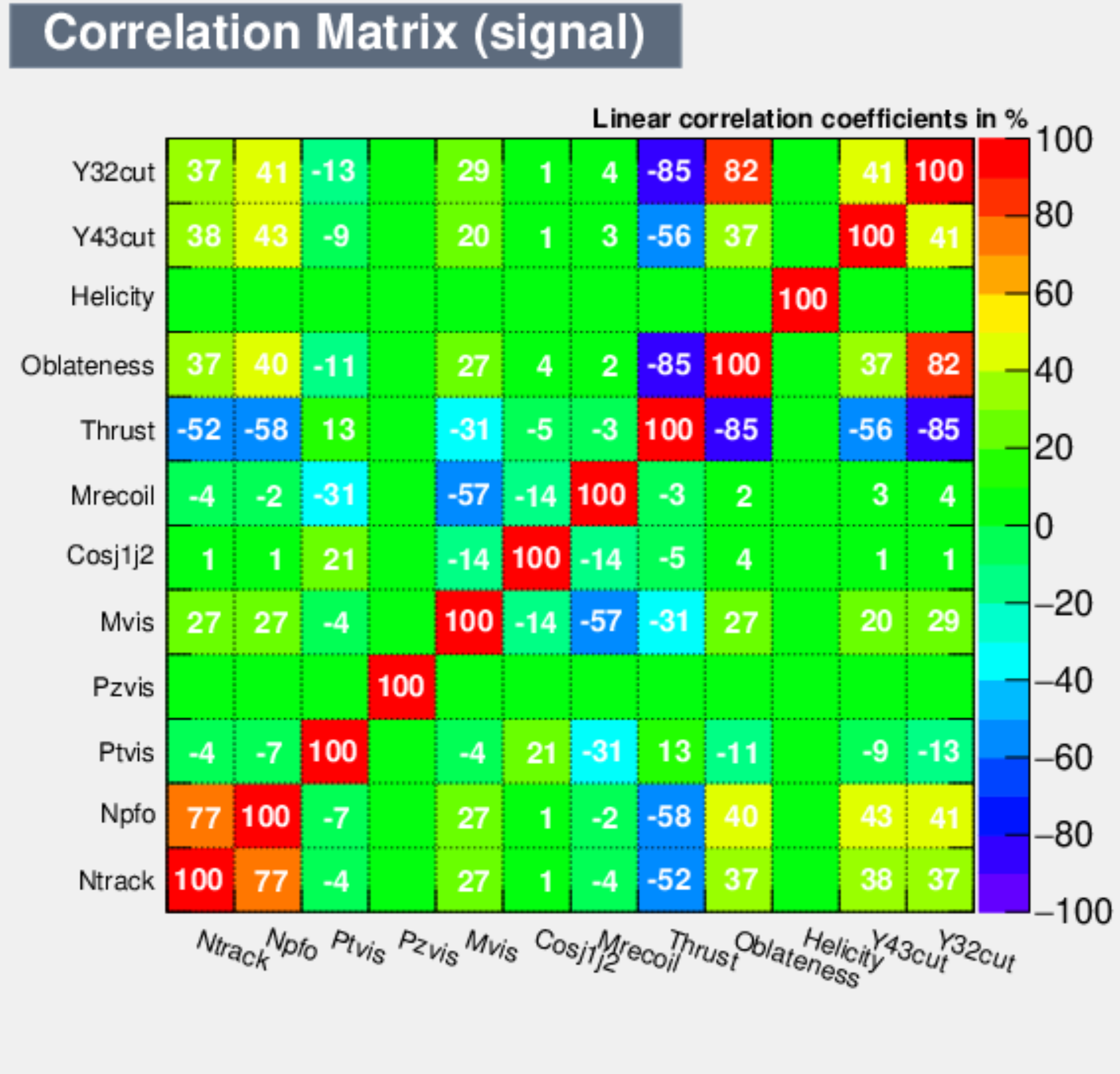}
\includegraphics[height=1.65in]{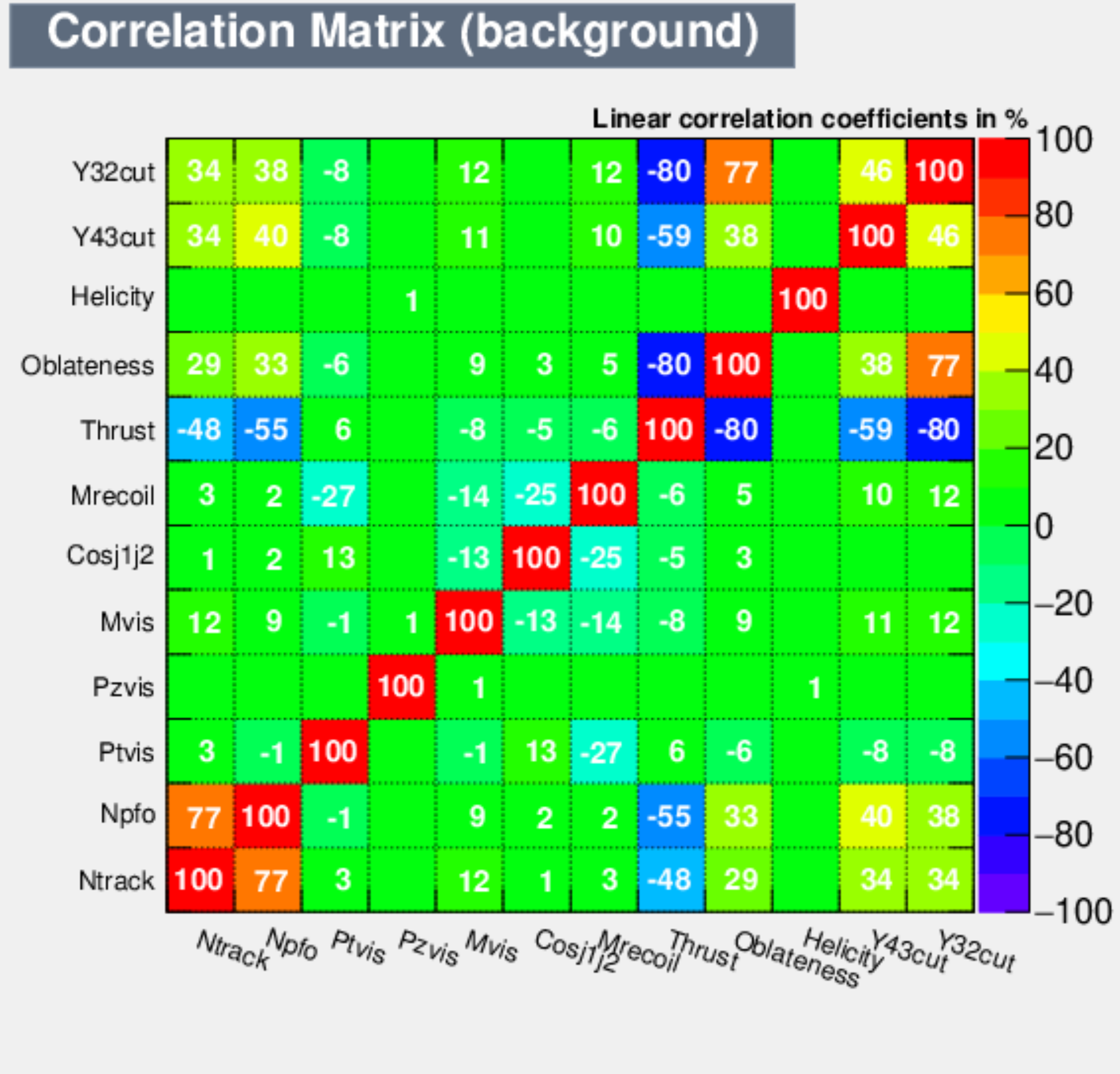}
\includegraphics[height=1.65in]{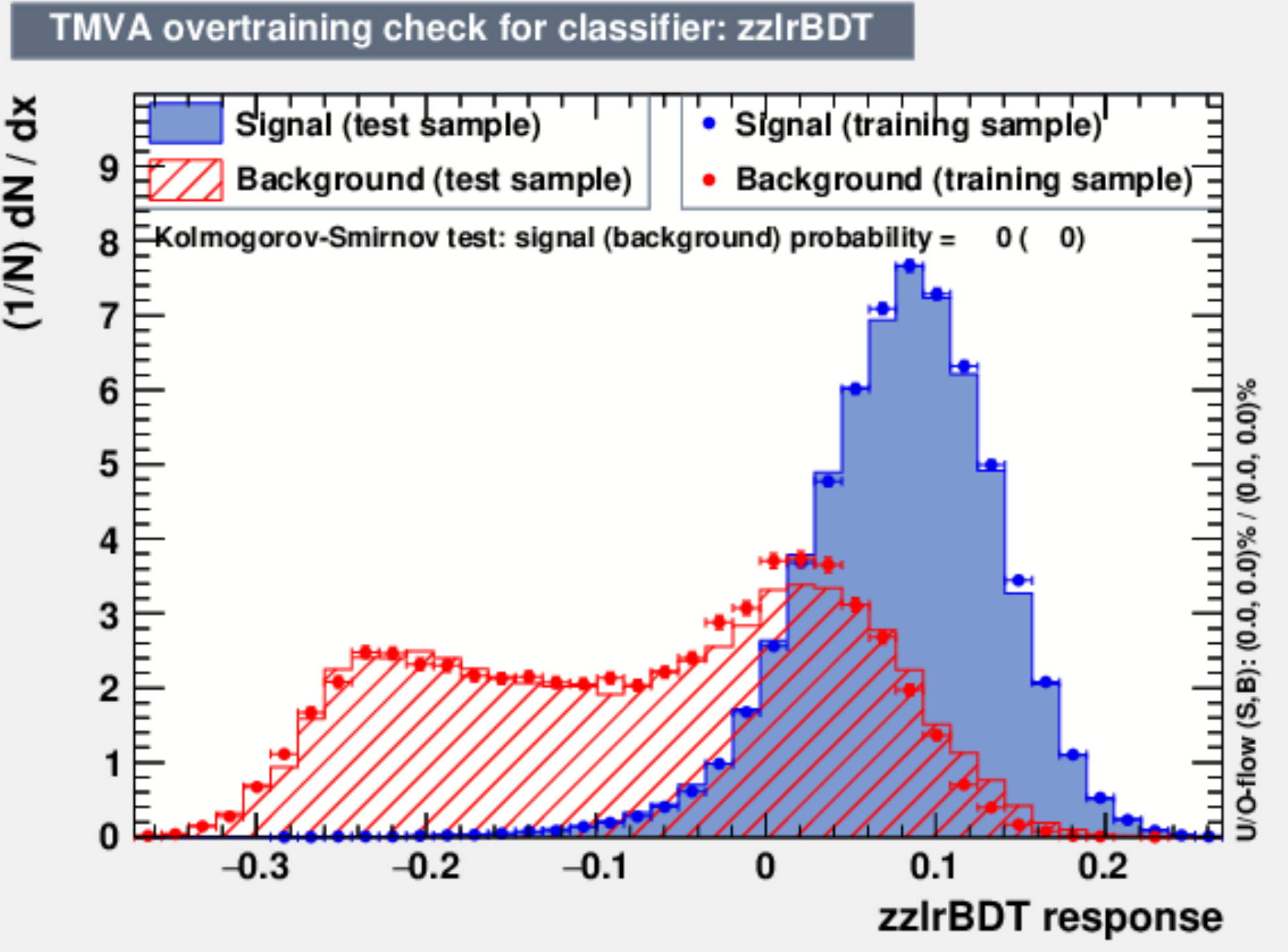}

\includegraphics[height=1.65in]{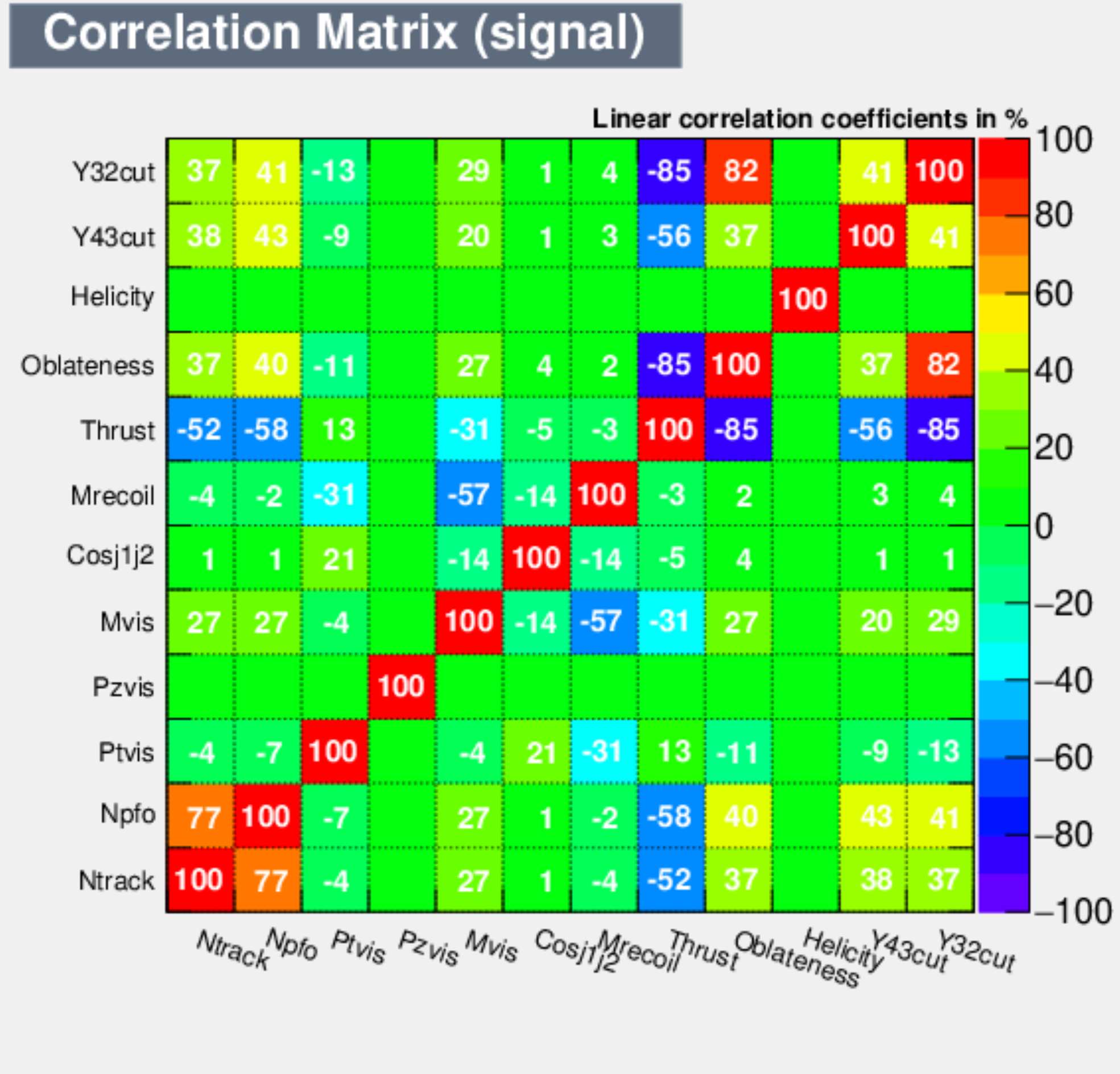}
\includegraphics[height=1.65in]{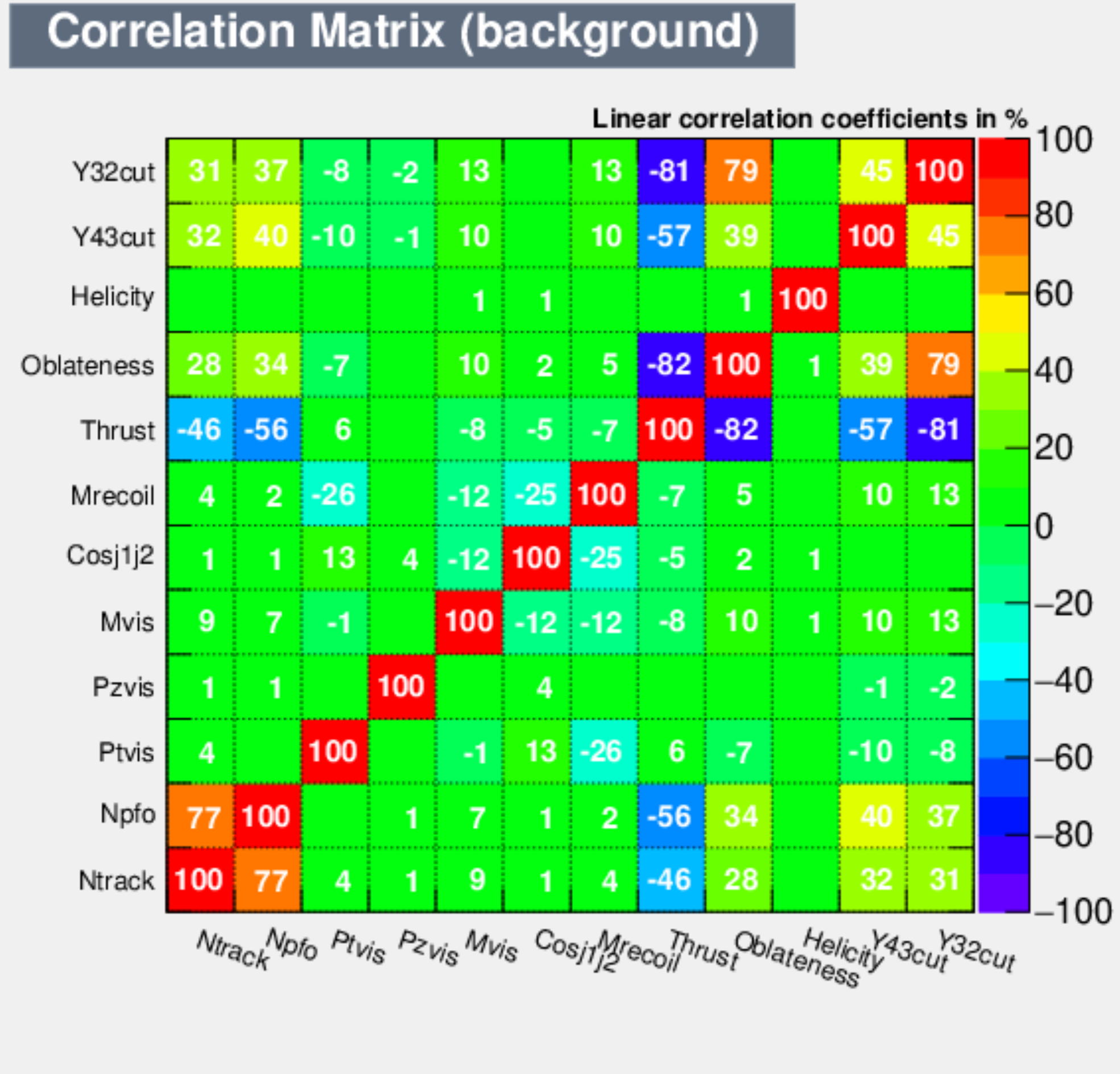}
\includegraphics[height=1.65in]{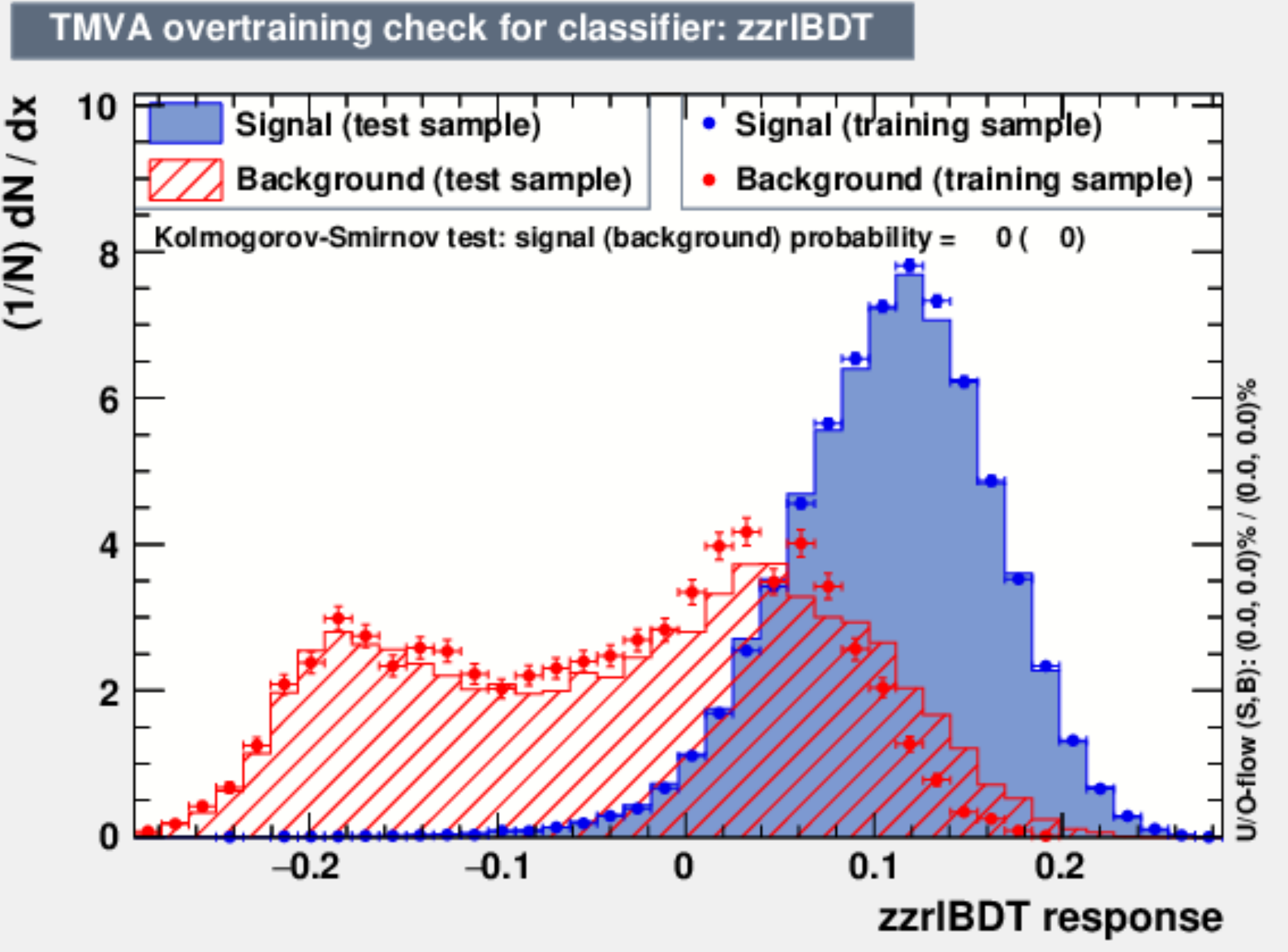}

\end{center}
\caption{BDT signal (left) and background (middle) inputs and correlations and outputs (right) for the hadron channel. From top to bottom are $e^{-}_{L}e^{+}_{R} \rightarrow \nu \nu Z$ (106571),  $e^{-}_{R}e^{+}_{L} \rightarrow \nu \nu Z$ (106572), $e^{-}_{L}e^{+}_{R} \rightarrow ZZ$ (106575),  and $e^{-}_{R}e^{+}_{L} \rightarrow ZZ$ (106576). In the BDT output distributions, test samples and training samples are plotted separately and show no evidence of overtraining.}
\label{fig:had3}
\end{figure}

Collected in this appendix are the BDT signal and background input variable correlations and the BDT output distributions. Figure \ref{fig:bdtbdt} shows these for the BDT of BDTs described in the text, Figure \ref{fig:lep} shows these for the electron and muon channel BDT, and Figures \ref{fig:had1}, \ref{fig:had2} and \ref{fig:had3} show these for the hadron channel BDT.

\end{document}